\documentclass[twocolumn,english,aps,prb,floatfix,amssymb,superscriptaddress,a4paper]{revtex4-1}
\usepackage{lipsum}
\usepackage{amsmath}
\usepackage{amssymb}
\usepackage{graphicx}
\usepackage{physics}
\usepackage{hyperref}
\usepackage[dvipsnames]{xcolor}
\usepackage{float}
\usepackage[caption=false]{subfig}
\hypersetup{
	colorlinks=true,
	hypertexnames=true
}

\newcommand{\Cbar}{\bar{C}}
\newcommand{\epin}{e_{\mathrm{pin}}}

\newcommand{\fpin}{f_{\mathrm{pin}}}

\newcommand{\alphaSP}{\alpha_\mathrm{sp}}

\newcommand{\vTH}{v_{\mathrm{th}}}

\newcommand{\teq}{t_\mathrm{eq}}

\newcommand{\xmjp}{x_-^{\mathrm{jp}}}
\newcommand{\xpjp}{x_+^{\mathrm{jp}}}

\newcommand{\Udp}{U_\mathrm{dp}}
\newcommand{\Up}{U_\mathrm{p}}
\newcommand{\omegaf}{\omega_\mathrm{f}}
\newcommand{\omegap}{\omega_\mathrm{p}}

\begin{document}

\title{Creep effects on the Campbell response in type II superconductors}

\author{Filippo Gaggioli}
\affiliation
{Institut f\"ur Theoretische Physik, ETH Z\"urich,
CH-8093 Z\"urich, Switzerland}

\author{Gianni Blatter}
\affiliation
{Institut f\"ur Theoretische Physik, ETH Z\"urich,
CH-8093 Z\"urich, Switzerland}

\author{Vadim B. Geshkenbein}
\affiliation
{Institut f\"ur Theoretische Physik, ETH Z\"urich,
CH-8093 Z\"urich, Switzerland}

\date{\today}

 \begin{abstract}
Applying the strong pinning formalism to the mixed state of a type II
superconductor, we study the effect of thermal fluctuations (or creep) on the
penetration of an ac magnetic field as quantified by the so-called Campbell
length $\lambda_{\rm \scriptscriptstyle C}$.  Within strong pinning theory,
vortices get pinned by individual defects, with the jumps in the pinning
energy ($\Delta e_\mathrm{pin}$) and force ($\Delta f_\mathrm{pin}$) between
bistable pinned and free states quantifying the pinning process.  We find that
the evolution of the Campbell length $\lambda_{\rm \scriptscriptstyle C}(t)$
as a function of time $t$ is the result of two competing effects, the change
in the force jumps $\Delta f_\mathrm{pin}(t)$ and a change in the trapping
area $S_\mathrm{trap}(t)$ of vortices; the latter describes the area around
the defect where a nearby vortex gets and remains trapped.  Contrary to
naive expectation, we find that during the decay of the critical state in a
zero-field cooled (ZFC) experiment, the Campbell length $\lambda_{\rm
\scriptscriptstyle C}(t)$ is usually nonmonotonic, first decreasing with time
$t$ and then increasing for long waiting times. Field cooled (FC) experiments
exhibit hysteretic effects in $\lambda_{\rm \scriptscriptstyle C}$;
relaxation then turns out to be predominantly monotonic, but its
magnitude and direction depends on the specific phase of the cooling--heating
cycle.  Furthermore, when approaching equilibrium, the Campbell length relaxes
to a finite value, different from the persistent current which vanishes at
long waiting times $t$, e.g., above the irreversibility line.  Finally,
measuring the Campbell length $\lambda_{\rm \scriptscriptstyle C}(t)$ for
different states, zero-field cooled, field cooled, and relaxed, as a function
of different waiting times $t$ and temperatures $T$, allows to
‘spectroscopyse’ the pinning potential of the defects.
\end{abstract}

\maketitle


\section{Introduction}

The phenomenological properties of type II superconductors subject to a
magnetic field $B$ are determined by vortices, linear topological defects that
guide the field through the material in terms of quantized fluxes
\cite{Abrikosov_1957} $\Phi_0 = hc/2e$.  The interaction of these flux tubes
with material defects has a decisive impact on the material's properties, as it
determines the amount of current density that the superconductor can transport
free of dissipation. This phenomenon, known under the name of vortex pinning
\cite{CampbellEvetts_2006}, has been studied extensively, both in theory and
experiment, as one important facet of vortex matter physics
\cite{Tinkham_2004, DeGennes_1999}.  Traditional tools to characterize vortex
pinning are measurements of critical current densities $j_c$ and full
current--voltage ($j$--$V$) characteristics \cite{CampbellEvetts_2006}, as
well as the penetration of a small ac magnetic test-field that is quantified
through the so-called Campbell penetration length \cite{Campbell_1969}
$\lambda_{\rm \scriptscriptstyle C}$. In characterizing the pinning properties
of the material's mixed state, the Campbell length $\lambda_{\rm
\scriptscriptstyle C}$ assumes a similar role as the skin depth $\delta$
(determining the resistivity $\rho$ in the normal state) or the London
penetration depth $\lambda_{\rm\scriptscriptstyle L}$ (determining the
superfluid density $\rho_s$ in zero magnetic field). Although well established
as an experimental tool, a quantitative calculation on the basis of strong
pinning theory \cite{Labusch_1969,Larkin_1979,Koopmann_2004} of the ac
Campbell response has been given only recently \cite{Willa_2015_PRL}.  In this
paper, we extend this `microscopic' description of the Campbell penetration to
include effects of thermal fluctuations, i.e., creep.

The penetration of an ac magnetic field into the mixed state of a
superconductor has been first analyzed by Campbell \cite{Campbell_1969}, see
also Refs.\ \onlinecite{Lowell_1972} and \onlinecite{Campbell_1978}. This
phenomenological theory relates the Campbell length $\lambda_{\rm
\scriptscriptstyle C} \propto B/\sqrt{\alpha}$ to the curvature $\alpha$ of
the pinning potential that is probed by small-amplitude oscillations of the
vortices. Applying strong pinning theory to this problem provides a lot of
insights on the pinning landscape: Within the strong pinning paradigm,
vortices exhibit bistable configurations in the presence of a defect.  These
bistable configurations describe pinned and unpinned (meta-)stable states, see
Fig.\ \ref{fig:geometry}, with a finite pinning force density $F_\mathrm{pin}$
resulting from an asymmetric occupation of the corresponding branches. While
the critical current density $j_c \propto \Delta e_\mathrm{pin}$ is determined
by the jumps in energy $\Delta e_\mathrm{pin}$ between pinned and unpinned
states at (de)pinning \cite{Labusch_1969,Larkin_1979,Koopmann_2004}, it turns
out \cite{Willa_2015_PRL, Willa_2016} that the Campbell length $\lambda_{\rm
\scriptscriptstyle C} \propto 1/\sqrt{\Delta f_\mathrm{pin}}$ is given by the
jumps in the pinning force $\Delta f_\mathrm{pin}$. Interestingly, the
relevant jumps $\Delta f_\mathrm{pin}$ determining $\lambda_{\rm
\scriptscriptstyle C}$ depend on the vortex state, e.g., the critical (or
zero-field cooled, ZFC) state first defined by Bean \cite{Bean_1962} or the
field cooled (FC) state. Even more, the pinned vortex state depends on the
time-trace of its experimental implementation, that leads to hysteretic
effects in $\lambda_{\rm \scriptscriptstyle C}$ as shown in Refs.\
\onlinecite{Willa_2015_PRL, Willa_2016} both theoretically and experimentally.

When including thermal fluctuations in the calculation of the pinning force
density $F_\mathrm{pin}$, different jumps $\Delta e_\mathrm{pin}(t)$ in the
pinning energy become relevant that depend on the time $t$ evolution of the
vortex state due to creep.  While this relaxational time dependence leads to
the decay of the persistent current density $j(t)$, the corresponding velocity
dependence leads to a rounding \cite{Buchacek_2018,Buchacek_2019} of the
transition \cite{Thomann_2012} between pinned and dissipative states in the
current--voltage characteristic; again the quantitative nature of the strong
pinning description allows for a detailed comparison of the
temperature-shifted and rounded excess-current characteristic predicted by
theory with experimental data on superconducting films
\cite{Buchacek_2019_exp}.

In the present paper, we determine the Campbell length $\lambda_{\rm
\scriptscriptstyle C}(t)$ including the effect of thermal fluctuations.  We
determine the relevant jumps $\Delta f_\mathrm{pin}(t)$ in the pinning force
which depend on the time $t$ during which the original, e.g., critical, state
has relaxed due to creep.  For this ZFC situation, we find that the evolution
of the Campbell length $\lambda_{\rm \scriptscriptstyle C}(t)$ is the result
of two competing effects, the change in the force jumps $\Delta
f_\mathrm{pin}(t)$ and, furthermore, an increase in the trapping area
$S_\mathrm{trap}(t)$ of vortices; the latter describes the area around the
defect where a nearby vortex gets and remains trapped, see Fig.\
\ref{fig:geometry}.  Contrary to expectation, we find that for intermediate
and very strong pinning, the Campbell length $\lambda_{\rm \scriptscriptstyle
C}(t)$ first {\it decreases} with time $t$ and then starts increasing for long
waiting times; at marginally strong pinning, we find $\lambda_{\rm
\scriptscriptstyle C}(t)$ decreasing.

Relaxation also appears for the case of field cooled states, as these
drop out of equilibrium upon changing temperature and relax when the cooling
or heating process is interrupted.  In a FC experiment, the Campbell
penetration exhibits hysteretic phenomena in a cooling--heating cycle.  The
relaxation of the Campbell length $\lambda_{\rm \scriptscriptstyle C}$ then
depends on the type of pinning and the location within the hysteresis loop. At
intermediate and large pinning, we find three different phases, one where
$\lambda_{\rm \scriptscriptstyle C}$ monotonously decreases with time, one
where it monotonously increases towards equilibrium, and a third phase where
relaxation is slow due to large creep barriers; at marginally strong pinning,
we find $\lambda_{\rm \scriptscriptstyle C}$ mainly decreasing in time.
Several of these findings have been observed in experiments
\cite{Prozorov_2003, Pasquini_2005} and we will discuss them below. We
conclude that measuring the Campbell length $\lambda_{\rm \scriptscriptstyle
C}$ for different states, ZFC, FC, and relaxed as a function of different
waiting times $t$, provides insights into the pinning mechanism and gives
access to the jumps $\Delta f_\mathrm{pin}$ at different locations of the
pinning curve, see Fig.\ \ref{fig:strong_pinning}(c); such measurements and
analysis then allow to `spectroscopyse' the pinning potential of defects.

In the following, we first recapitulate relevant aspects of the strong pinning
theory and of creep (Secs.\ \ref{sec:strong_pinning} and \ref{sec:transport}
that focuses on transport) and then proceed with the calculation of the
Campbell penetration depth $\lambda_{\rm \scriptscriptstyle C}$ in the
presence of thermal fluctuations, see Sec.\ \ref{sec:ac}. We first focus on
the critical (or zero-field cooled, ZFC) state in Secs.\ \ref{sec:ZFC} and
\ref{sec:creep_l_C_th} and then extend the analysis to the case of field
cooling (FC) in Secs.\ \ref{sec:FC} and \ref{sec:creep_FC}.  Section
\ref{sec:sum} summarizes and concludes our work.

\section{Strong pinning theory}\label{sec:strong_pinning}

\begin{figure}[t]
        \includegraphics[width = 1.\columnwidth]{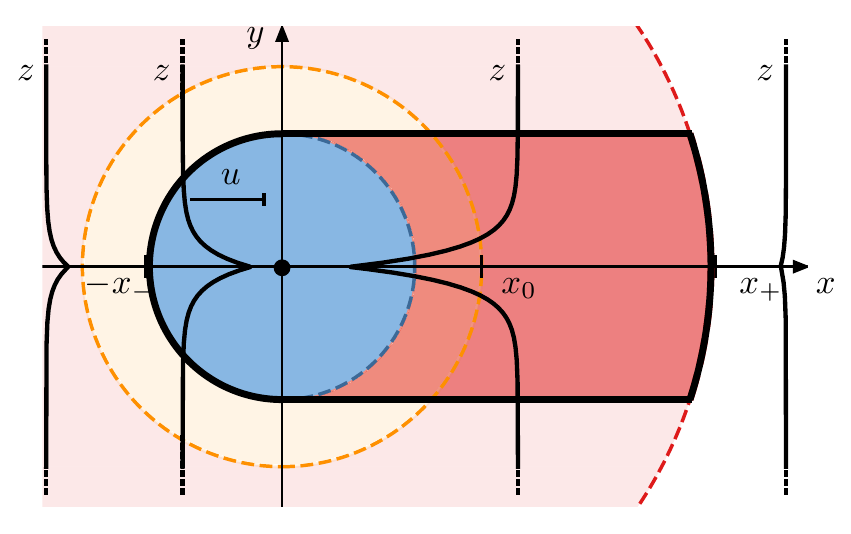}
        \caption{Vortex- and trapping geometries under strong pinning
        conditions; note the different meaning of vertical axes referring to
        the vortices ($z$) and the trapping area ($y$), respectively.  Shown
        are vortex configurations $u(z)$ (black solid lines) on approaching
        the defect head-on (on the $x$-axis) near the bistable interval
        $[-x_-,x_+]$. Four stages are highlighted, the (weakly deformed) free
        state before pinning at $x < -x_-$, pinned on the left at $-x_- <x <
        0$, pinned on the right at $0 < x < x_+$, both strongly deformed, and
        free at $x > x_+$ after depinning.  Note the asymmetry in the pinning
        (at $-x_-$) and depinning (at $x_+$) processes. In the \textrm{ZFC}
        state, the trapping area $S_\mathrm{trap}$ extends over $t_\perp =
        2x_-$ along the transverse direction $y$ and over $x_- + x_+$ in the
        longitudinal one.  The total trapping area (with unit branch
        occupation in the absence of thermal fluctuations) is the sum of the
        blue and red regions enclosed by the black solid line in the figure.
        At equilibrium, the branch occupation is radially symmetric with jumps
        at $R = x_0$, where $x_0$ denotes the branch crossing point, see Fig.\
        \ref{fig:strong_pinning}; the trapping region (light orange) is
        enclosed by the orange dashed circumference with radius $R = x_0$. In
        the \textrm{FC} state, the branch occupation is again radially
        symmetric and the position of the jump depends on the state
        preparation. The trapping region is circular, with a radius $x_- \leq
        R \leq x_+$; the two extreme cases coincide with the dashed blue
        circle with $R = x_-$ (phase $b$ in Fig.\ \ref{fig:hysteresis_scheme})
        and the dashed red circle with $R = x_+$ (phase $b'$ in Fig.\
        \ref{fig:hysteresis_scheme}).}
    \label{fig:geometry}
\end{figure}

A complete derivation of strong pinning theory starting from an elastic
description of the vortex lattice (we assume a lattice directed along $z$ with
a lattice constant $a_0$ determined by the induction $B = \Phi_0/a_0^2$) that
is interacting with a random assembly of defects of density $n_p$ has been
given in several papers; here, we make use of the discussion and notation in
Refs. \onlinecite{Koopmann_2004, Thomann_2017, Willa_2016, Buchacek_2019}, see
also Refs.\ \onlinecite{Kwok_2016, Willa_2018a, Willa_2018b} for numerical
work on strong pinning.  It turns out, that in the low density limit $n_p a_0
\xi^2 \kappa \ll 1$, where $\xi$ denotes the coherence length and $\kappa$ is
the Labusch parameter, see Eq.\ \eqref{eq:Lab_par} below, this complex
many-body problem can be reduced to an effective single-vortex--single-pin
problem. The latter involves an individual flux line with an effective
elasticity $\bar{C} \approx \nu \varepsilon (a_0^2/\lambda)
\sqrt{c_{66}c_{44}(0)} \sim \varepsilon \varepsilon_0/a_0$ that accounts for
the presence of other vortices. Here, $\varepsilon_0 =
(\Phi_0/4\pi\lambda_{\rm\scriptscriptstyle L})^2$ is the vortex line energy,
$\lambda_{\rm\scriptscriptstyle L}$ denotes the London penetration depth,
$\varepsilon < 1$ is the anisotropy parameter for a uniaxial material
\cite{Blatter_1994}, and $\nu$ is a numerical, see Refs.\
\onlinecite{Kwok_2016, Willa_2018b}; the result derives from the elastic
Green's function $G_{\alpha\beta}(\mathbf{k})$ of the vortex lattice, see
Ref.\ \onlinecite{Blatter_1994}, involving shear ($c_{66}$) and dispersive
tilt ($c_{44}(\mathbf{k}))$ and assuming a field that is aligned with the
material's axis.  Second, the problem involves the pinning potential
$e_\mathrm{def}(\mathbf{r})$ of individual defects; for a point-like defect,
$e_\mathrm{def}(\mathbf{r}) = e_p(\mathbf{R})\delta(z)$ is determined by the
form of the vortex core with $e_p(R) = -e_p/(1+ R^2/2\xi^2)$ taking a
Lorentzian shape and $\mathbf{R} = (x,y)$ denoting the in-plane coordinate.
Below, we will consider a potential of general form whenever possible and
focus on Lorentzian-shaped defect potentials $e_p(R)$ in order to arrive at
numerically accurate results; results for non-Lorentzian shaped potentials
remain qualitatively the same.

The generic setup involves a vortex line driven along $x$ with asymptotic
position $\mathbf{R}_v(z \to \pm \infty) = \mathbf{R}_\infty$ that impacts on
the defect located, say, at the origin. The simplest geometry is that of a
head-on collision with $\mathbf{R}_\infty = (x,0)$ and increasing $x$ for an
impact from the left; given the rotational symmetry of the defect potential
$e_p(R)$, the geometry for the collision at a finite impact parameter $y$,
$\mathbf{R}_\infty = (x,y)$ follows straightforwardly.  Assuming a head-on
collision to begin with, the geometry simplifies considerably and involves the
asymptotic vortex position $x$ and the deformation $u(z)$ of the vortex, see
Fig.\ \ref{fig:geometry}; it turns out, that the problem is fully
characterized by its value $u = u(z=0)$ at the pin, with the vortex line
smoothly joining the tip position $r = x+u$ at $z = 0$ with the asymptotic
position $x$ at $z \to \pm \infty$, see Fig.\ \ref{fig:geometry}.  The
detailed shape $x+u(z)$ of the vortex line then follows from a simple
integration \cite{Koopmann_2004, Willa_2016}.  The cusp at $z=0$ is a measure
of the pinning strength.

\begin{figure}
        \includegraphics[width = 1.\columnwidth]{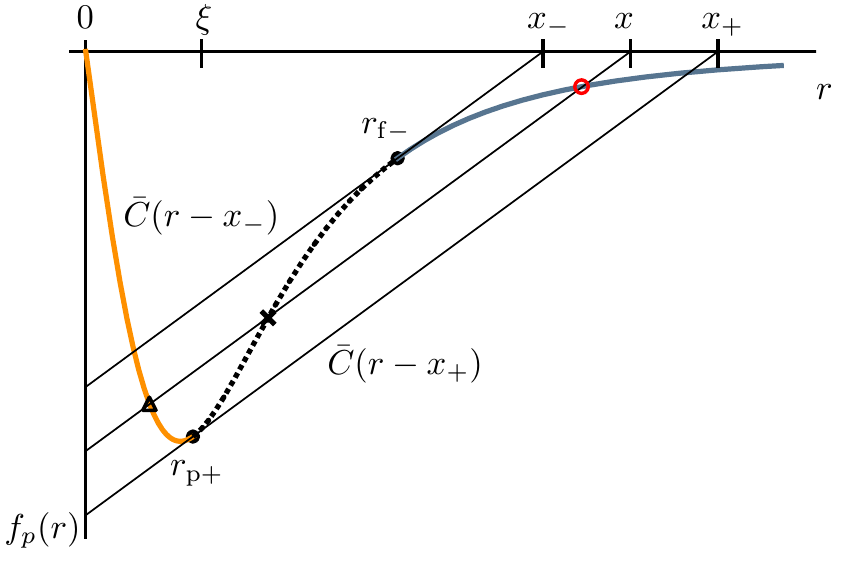}
	\caption{Graphical illustration\cite{Buchacek_2019} of the
	self-consistent solution of the microscopic force-balance equation
	Eq.\ \eqref{eq:force_balance} for a Lorentzian potential with $\kappa
	= 2.5$.  When moving the asymptotic vortex position $x$ across the
	bistable interval $[x_-,x_+]$, we obtain three solutions describing
	pinned $r_\mathrm{p} < \xi$ (orange), free $r_\mathrm{f}$ (blue), and
	unstable $r_\mathrm{us}$ (black dotted) states.  At the edges of the
	bistable interval, we define the limits $r_\mathrm{p}(x_+) \equiv
	r_\mathrm{p+}$ and $r_\mathrm{f}(x_-) \equiv r_\mathrm{f-}$ with
	$f^\prime_p (r_\mathrm{p+}) = f^\prime_p (r_\mathrm{f-}) = \Cbar$
	(black solid dots). The tip positions for the pinned
	($r_\mathrm{p}(x)$, open black triangle) and free ($r_\mathrm{f}(x)$,
	open red circle) branches increase with $x$, while the unstable one
	($r_\mathrm{us}(x)$, black cross) decreases.}
    \label{fig:self-cons-sol}
\end{figure}

\begin{figure}
        \includegraphics[width = 1.\columnwidth]{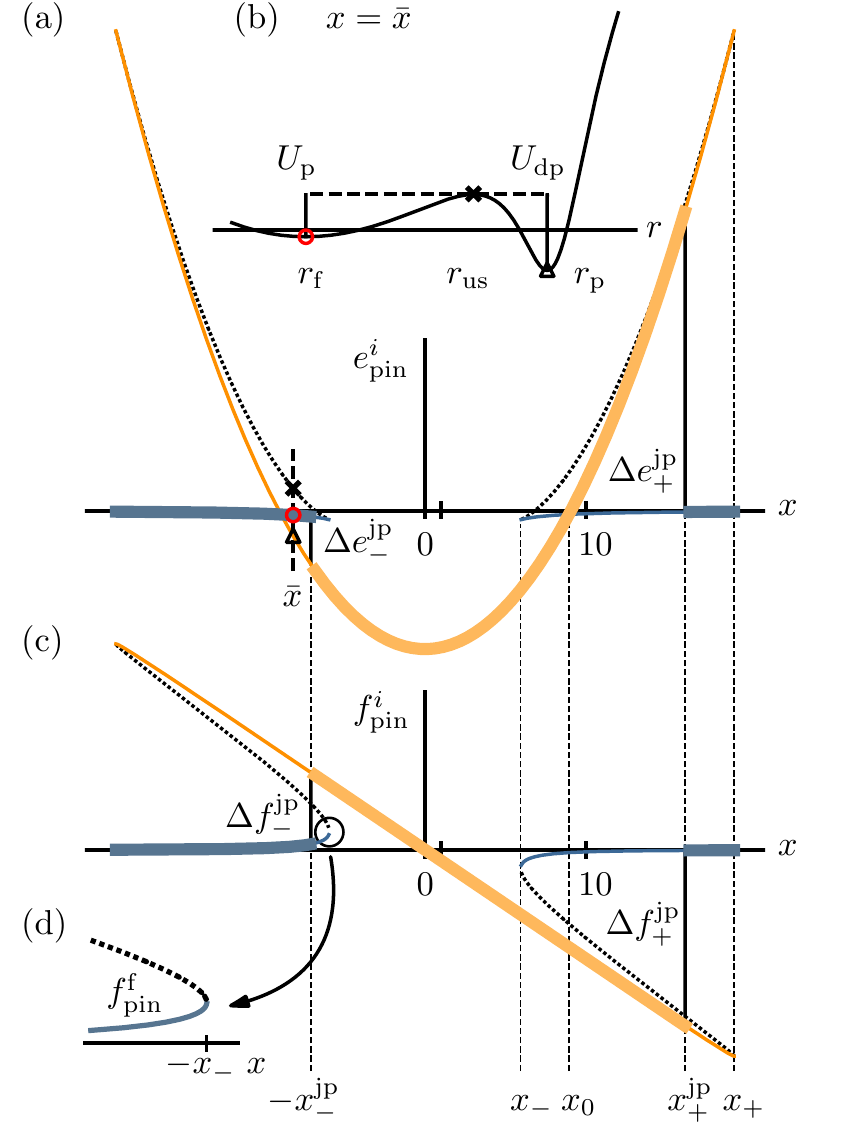}
	\caption{(a) Multi-valued pinning energy landscape $\epin^i(x)$, with
	$i=\mathrm{p,f,us}$ corresponding to the pinned (orange), free (blue),
	and unstable (dotted) branches for $\kappa=10$. The vortex coordinate
	$x$ is expressed in units of $\xi$. The bistability extends over the
	intervals $|x|\in\left[x_-,x_+\right]$ where the different branches
	coexist; pinned and unpinned vortex branches cut at the branch
	crossing point $x=x_0$. (b) Total energy $\epin(\bar{x};r)$ versus
	vortex tip position $r$ for a fixed vortex position $x=\bar{x}$
	(dashed vertical line in $(a)$).  The points $r_\mathrm{f}$ (red dot),
	$r_\mathrm{p}$ (black triangle), and $r_\mathrm{us}$ (black cross)
	mark the free, pinned, and unstable solutions of the force-balance
	equation \eqref{eq:force_balance}. The barriers $\Up$ and $\Udp$
	stabilize the free and pinned states against thermal fluctuations;
	they coincide in size at the branch crossing point $x_0$.  The maximal
	pinning force density $F_\mathrm{pin} = F_c$ is realized for a
	maximally asymmetric pinned-branch occupation $p_c(-x_-<x <x_+) = 1$;
	for the symmetric equilibrium occupation $p_\mathrm{eq}(-x_0 <x < x_0)
	= 1$ the pinning force vanishes. Shown in the figure is the
	pinned-branch occupation at finite temperatures $T$ (thick colored
	lines) where thermal fluctuations allow for early pinning at $-\xmjp <
	- x_-$ and early depinning at $\xpjp < x_+$. The corresponding (total)
	energy jump $\Delta e_\mathrm{pin}^\mathrm{jp} = \Delta
	e^\mathrm{jp}_+ + \Delta e^\mathrm{jp}_-$ (vertical black solid lines)
	determining the pinning force density $F_\mathrm{pin}(T)$ is reduced
	with respect to its critical value (with jumps at $-x_-$ and $x_+$).
	(c) Pinning force $\fpin^i(x)$ corresponding to pinned (orange), free
	(blue), and unstable (dotted) states.  The sum of force jumps $\Delta
	f_\mathrm{pin} = \Delta f^\mathrm{jp}_- +\Delta f^\mathrm{jp}_+$
	(vertical black solid lines) determines the Campbell length
	$\lambda_{\rm \scriptscriptstyle C}$.  (d) The inset shows a zoom of
	the pinning force $f_\mathrm{pin}^\mathrm{f}$ near $-x_-$ that
	contributes to $\lambda_{\rm \scriptscriptstyle C}$ with a steep
	square-root change in the force jump in the presence of creep, see
	discussion in Sec.\ \ref{sec:st}.}
    \label{fig:strong_pinning}
\end{figure}

The energy (or Hamiltonian) of this setup involves elastic and pinning
energies and is given by
\begin{equation}\label{eq:en_pin_tot}
   \epin(x,r) = \frac{1}{2}\Cbar(r-x)^2 + e_p(r).
\end{equation}
Minimizing this energy with respect to $r$ at fixed asymptotic position $x$,
we find the vortex tip position $r(x)$ by solving the nonlinear problem
\begin{equation}\label{eq:force_balance}
   \Cbar(r-x)=-\partial_r e_p = f_p(r),
\end{equation}
see Fig.\ \ref{fig:self-cons-sol} for a graphical solution of this
self-consistency problem.  This (microscopic) force-balance equation develops
multiple solutions when the pin is sufficiently strong, as quantified by the
conditions
\begin{equation}\label{eq:inflection_point}
   \partial_r^2\epin = \Cbar - f'_p(r) = 0  ~\textrm{ and }~
   \partial_r \epin = 0 
\end{equation}
for the appearance of a local maximum in $\epin(x,r)$, see Fig.\
\ref{fig:strong_pinning}(b).  The condition \eqref{eq:inflection_point}
defines the Labusch parameter
\begin{equation}\label{eq:Lab_par}
   \kappa = \max_r \frac{f'_p(r)}{\Cbar} = \frac{f'_p(r_m)}{\Cbar}
\end{equation}
(with $f_p''(r_m) = 0$ providing the maximal force derivative at $r_m$) that
determines the Labusch criterion
\begin{equation}\label{eq:labusch_criterion}
   \kappa > 1
\end{equation}
for strong pinning. Defining the force scale $f_p \equiv e_p/\xi$ and
estimating the force derivative or curvature $f_p^\prime  =
-e_p^{\prime\prime} \sim f_p/\xi$ produces a Labusch parameter $\kappa \sim
e_p/\Cbar\xi^2$, hence, strong pinning is realized for either large pinning
energy $e_p$ or small effective elasticity $\Cbar$. For the Lorentzian
potential, we obtain a maximal force derivative $f_p^\prime(r_m) = e_p /4
\xi^2$ at $r_m = \sqrt{2}\,\xi$ and hence $\kappa = e_p/4\Cbar\xi^2$.

Within the (symmetric) bistable regions $[-x_+, -x_-]$ and $[x_-, x_+]$
opening up at $\kappa > 1$, the force-balance equation Eq.\
\eqref{eq:force_balance} exhibits multiple solutions $r(x)$ corresponding to
free ($r_\mathrm{f}$, elasticity dominated) and pinned ($r_\mathrm{p}$,
pinning dominated) solutions, see Fig.\ \ref{fig:self-cons-sol}, as well as an
unstable solution $r_\mathrm{us}$ that sets the barrier for creep, see below.

A vortex approaching the defect from the left gets trapped by the pin at
$-x_-$ and is dragged towards the pinning center. Upon leaving the defect, the
vortex gets strongly deformed, see Fig.\ \ref{fig:geometry} and depins at
$x_+$. Inserting the solutions $r_\mathrm{f}(x)$, $r_\mathrm{p}(x)$, and
$r_\mathrm{us}(x)$ of Eq.\ \eqref{eq:force_balance} back into Eq.\
\eqref{eq:en_pin_tot}, we obtain the pinning energy landscape
\begin{equation}\label{eq:e_pin_i}
   e^\mathrm{i}_\mathrm{pin}(x) = e_\mathrm{pin}[x,r_\mathrm{i}(x)]
\end{equation}
with its multiple branches $\mathrm{i} = \mathrm{f,p,us}$ shown in Fig.\
\ref{fig:strong_pinning}(a). The same way, we find the pinning force
$f_p[r(x)]$ acting on the vortex tip; inserting the different solutions
$r_\mathrm{f}(x)$, $r_\mathrm{p}(x)$, and $r_\mathrm{us}(x)$, we obtain the
pinning force $f^\mathrm{i}_\mathrm{pin}(x) \equiv f_p[r_\mathrm{i}(x)]$ with
its multiple branches $\mathrm{i} = \mathrm{f,p,us}$ as shown in Fig.\
\ref{fig:strong_pinning}(c). Note that the pinning force $f_\mathrm{pin}(x)$
can be written as the total derivative of the energy $e_\mathrm{pin}[x,r(x)]$,
\begin{equation}\label{eq:fo_pin_tot}
   f_\mathrm{pin}(x) = f_p[r(x)] = -\frac{d e_\mathrm{pin}[x,r(x)]}{dx},
\end{equation}
where we have used the force-balance equation \eqref{eq:force_balance} to
arrive at the last relation.

The energy $e_\mathrm{pin}(x)$ and force $f_\mathrm{pin}(x)$ experienced by
the vortex are shown in Fig.\ \ref{fig:strong_pinning}. Due to the presence of
multiple branches, we see that a right-moving vortex undergoes jumps in energy
$\Delta e_\mathrm{pin}$ and force $\Delta f_\mathrm{pin}$ at the edges $-x_-$
and $x_+$ of the bistable intervals (for a left moving vortex, corresponding
jumps appear at $x_-$ and $-x_+$).  These jumps are the hallmark of strong
pinning and determine physical quantities such as the critical current density
$j_c$ or the Campbell penetration depth $\lambda_{\rm \scriptscriptstyle C}$.
In the following, we evaluate the characteristic quantities defining the
pinning landscape of Fig.\ \ref{fig:strong_pinning} in the limits of very
strong ($\kappa \gg 1$) and marginal ($\kappa - 1 \ll 1$) pinning.

\subsubsection{Bistable interval $[x_-,x_+]$ and extremal tip positions}

The extent of the bistable interval $[x_-,x_+]$ is easily found in the very
strong pinning limit with $\kappa \gg 1$: With reference to Fig.\
\ref{fig:self-cons-sol}, we approximate $f_p[r_\mathrm{p}(x_+)] \approx
f_{p,\mathrm{max}}$ and drop $r_\mathrm{p}(x_+) < \xi$ against $x_+ \sim
\kappa\xi$ in \eqref{eq:force_balance} to find
\begin{equation}\label{eq:x_+}
   x_+ \approx f_{p,\mathrm{max}}/\Cbar \sim \kappa \xi.
\end{equation}
The lower boundary $x_-$ is conveniently obtained from the condition
$f_p^\prime[r_\mathrm{f}(x_-)] = \Cbar$. For large $\kappa$, we have
$r_\mathrm{f}(x_-) \gg \xi$ residing in the tail of the pinning potential;
assuming a defect potential decaying as $e_p(R) \sim -2 e_p (\xi/R)^n$, we
obtain  
\begin{equation}\label{eq:r_f-}
   r_\mathrm{f}(x_-) \approx \xi \left[\frac{2 n (n\!+\!1)e_p}
   {\Cbar \xi^2}\right]^{1/(n+2)}\!\!\!\!
   \sim \xi \kappa^{1/(n+2)}.
\end{equation}
Inserting this result back into Eq.\ \eqref{eq:force_balance}, we find that
\begin{equation}\label{eq:x_-}
   x_- \approx \frac{n+2}{n+1}\> r_\mathrm{f}(x_-).
\end{equation}
For a Lorentzian potential, we have $f_{p,\mathrm{max}} = (3/2)^{3/2}\,
e_p/4 \xi$ and $r_\mathrm{p}(x_+) \approx \sqrt{2/3} \,\xi$ and hence
\begin{equation}\label{eq:x_+L}
   x_+ \approx (3/2)^{3/2} \, \kappa \xi.
\end{equation}
The lower boundary $x_-$ relates to $r_\mathrm{f}(x_-)$ via $x_- \approx (4/3)
r_\mathrm{f}(x_-)$ and with $r_\mathrm{f}(x_-) \approx 2 (3\kappa)^{1/4} \xi$,
we obtain
\begin{equation}\label{eq:x_-L}
   x_- \approx (8/3) (3\kappa)^{1/4} \xi.
\end{equation}

The marginally strong pinning case $\kappa \gtrsim 1$ can be quantitatively
described via an expansion of the pinning force $f_p(r)$ around the inflection
point $r_m$ defined through $f_p^{\prime\prime}(r_m) = 0$ and using the Labusch
parameter in the form $f_p'(r_m) = \kappa \Cbar$,
\begin{equation}\label{eq:f_p_exp_sk}
   f_p(r_m + \delta r) \approx f_p(r_m) + \kappa \Cbar \> \delta r
   - \gamma \, (e_p /3\xi^4) \delta r^{\> 3}.
\end{equation}
We use $\kappa -1 \ll 1$ as our small parameter and set $\kappa \approx 1$
otherwise (however, beware of additional corrections in $\kappa - 1$ through
$\kappa \approx 1 + (\kappa - 1)$). For a Lorentzian potential, the shape
parameter $\gamma$ assumes the value $\gamma = 3/8$.  The cubic expansion
\eqref{eq:f_p_exp_sk} is antisymmetric about the inflection point $r_m$, thus
producing symmetric results for pinning and depinning. 

The tip locations 
\begin{equation}\label{eq:rp+f-}
   r_\mathrm{p}(x_+) = r_m - \delta r_\mathrm{max}, ~~~
   r_\mathrm{f}(x_-) = r_m + \delta r_\mathrm{max}
\end{equation}
at (de)pinning are defined
by the conditions $f_p[r_\mathrm{f}(x_-)] = f_p[r_\mathrm{p}(x_+)] = \Cbar$, see
Fig.\ \ref{fig:self-cons-sol}; making use of the expansion \eqref{eq:f_p_exp_sk},
we find
\begin{equation}\label{eq:delta_r_sk}
   \delta r_\mathrm{max} \approx \frac{\xi}{2}
   \sqrt{\frac{4\Cbar \xi^2}{\gamma e_p}} (\kappa-1)^{1/2}.
\end{equation}
Inserting this result into the force-balance equation \eqref{eq:force_balance}
and using \eqref{eq:f_p_exp_sk}, we find the boundaries 
\begin{equation}\label{eq:x_pm_sk}
   x_\pm = x_m \pm \delta x_\mathrm{max}
\end{equation}
of the bistable region with
\begin{equation}\label{eq:dx_max}
   \delta x_\mathrm{max} \approx
   \frac{\xi}{3}\sqrt{\frac{4\Cbar \xi^2}{\gamma e_p}}(\kappa-1)^{3/2}.
\end{equation}
The pair $x_m$ and $r_m$ of asymptotic and tip positions depends on the
details of the potential; while $r_m$ derives solely from the shape $e_p(R)$
and thus does not depend on the elasticity $\Cbar$, $x_m$ as given by
\eqref{eq:force_balance} involves $\Cbar$ and shifts $\propto (\kappa - 1)$.
For a Lorentzian potential, we have
\begin{equation}\label{eq:r_m-x_m}
   r_m = \sqrt{2}\xi, 
   \quad x_m  = 2\sqrt{2} \xi + \sqrt{2}\xi (\kappa - 1),
\end{equation}
and
\begin{eqnarray}\label{eq:dr_m-dx_m}
   \delta r_\mathrm{max} &\approx& \xi \, [2(\kappa-1)/3]^{1/2}, \\ \nonumber
   \delta x_\mathrm{max} &\approx& \xi \, [2(\kappa-1)/3]^{3/2}.
\end{eqnarray}

Besides the tip positions $r_\mathrm{p}(x_+)$ and $r_\mathrm{f}(x_-)$ at
(de)pinning, we also need the tip positions $r_\mathrm{f}(x_+)$ and
$r_\mathrm{p}(x_-)$ that are not associated with a special point on the free
and pinned branches. They are obtained by solving the force-balance equation
\eqref{eq:force_balance} at $x_\pm = x_m \pm \delta x_\mathrm{max}$ using
the expansion \eqref{eq:f_p_exp_sk} with the ansatz $r_\mathrm{f}(x_+) =
r_m + \mu \delta r_\mathrm{max}$ and $r_\mathrm{p}(x_-) = r_m - \mu \delta
r_\mathrm{max}$; the resulting equation for $\mu$,
\begin{equation}\label{eq:mu}
    \mu + 2/3 - \mu^3/3 = 0,
\end{equation}
is solved by $\mu = -1$ ($\to$ the result \eqref{eq:delta_r_sk} obtained before)
and $\mu = 2$, hence 
\begin{eqnarray}\label{eq:rf+}
   r_\mathrm{f}(x_+) - r_m = r_m - r_\mathrm{p}(x_-) \approx 2 \delta
   r_\mathrm{max},
\end{eqnarray}
with $\delta r_\mathrm{max}$ given in \eqref{eq:delta_r_sk}.

\subsubsection{Branch crossing point $x_0$}

At very strong pinning, the bistable region is arranged asymmetrically around
the branch crossing point $x_0$, see Fig.\ \ref{fig:strong_pinning}; we
find the latter by equating the pinning energies \eqref{eq:en_pin_tot} for the
free and pinned branches: with $x_0 \gg \xi$ at large $\kappa$, we have
the free and pinned vortex tip positions 
\begin{equation}\label{eq:r_f-p}
   r_\mathrm{f}(x_0) \approx x_0~~\textrm{ and }~~r_\mathrm{p}(x_0) \ll \xi,
\end{equation}
as follows from the force-balance equation $x_0 - r_\mathrm{f}(x_0) = -
f_p[r_\mathrm{f}(x_0)]/\Cbar$ (dropping the force term
$f_p[r_\mathrm{f}(x_0)]$) and Fig.\ \ref{fig:self-cons-sol}. With
$e_\mathrm{pin}^\mathrm{f}(x_0) \approx 0$ and $e_\mathrm{pin}^\mathrm{p}
(x_0) \approx \Cbar x_0^2/2 - e_p$, we find that 
\begin{equation}\label{eq:x_0}
   x_0\approx 2 \sqrt{2}\xi \left(\frac{e_p}{4\Cbar\xi^2}\right)^{1/2} \!\!\!\!\! \sim
   \sqrt{\kappa} \xi.
\end{equation}
For the Lorentzian potential, we find $x_0 \approx 2\sqrt{2\kappa}\xi$.

When strong pinning is marginal, $\kappa - 1 \ll 1$, the branch crossing point
$x_0$ coincides with $x_m$. Its location depends on the detailed shape of the
potential; for a Lorentzian, we have (see Eq.\ \eqref{eq:r_m-x_m})
\begin{equation}\label{eq:x_0_sk}
    x_0 \approx x_m = 2\sqrt{2} \xi + \sqrt{2}\xi (\kappa - 1).
\end{equation}

\subsubsection{Activation barrier $U_0$}

Finally, we briefly discuss the barriers for thermal activation between
bistable branches, specifically, the barrier scale $U_0$ at the branch crossing point.
The latter is given by
\begin{align}\label{eq:U0_def}
   U_0 &= \epin[x_0,r_\mathrm{us}(x_0)] - \epin[x_0,r_\mathrm{f}(x_0)]\nonumber\\
   &=\epin[x_0,r_\mathrm{us}(x_0)] - \epin[x_0,r_\mathrm{p}(x_0)]
\end{align}
and therefore depends on the unstable and pinned/free tip position at $x_0$.
At large $\kappa \gg 1$, the vortex free and pinned vortex tip positions are
given in \eqref{eq:r_f-p}. We find the unstable solution $r_\mathrm{us}$ by
using the asymptotic decay $f_p(R) \approx - 2 n\, f_p (\xi/R)^{n+1}$ and
dropping the term $r_\mathrm{us}(x_0)$ against $x_0$ in the force-balance
equation \eqref{eq:force_balance}, with the result that
\begin{equation}\label{eq:r_us}
   r_\mathrm{us}(x_0) \approx \xi
   \left(\frac{2 n^2 e_p}{\Cbar \xi^2}\right)^{1/2(n+1)}\!\!\!\!\!\!\!\!\!,
\end{equation}
for a Lorentzian potential, $r_\mathrm{us}(x_0) \approx 2(\kappa/2)^{1/6}\xi$;
indeed, the ratio $r_\mathrm{us}(x_0)/x_0 \approx (1/4\kappa)^{1/3} \ll 1$ is
parametrically small at large $\kappa$. The barrier scale $U_0$ then evaluates
to
\begin{align}\label{eq:U0_lk}
   U_0 \approx \frac{\Cbar}{2}x_0^2\approx e_p
\end{align}
with small corrections $\propto 1/\kappa^{n/2(n+1)}$.

In the marginally strong pinning case, we find the tip positions $r = r_m
+\delta r$ by solving the force-balance equation \eqref{eq:force_balance} with
the expansion \eqref{eq:f_p_exp_sk} at $x = x_0 = x_m$, with the three
solutions $\delta r = 0$, providing the unstable solution $r_\mathrm{us}(x_0) =
r_m$, and the free and pinned meta-stable solutions $r_\mathrm{f,p} (x_0)$
arranged symmetrically with $\delta r = \pm \sqrt{3}\,\delta r_\mathrm{max}$,
\begin{equation}\label{eq:delta_r0}
   r_\mathrm{f}(x_0) - r_m = r_m - r_\mathrm{p}(x_0) = 
   \sqrt{3}\,\delta r_\mathrm{max}.
\end{equation}
Making use of these results in the definition \eqref{eq:U0_def} for $U_0$ and
expanding $e_\mathrm{pin}(x_0,r_\mathrm{f} = r_m + \sqrt{3}\,\delta
r_\mathrm{max})$ to fourth order in $\delta r_\mathrm{max}$, we find that
\begin{equation}\label{eq:U0_sk}
  U_0 \approx \frac{3}{4}\frac{\Cbar^2\xi^4}{\gamma e_p} (\kappa - 1)^2
   = \frac{e_p}{8}(\kappa - 1)^2,
\end{equation}
where the last equation applies to the Lorentzian shaped potential.
In deriving \eqref{eq:U0_sk}, we have used the expansion
\eqref{eq:f_p_exp_sk} as well as the force balance equation
\eqref{eq:force_balance} to convert the elastic energy $\Cbar (r_m - x_0)
\delta r_\mathrm{max}$ to a pinning energy $f_p(r_m) \delta r_\mathrm{max}$.

\section{Transport}\label{sec:transport}

One of the central features of superconductivity is dissipation-free
transport. We briefly discuss the results of strong pinning theory for
critical current densities $j_c$ and the effect of thermal fluctuations
resulting in a slowly decaying `persistent' current.

The transport properties of a type II superconducting material is determined
by the vortex dynamics as described by the (macroscopic) force-balance
equation
\begin{equation}\label{eq:macroscopic_force_balance}
  \eta \mathbf{v} = \mathbf{F}_{\rm \scriptscriptstyle
  L}(\mathbf{j})-\mathbf{F}_\mathrm{pin}(\mathbf{v},T),
\end{equation}
a non-linear equation for the mean vortex velocity $\mathbf{v}$, with $\eta =
B H_{c2}/\rho_n c^2$ the Bardeen-Stephen viscosity \cite{Bardeen_1965} (per
unit volume; $\rho_n$ is the normal state resistivity) and $\mathbf{F}_{\rm
\scriptscriptstyle L} = \mathbf{j} \times \mathbf{B}/c$ the Lorentz force.
The pinning force density $\mathbf{F}_\mathrm{pin}$ is directed along
$\mathbf{v}$; it depends on the velocity $v$, that turns finite beyond the
critical force density $F_c$, and on the temperature $T$ driving thermal
fluctuations, i.e., creep---we will discuss these effects shortly.

The pinning force density $F_\mathrm{pin}$ is given by the sum over all force
contributions $f_\mathrm{pin}$; assuming a uniform distribution of defects, we
have to take the average $F_\mathrm{pin} = n_p \langle f_\mathrm{pin} \rangle$
with the appropriate branch occupation of vortices.  For a vortex approaching
the defect head-on along $x$, the free branch terminates at $-x_-$ and the
vortex jumps to the pinned branch, gaining the energy $\Delta
e^\mathrm{fp}_\mathrm{pin} = e^\mathrm{f}_\mathrm{pin}(-x_-) -
e^\mathrm{p}_\mathrm{pin}(-x_-)>0$ (denoted as $\Delta e^\mathrm{jp}_-$ in
Fig.\ \ref{fig:strong_pinning}). Moving forward, the vortex remains pinned
until the branch ends at $x_+$, where the jump to the free branch involves the
energy $\Delta e^\mathrm{pf}_\mathrm{pin} = e^\mathrm{p}_\mathrm{pin}(x_+) -
e^\mathrm{f}_\mathrm{pin}(x_+)>0$ (denoted as $\Delta e^\mathrm{jp}_+$ in
Fig.\ \ref{fig:strong_pinning}). The critical pinned-branch occupation for
head-on trajectories then is $p_c(x) \equiv \Theta(x+x_-) -\Theta(x-x_+)$,
while for a finite impact factor $y$, the branch occupation $p_c(\mathbf{R})$
coincides with characteristic function of the trapping area shown in Fig.\
\ref{fig:geometry}. The critical branch occupation is maximally asymmetric,
what produces the largest possible pinning force. Other branch occupations
produce different pinning forces, e.g., the radially symmetric
equilibrium occupation $p_\mathrm{eq}(\mathbf{R}) = \Theta(x_0-R)$, with
$x_0$ the branch cutting point shown in Fig.\ \ref{fig:strong_pinning}, leads
to a vanishing pinning force.

Averaging the pinning force $\mathbf{f}_\mathrm{pin}$ over $x$ and $y$ with
the vortex population described by the critical branch occupation
$p_c(\mathbf{R})$, we obtain the critical pinning force density $\mathbf{F}_c$
(we exploit the anti-symmetry of $\mathbf{f}_\mathrm{pin}(\mathbf{R})$)
\begin{align}\label{eq:F_pin_vec}
  \mathbf{F}_c &= - n_p \int \frac{d^2\mathbf{R}}{a_0^2}\bigl[
   p_{c}(\mathbf{R})\mathbf{f}^\mathrm{p}_\mathrm{pin}(\mathbf{R}) + (1-p_{c}(\mathbf{R}))
   \mathbf{f}^\mathrm{f}_\mathrm{pin}(\mathbf{R})\bigr]\nonumber\\
   &= - n_p \int \frac{d^2\mathbf{R}}{a_0^2} p_{c}(\mathbf{R})
   [\partial_x\Delta e^\mathrm{fp}_\mathrm{pin}(\mathbf{R})]\mathbf{e}_x,
\end{align}
with the energy difference $\Delta
e^\mathrm{fp}_\mathrm{pin}(\mathbf{R})=e^\mathrm{f}_\mathrm{pin}(\mathbf{R}) -
e^\mathrm{p}_\mathrm{pin}(\mathbf{R})$ and $\mathbf{e}_x$ the unit vector
along $x$; the $y$-component of the pinning force density vanishes due to the
antisymmetry in $f_{\mathrm{pin},y}$.  Following convention, we have included
a minus sign in the definition of $F_c$. The branch-occupation
$p_c(\mathbf{R})$ restricts the integral to the trapping area shown in
Fig.\ \ref{fig:geometry}; the integration over $x$ brings forward the constant
energy jumps at the two semi-circular boundaries, hence
\begin{align}\label{eq:F_pin}
   F_c &= -n_p \int^{x_-}_{-x_-} \frac{dy}{a_0}
   \frac{\Delta e^\mathrm{fp}_\mathrm{pin}(x,y)}{a_0}
   \Big|_{x=-\sqrt{x_-^2-y^2}}^{x=+\sqrt{x_+^2-y^2}}\nonumber\\
   &= n_p \frac{\Delta e^\mathrm{fp}_\mathrm{pin}+ 
   \Delta e^\mathrm{pf}_\mathrm{pin}}{a_0}\int^{x_-}_{-x_-} \frac{dy}{a_0}\nonumber\\
   &= n_p \frac{t_\perp}{a_0} \frac{\Delta e^\mathrm{fp}_\mathrm{pin}+ 
   \Delta e^\mathrm{pf}_\mathrm{pin}}{a_0},
\end{align}
where we have defined the transverse trapping length $t_\perp = 2 x_-$. The
result \eqref{eq:F_pin} for the pinning force density shows that all vortices
hitting the left-side semi-circle of diameter $t_\perp$ get pinned, see Fig.\
\ref{fig:geometry}, and contribute equally to the pinning force density, a
consequence of the rotationally symmetric pinning potential $e_p(R)$. We
confirm that the multi-valued energy landscape in Fig.\
\ref{fig:strong_pinning} is central for obtaining a finite pinning force
density $F_\mathrm{pin} \propto n_p$; for $\kappa < 1$ jumps are absent and
the integral over the corresponding smooth periodic function
$f_\mathrm{pin}(x)$ in Eq.\ \eqref{eq:F_pin} vanishes.  This is the realm of
weak pinning with a mechanism that is collective, resulting in a density
scaling \cite{Koopmann_2004} $F_\mathrm{pin} \propto n_p^2$.

\subsection{Critical current density $j_c$}\label{sec:j_c}

We obtain the critical current density $j_c$ from the force balance
\eqref{eq:macroscopic_force_balance} by setting $v = 0$ and choosing the
maximal pinning force density $F_c$ associated with the most asymmetric branch
occupation $p_c(\mathbf{R})$,
\begin{equation}\label{eq:j_c}
   j_c = c F_c/B = (c/\Phi_0) n_p t_\perp \Delta e_\mathrm{pin}
\end{equation}
with $\Delta e_\mathrm{pin} = \Delta e^\mathrm{fp}_\mathrm{pin}|_{-x_-} +
\Delta e^\mathrm{pf}_\mathrm{pin}|_{x_+}$ the sum of (positive) jumps in the
pinning energy $e_\mathrm{pin}(x)$ and $t_\perp = 2 x_-$.  Note that strong
pinning does not necessarily imply a large critical current density $j_c$, as
our approximation of independent pins requires a small density $n_p$.

\subsection{Creep effects on transport: persistent current}\label{sec:creep_pc}

Starting with a non-equilibrium initial state at time $t = 0$, thermal
fluctuations (or creep) drive the system towards equilibrium. To fix ideas, we
start from a critical or ZFC state (and a head-on collision) characterized by
the critical pinned-branch occupation $p_c(x) = \Theta(x + x_-) -\Theta(x -
x_+)$ and let it decay through creep; the extension of the result to the 2D
situation is straightforward.  The presence of thermal fluctuations then
increases the probabilities for pinning near $-x_-$ and depinning near $x_+$,
that leads to a reduction of the pinning force density $F_\mathrm{pin} < F_c$.
We account for such thermal hops of vortices into and out of the pin through
proper calculation of the thermal pinned-branch occupation probability
$p_\mathrm{th}(x;t,T)$ via solution of the rate equation
\cite{BrazovskiLarkin_1999, BrazovskiiNattermann_2004, Buchacek_2019} (we set
the Boltzmann constant to unity, $k_{\rm\scriptscriptstyle B} = 1$)
\begin{equation}\label{eq:rate_equation}
   \frac{dp_\mathrm{th}}{dt} = -\omegap\,p_\mathrm{th}\,e^{-\Udp/T} 
   + \omegaf\,(1-p_\mathrm{th})\, e^{-\Up/T},
\end{equation}
where 
\begin{eqnarray}\label{eq:U_creep}
   \Up(x) &=& e_\mathrm{pin}^\mathrm{us}(x) - e_\mathrm{pin}^\mathrm{f}(x),
   \\ \nonumber
   \Udp(x) &=& e_\mathrm{pin}^\mathrm{us}(x) - e_\mathrm{pin}^\mathrm{p}(x),
\end{eqnarray}
denote the barriers for pinning and depinning (cf.\ Eq.\ \eqref{eq:e_pin_i})
and $\omegap(x)$, $\omegaf(x)$ are the corresponding attempt frequencies. It
follows from Fig.\ \ref{fig:strong_pinning}(a) that the barriers $\Up(x \to
-x_-)$ and $\Udp(x \to x_+)$ for pinning and depinning vanish, implying that
modifications of the pinned-branch occupation probability are largest near
$-x_-$ and $x_+$ where we can simplify the rate equation
\eqref{eq:rate_equation} by dropping one of the terms.  One finds
\cite{Buchacek_2019}, that after a finite waiting time $t$, thermal
fluctuations produce a shift in the jump positions for pinning and depinning
(and a small rounding of the steps in $p_\mathrm{th}(x)$ that we can ignore):
the jump from the free to the pinned branch appears earlier at
$-x^\mathrm{jp}_- < -x_-$ and so does the location of depinning,
$x^\mathrm{jp}_+ < x_+$, with the solution of the rate equation
\eqref{eq:rate_equation} well approximated by the step function
$p_\mathrm{th}(x;t,T) \approx \Theta[x + x^\mathrm{jp}_-(t,T)] - \Theta[x -
x^\mathrm{jp}_+(t,T)]$.  

The renormalized jump positions $-x^\mathrm{jp}_-(t,T)$ and
$x^\mathrm{jp}_+(t,T)$ are determined by the relations \cite{Buchacek_2019}
\begin{equation}\label{eq:barriers}
   \Up(-x_-^\mathrm{jp}) \approx \Udp(x_+^\mathrm{jp})
                         \approx T \ln (t/t_0),
\end{equation}
with the diffusion time $\tau_0 = \pi j_c d^2/2c\vTH B$ ($d$ is the sample
dimension) and $t_0$ to be determined self-consistently \cite{Buchacek_2019}
from $t_0 = \tau_0 T/ (j_c|\partial_j U|)$. Equation \eqref{eq:barriers} tells
us, that thermal fluctuations driven by the temperature $T$ can overcome
(de)pinning barriers $U_\mathrm{p(dp)}$ of size $T \, \ln (t/t_0)$ after a
waiting time $t$. As a result, waiting a time $t$ at temperature $T$, the
pinned-branch occupation probability changes from $p_c(x) \approx 0$ to
$p_\mathrm{th}(x) \approx 1$ at all positions $x$ within the intervall
$[-x_-^\mathrm{jp}(t),-x_-]$ and drops from $p_c(x) \approx 1$ to
$p_\mathrm{th}(x) \approx 0$ for $x \in [x_+^\mathrm{jp}(t),x_+]$, thereby
reducing the asymmetry of the critical occupation probability $p_c(x)$.

The waiting time $t$ then determines the shape of the pinned-branch occupation
probability $p_\mathrm{th}(x;t,T)$: at short times, thermal relaxation is weak
and $p_\mathrm{th}(x;t,T)$ remains close to $p_c(x)$. On the other hand, for
finite $T$ and long waiting times $t \lesssim t_\mathrm{eq} \equiv t_0
\exp(U_0/T)$, with $U_0= \Up(x_0)=\Udp(x_0)$ the barrier at the branch cutting
point $x_0$, see Fig.\ \ref{fig:strong_pinning}, relaxation is strong and
$p_\mathrm{th}(x;t,T)$ approaches the symmetric equilibrium occupation
$p_\mathrm{eq}(x) = \Theta(x+x_0) -\Theta(x-x_0)$. Going to very long times
$t$ beyond $t_\mathrm{eq}$, both of the terms in \eqref{eq:rate_equation}
accounting for pinning and depinning hops near $x_0$ become equally important
in establishing the precise equilibrium shape of the pinned-branch occupation
probability.

Generalizing from the head-on collision to a finite-impact geometry is
straightforward; evaluating the pinning force density Eq.\ \eqref{eq:F_pin}
with $p_c$ replaced by $p_\mathrm{th}$, we obtain the result
\begin{equation}\label{eq:F_pin_tT_num}
   F_\mathrm{pin}(t,T) = n_p\frac{t^\mathrm{jp}_\perp(t,T)}{a_0}
   \frac{\Delta e^\mathrm{jp}_\mathrm{pin}(t,T)}{a_0},
\end{equation}
that depends on the temperature $T$ and the waiting time $t$.  The premature
pinning and depinning processes at $-x_-^\mathrm{jp}$ and $x_+^\mathrm{jp}$
modify the trapping length $t_\perp^\mathrm{jp} = 2 x_-^\mathrm{jp} > t_\perp$
and reduce the (sum of) jumps in the pinning energy, $\Delta e_\mathrm{pin}
\to \Delta e^\mathrm{jp}_\mathrm{pin} = \Delta
e^\mathrm{fp}_\mathrm{pin}|_{-\xmjp} + \Delta
e^\mathrm{pf}_\mathrm{pin}|_{\xpjp}$.  For times $t \gg t_0$, the pinning
force density \eqref{eq:F_pin_tT_num} takes the analytical form
\cite{Buchacek_2019}
\begin{equation}\label{eq:F_pin_tT}
   F_\mathrm{pin}(t,T) = F_c\> \bigl[1- g(\kappa)\> \mathcal{T}^{2/3} +
   \mathcal{O}(\mathcal{T}^{4/3}) \bigr],
\end{equation}
with the dimensionless creep parameter $\mathcal{T}$
\begin{equation}\label{eq:def_U}
   \mathcal{T}(t,T) \equiv \frac{T}{e_p}\ln\frac{t}{t_0}.
\end{equation}
The exponent $2/3$ derives from the vanishing of barriers on approaching the
boundaries of the bistable region, $U_\mathrm{dp,p}(x) \propto
|x-x_\pm|^{3/2}$, with the value $3/2$ universal for a smooth pinning
potential $e_p(R)$; higher-order terms relevant away from the edges $x_\pm$
produce the corrections $\propto \mathcal{T}^{4/3}$ in \eqref{eq:F_pin_tT}.
The coefficient $g(\kappa)$ subsumes all dependencies on the Labusch parameter
$\kappa$ and has been calculated in Ref.\ \onlinecite{Buchacek_2019}; it
involves the competing effects of an increasing trapping length
$t_\perp^\mathrm{jp}$ and a decreasing jump in the total pinning energy
$\Delta e_\mathrm{pin}^\mathrm{jp}$. As the latter is the dominating one for
not too strong pinning parameters below $\kappa \sim 10^2$, the pinning force
density $F_\mathrm{pin}(t,T)$ usually decreases under the influence of creep.
The relative importance of these two effects will be modified in the analysis
of the Campbell penetration depth below, where the role of $\Delta
e_\mathrm{pin}^\mathrm{jp}$ is replaced by $\Delta
f_\mathrm{pin}^\mathrm{jp}$.

Inserting the result \eqref{eq:F_pin_tT} back into the force-balance equation
\eqref{eq:macroscopic_force_balance}, we immediately obtain the persistent
current density: in a typical relaxation experiment (i.e., after a short
initial waiting time), we can neglect the dissipative term $\eta v$ in Eq.\
\eqref{eq:macroscopic_force_balance} and we arrive at the persistent current
density in the form
\begin{equation}\label{eq:pers_curr}
   j(t,T) \approx c F_c \bigl[1- g(\kappa)\> \mathcal{T}^{2/3}(t,T) \bigr] /B.
\end{equation}
The result \eqref{eq:pers_curr} is valid for times $t \ll t_\mathrm{eq}$.  For
large times beyond $t_\mathrm{eq}$, we go over to the TAFF region (thermally
assisted flux flow \cite{Kes_1989}) where the creep dynamics governed by the
slow $\ln(t/t_0)$ behavior turns into a diffusive vortex motion (and thus
ohmic response). The vortex front at $R_\mathrm{vf}$ then moves into the
sample following the diffusion law $R_\mathrm{vf}(t) \sim \mathrm{const.} -
\sqrt{D_{\rm\scriptscriptstyle TAFF} t}$ with the diffusion constant
\cite{Blatter_1994} $D_{\rm\scriptscriptstyle TAFF} \sim c^2
\rho_{\rm\scriptscriptstyle TAFF}$ and $\rho_{\rm\scriptscriptstyle TAFF}
\propto \rho_n(B/H_{c2}) \exp(-U_0/T)$. The current decays algebraically, $j
\propto 1/\sqrt{t}$ until the sample (of size $d$) is fully penetrated at
$t_\mathrm{fp} \sim d^2/ D_{\rm\scriptscriptstyle TAFF}$. Thereafter, the
remaining persistent current decays exponentially, $j(t > t_\mathrm{fp})
\propto \exp(-t/t_\mathrm{fp})$.

The above scenario applies to the strong pinning paradigm where barriers
saturate in the limit of vanishing currents, $j \to 0$. In reality,
correlations between different pinning centers are expected to become relevant
at very small drives $j$, implying growing barriers and glassy response instead.

Below, we will study the influence of creep on the linear response under a
small external ac magnetic field, that is, again a typical relaxation
experiment involving the waiting time $t$ determining the evolution of the
vortex state.  It will be interesting to see that creep affects the persistent
current and the ac penetration depth very differently, with $j(t,T)$ vanishing
at long times while $\lambda_{\rm\scriptscriptstyle C}(t,T)$ remains finite.

\section{ac linear response}\label{sec:ac}

Probing the superconductor with a small ac field $\delta B = h_\mathrm{ac} \,
\exp(-i\omega t) \ll B_0$ on top of the (large) dc external field $B_0$
provides us with valuable information on the pinning landscape. Rather than
telling about the jumps $\Delta e_\mathrm{pin}$ in the energy landscape when
measuring $j_c$, the Campbell penetration depth
$\lambda_{\rm\scriptscriptstyle C}$ informs us about the force landscape,
specifically, the jumps $\Delta f_\mathrm{pin}$, see Fig.\
\ref{fig:strong_pinning}.

Solving the force-balance equation \eqref{eq:macroscopic_force_balance} for
the displacement field $U(X,t)$ (we denote coarse grained quantities averaging
over many vortices with capital letters, see Ref.\
\onlinecite{Willa_2015_PRB}) assuming a phenomenological Ansatz
\cite{Campbell_1978} for the pinning force density $F_\mathrm{pin} = F_0
-\alpha U$, one finds that the ac field penetrates the superconductor over a
distance given by $\lambda_{\rm\scriptscriptstyle C}^2 (\omega) =
B_0^2/[4\pi(\alpha - i\omega\eta)]$.  At the low frequencies typical of such
penetration experiments, we can drop the dissipative contribution $\propto
\eta\omega$ and obtain the phenomenological result
\begin{equation}\label{eq:l_C_phen}
   \lambda_{\rm\scriptscriptstyle C}(\omega) = \biggl[\frac{B_0^2}{4\pi\> \alpha}
   \biggr]^{1/2}
\end{equation}
due to Campbell \cite{Campbell_1978}. In the following, we discuss the
Campbell penetration physics within the strong pinning paradigm, first for the
zero-field cooled (ZFC) or critical state and subsequently for the field
cooled (FC) situation, including also hysteretic effects appearing upon
cycling the temperature up and down in the experiment.

\subsection{Campbell penetration depth $\lambda_{\rm\scriptscriptstyle C}$ in
\textrm{ZFC} state}\label{sec:ZFC}

Within our quantitative strong pinning theory, the action of the ac field on
the zero-field cooled state is to reshuffle vortices at the boundaries $-x_-$
and $x_+$, producing a restoring force density proportional to the
displacement $\mathbf{U}$ of the vortices. We compute the change in the
pinning force density $\delta \mathbf{F}_\mathrm{pin}(\mathbf{U})$ by
subtracting from \eqref{eq:F_pin_vec} the expression with the displaced branch
occupation $p_c(\mathbf{R} - \mathbf{U})$, 
\begin{align}\label{eq:dFpin_ZFC}
   \delta \mathbf{F}_\mathrm{pin}(\mathbf{U}) 
   & = - n_p \int \frac{d^2\mathbf{R}}{a_0^2}
   [p_{c}(\mathbf{R}) - p_{c}(\mathbf{R}-\mathbf{U})]\nonumber\\
   &\qquad\qquad\qquad\qquad
   \times[\mathbf{f}^\mathrm{p}_\mathrm{pin}(\mathbf{R})
   - \mathbf{f}^\mathrm{f}_\mathrm{pin}(\mathbf{R})]\nonumber\\
   &\approx - n_p \int \frac{d^2\mathbf{R}}{a_0^2}
   \left[\mathbf{\nabla}p_{c}(\mathbf{R})\cdot\mathbf{U}\right]
   \Delta \mathbf{f}^\mathrm{pf}_\mathrm{pin}(\mathbf{R}).
\end{align}
With $\mathbf{U}$ directed along $x$, the scalar product in the last line of
\eqref{eq:dFpin_ZFC} is non-vanishing only along the circular sections of
the trapping area in Fig.\ \ref{fig:geometry}; furthermore, the gradient
$\mathbf{\nabla}p_{c}(\mathbf{R})$ is strongly peaked (with unit weight) on
the circular boundaries and directed parallel to $\mathbf{e}_R$, the radial
unit vector. The scalar product then evaluates to
\begin{eqnarray}\nonumber
   &&-\mathbf{\nabla}p_{c}(\mathbf{R})\cdot\mathbf{U}=
    \bigl[
   \delta(R-x_-)(\Theta(\phi - \pi/2) - \Theta(\phi -3\pi/2))
   \\ \label{eq:s_prod_ZFC}
   &&\qquad +\, \delta(R-x_+)(\Theta(\phi + \phi_+) - \Theta(\phi - \phi_+))
   \bigr] U \cos \phi
\end{eqnarray}
with the polar angle $\phi$ restricted to angles $\pi/2 < \phi < 3\pi/2$ on
the left circular segment of the trapping boundary and $-\phi_+ < \phi <
\phi_+$ with $\phi_+=\arcsin(x_-/x_+)$ on the right one.  Inserting the
expression \eqref{eq:s_prod_ZFC} into \eqref{eq:dFpin_ZFC} and writing 
$\Delta \mathbf{f}^\mathrm{pf}_\mathrm{pin}(\mathbf{R}) \equiv - \Delta
f(R)\,\mathbf{e}_R$ (with $\Delta f(R)$ the modulus of $\Delta
\mathbf{f}^\mathrm{pf}_\mathrm{pin}(\mathbf{R})$) directed along the radial
coordinate, the change in the pinning force density can be evaluated as
\begin{align}\nonumber
   &\delta \mathbf{F}_\mathrm{pin}
   \approx - n_p\,U
   \biggl[
   \frac{x_- \Delta f(x_-)}{a_0^2}
   \int_{1}^{-1}  \!\!\! d\,\sin\phi\,[\cos\phi, \sin\phi]
   \\ \label{eq:dFpin_ZFC_bis}
   &\quad\quad\quad
   + \frac{x_+ \Delta f(x_+)}{a_0^2}
   \int_{-x_-/x_+}^{x_-/x_+} \!\!\!\!\!\!\!\! d\,\sin \phi\,[\cos\phi, \sin\phi]
   \biggr]
   \\ \nonumber
   &= -\frac{n_p\,U}{2a_0^2} \big[\pi\, x_- \Delta f(x_-) 
   + \theta_+\, x_+ \Delta f(x_+),0\big],
\end{align}
with the effective angle $\theta_+ = 2(\phi_+ + \sin \phi_+\cos \phi_+)$.  The
expression \eqref{eq:dFpin_ZFC_bis} is originally calculated for a left-shift
$U < 0$ of the vortex critical state; after a short initialization period
\cite{Willa_2015_PRB}, the same result applies for positive displacements as
well, and $\delta F_\mathrm{pin}(U) = -\alpha_\mathrm{sp} (U-U_0)$ with
$U_0(X,t) = \max_{t'<t} U(X,t')$, not to be confused with the barrier $U_0$
at the branch-cutting point $x_0$.

In the large $\kappa$ limit, $x_+\gg x_-$, see Eqs.\ \eqref{eq:x_+L} and
\eqref{eq:x_-L}, and the curvature in the boundary of the trapping region at
$R = x_+$ becomes negligible, see Fig.\ \ref{fig:geometry}. We can then
approximate $\phi_+\approx\sin \phi_+\approx x_-/x_+\ll 1$, and the strong
pinning expression for the effective curvature $\alpha_\mathrm{sp}$ in the
\textrm{ZFC} state reads
\begin{align}\label{eq:alpha_sp_lk}
   \alpha_\mathrm{sp} &\approx n_p\frac{t_\perp}{a_0}
   \frac{\frac{\pi}{4} \Delta f^\mathrm{pf}_\mathrm{pin}
   + \Delta f^\mathrm{fp}_\mathrm{pin}}{a_0},
\end{align}
with the force jumps $\Delta f^\mathrm{pf}_\mathrm{pin} =
f^\mathrm{p}_\mathrm{pin}(-x_-) - f^\mathrm{f}_\mathrm{pin}(-x_-) >0$ and
$\Delta f^\mathrm{fp}_\mathrm{pin} = f^\mathrm{f}_\mathrm{pin}(x_+) -
f^\mathrm{p}_\mathrm{pin}(x_+)>0$, where we have returned to the original
notation for the two jumps at $-x_-$ and $x_+$ for convenience (i.e., the
difference $\Delta f^\mathrm{fp}_\mathrm{pin}$ is equal to the modulus $\Delta
f(x_+)$ in Eq.\ \eqref{eq:dFpin_ZFC_bis}).  The factor $\pi/4$ in the
numerator of \eqref{eq:alpha_sp_lk} has its origin in the different geometries
of the circular boundaries at $-x_-$ and at $x_+$.

In the opposite limit of marginally strong pinning with $\kappa-1 \ll 1$,
$x_-\lesssim x_0 \lesssim x_+$ and the trapping region acquires an approximately
circular geometry. The angle $\phi_+$ can be expanded as  $\phi_+\approx \pi/2
- \delta \phi_+/2$, with $\delta \phi_+ \ll \pi/2$, allowing to approximate
the effective angle $\theta_+$ as
\begin{equation}\label{eq:angle_pi/2}
   \theta_+ \approx \pi + \mathcal{O}(\delta \phi_+^3).
\end{equation}
The bistable region is symmetric around $x_0$, see Eq.\ \eqref{eq:x_pm_sk},
and we have $\Delta f(x_-) \approx \Delta f(x_+) = \Delta
f^\mathrm{fp}_\mathrm{pin}(x_+)$ as well as $x_0 \approx (x_+ + x_-)/2$, that
produces the following simple expression for the Campbell curvature
\begin{align}\label{eq:alpha_sp_sk}
   \alpha_\mathrm{sp} &\approx n_p \frac{\pi x_0}{a_0} 
   \frac{\Delta f^\mathrm{fp}_\mathrm{pin}(x_+)} {a_0},
\end{align}
where we have again returned to the original notation for the jump at $x_+$,
$\Delta f^\mathrm{fp}_\mathrm{pin}(x_+) = f^\mathrm{f}_\mathrm{pin}(x_+) -
f^\mathrm{p}_\mathrm{pin}(x_+)>0$.  The above results differ from those
in Ref.\ \onlinecite{Willa_2016} in the more accurate handling of the geometry
in the trapping boundary, leading to the appearance of $\theta_+$ and factors
$\pi/4$.

The derivation \eqref{eq:dFpin_ZFC_bis} applies to the critical state---the
corresponding result for other vortex states is obtained by the proper
replacement $p_c(\mathbf{R}) \to p(\mathbf{R})$. E.g., for the equilibrium
distribution $p_\mathrm{eq}(\mathbf{R})$ with a radially symmetric jump at $R
\approx x_0$, the result reads
\begin{equation}\label{eq:alpha_0}
   \alpha_0 = n_p \frac{\pi x_0}{a_0} \frac{\Delta f^\mathrm{fp}_\mathrm{pin}(x_0)}{a_0}.
\end{equation}
Note that, at $\kappa - 1 \ll 1$, the jump $\Delta
f^\mathrm{fp}_\mathrm{pin}(x_0) = f^\mathrm{f}_\mathrm{pin}(x_0) -
f^\mathrm{p}_\mathrm{pin}(x_0)$ at $x_0$ is larger (by a factor
$2/\sqrt{3}$) than the jumps $\Delta f^\mathrm{fp}_\mathrm{pin}$ at $x_+$ or
$\Delta f^\mathrm{pf}_\mathrm{pin}$ at $x_-$.

Physically, the expressions $\eqref{eq:alpha_sp_lk}-\eqref{eq:alpha_0}$
describe the {\it average curvature} in the pinning landscape that, upon
integration, is given by the sum of jumps in the pinning force
$f_\mathrm{pin}$. This should be compared with the {\it average pinning force}
in Eq.\ \eqref{eq:F_pin} that provides the critical force density $F_c$ and is
given by the sum of jumps in the pinning energy $e_\mathrm{pin}$.
Furthermore, the results for the Campbell curvature
$\eqref{eq:alpha_sp_lk}-\eqref{eq:alpha_0}$ involve the precise geometry of
the trapping area with its circular boundaries, while the pinning force
density \eqref{eq:F_pin} depends only on the total width $t_\perp = 2x_-$.
The above interpretation of $\alphaSP \propto \Delta f_\mathrm{pin}$ in terms
of the {\it average} curvature naturally relates the strong pinning result to
the phenomenological derivation of $\lambda_{\rm\scriptscriptstyle C}$ by
Campbell \cite{Campbell_1978}.  Finally, we obtain the microscopic expression
for the Campbell penetration depth within strong pinning theory,
\begin{equation}\label{eq:l_C_spt}
   \lambda_{\rm\scriptscriptstyle C}(\omega) = \biggl[\frac{B_0^2}
   {4\pi\alpha_\mathrm{sp}}\biggr]^{1/2}.
\end{equation}

\subsection{Creep effects on $\lambda_{\rm\scriptscriptstyle C}$ in \textrm{ZFC} state}
\label{sec:creep_l_C_th}

At finite temperatures, the analysis of the vortex system's linear response
proceeds in a manner analogous to the one above ignoring thermal fluctuations,
with the following modifications: {\it i}) the oscillations in the vortex
lattice induced by the small $ac$ field now reshuffle those vortex lines close
to the thermal jump points at $-\xmjp$ and $\xpjp$, implying that the relevant
jumps in force are $\Delta f^\mathrm{pf,\,jp}_\mathrm{pin} =
f^\mathrm{p}_\mathrm{pin}(-\xmjp) - f^\mathrm{f}_\mathrm{pin}(-\xmjp) > 0$ and
$\Delta f^\mathrm{fp,\,jp}_\mathrm{pin} = f^\mathrm{f}_\mathrm{pin}(\xpjp) -
f^\mathrm{p}_\mathrm{pin}(\xpjp) > 0$, and  {\it ii}) the trapping area
involves the renomalized jump locations $x_\pm^\mathrm{jp}$, producing the
thermally renormalized angles $\phi_+^\mathrm{jp} =
\arcsin \bigl(x_-^\mathrm{jp} /x_+^\mathrm{jp}\bigr)$ and
$\theta_+^\mathrm{jp}=2(\phi_+^\mathrm{jp} + \sin \phi_+^\mathrm{jp}\cos
\phi_+^\mathrm{jp})$.  After a short initialization period, that is not
relevant for the present discussion, the field penetration is determined by
the effective curvature
\begin{multline}\label{eq:alpha_finite_T}
   \alpha_\mathrm{sp}(t,T)=n_p\bigg(
   \frac{\pi}{2}\frac{x_-^\mathrm{jp}}{a_0} 
   \frac{\Delta f^\mathrm{pf,\,jp}_\mathrm{pin}}{a_0}
   + \frac{\theta_+^\mathrm{jp}}{2}
   \frac{x_+^\mathrm{jp}}{a_0} \frac{\Delta f^\mathrm{fp,\,jp}_\mathrm{pin}}{a_0}
   \bigg).
\end{multline}
Equation \eqref{eq:alpha_finite_T} is the central result of this work; it
allows us to trace the evolution of the Campbell penetration length
$\lambda_{\rm \scriptscriptstyle C}(t,T) \propto 1/\sqrt{\alpha_\mathrm{sp}}$
as a function of time $t$ during a relaxation experiment. The expression
\eqref{eq:alpha_finite_T} fully characterizes the dependence of
$\alpha_\mathrm{sp}$ on the pinning parameters for times $t \gg t_0$. 

At short times and very strong pinning $\kappa\gg1$, the branch occupation is
highly asymmetric and $x_+^\mathrm{jp}\gg x_-^\mathrm{jp}$, leading to
$\theta_+^\mathrm{jp}\approx 4\phi_+^\mathrm{jp}\approx
4x_-^\mathrm{jp}/x_+^\mathrm{jp}$. The Campbell curvatures then reads
\begin{equation}\label{eq:alpha_finite_T_lk}
   \alpha_\mathrm{sp}(t,T) \approx n_p\frac{t^\mathrm{jp}_\perp}{a_0}
   \frac{\frac{\pi}{4} \Delta f^\mathrm{pf,jp}_\mathrm{pin} 
   +\Delta f^\mathrm{fp,jp}_\mathrm{pin}}{a_0},
\end{equation}
with the thermally enhanced trapping length $t^\mathrm{jp}_\perp =
2x^\mathrm{jp}_-$. In the marginally strong pinning limit, we have
$\kappa-1\ll 1$ and $x_-^\mathrm{jp}\lesssim x_0\lesssim x_+^\mathrm{jp}$,
leading to a saturation of the effective angle $\theta_+^\mathrm{jp}\to \pi$.
In this regime, relaxation behaves symmetrically on both sides of the bistable
region with $\Delta f^\mathrm{pf,jp}_\mathrm{pin} \approx \Delta
f^\mathrm{fp,jp}_\mathrm{pin}$, see Eq.\ \eqref{eq:f_p_exp_sk}. Using $x_0
\approx (x_-^\mathrm{jp} + x_+^\mathrm{jp})/2$, the Campbell curvature takes a
simple form analogous to Eq.\ \eqref{eq:alpha_sp_sk},
\begin{equation}\label{eq:alpha_finite_T_sk}
   \alpha_\mathrm{sp}(t,T) \approx n_p \frac{\pi x_0}{a_0} 
   \frac{\Delta f^\mathrm{fp,jp}_\mathrm{pin}} {a_0}.
\end{equation}

\begin{figure*}
        \includegraphics[width = 1.\textwidth]{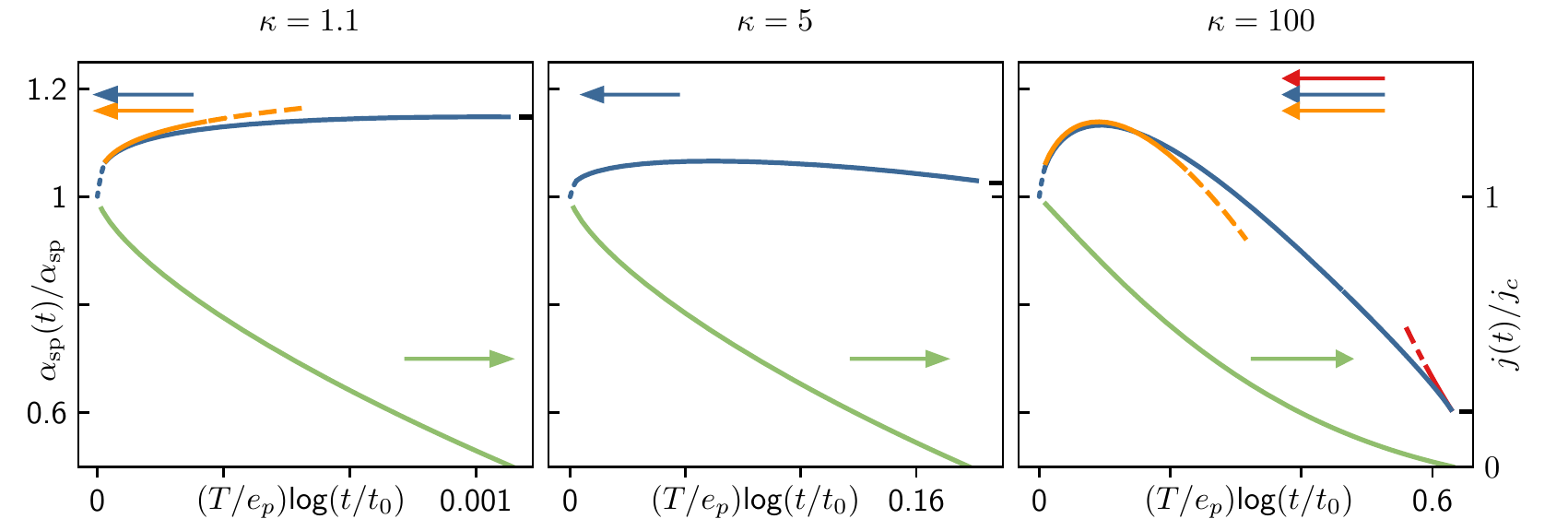}
	\caption{Comparison of the numeric (blue) and analytic (orange and
	red) results for the scaled Cambpell curvature (upper part) and the
	scaled persistent current density $j(t,T)$, Eq.\ \eqref{eq:pers_curr}
	(green, lower part), as a function of the creep parameter in the form
	of the scaled logarithmic time $(T/e_p)\log(t/t_0)$.  We have assumed
	a Lorentzian pinning potential with different values of $\kappa$.  For
	marginally strong pinning, $\kappa = 1.1$, $\alphaSP(t,T)/\alphaSP$
	grows monotonically as a function of time, with a satisfactory
	agreement between numeric and analytic (Eq.\
	\eqref{eq:resc_curv_small_kappa}) results; throughout the domain of
	applicability, the result is dominated by the positive contribution of
	the curvature in the pinning force branches close to $-x_-$ and $x_+$.
	For very strong pinning, $\kappa = 100$, the increase away from zero
	is dominated by the enhanced trapping length $t_\perp^\mathrm{jp}$; at
	larger times, the negative contribution to the force jump dominates
	the behavior, producing a pronounced maximum in the Campbell curvature
	$\alphaSP(t,T)$ that is visible in both the numeric (blue) and
	analytic (orange, Eq.\ \eqref{eq:resc_curv_large_kappa_lT}) results.
	The precision in the large $\kappa$ result is limited in time with
	deviations showing up when the vortex state approaches equilibrium
	where the force jumps are close to $x_0$ rather than $\pm x_\pm$. The
	red curve is the result Eq.\ \eqref{eq:resc_curv_large_kappa_slow} of
	an expansion around equilibrium that agrees well with the numeric
	curve.  The thick ticks at large times mark the asymptotic value
	$\alpha_0/\alphaSP$ close to equilibrium $t \sim t_\mathrm{eq}$ before
	entering the TAFF region, see text. At intermediate values of the
	pinning strength, $\kappa = 5$, the increase in $\alphaSP(t,T)
	/\alphaSP$, originally due to the curvature term in the force jump,
	gets further enhanced by the increase in trapping length.  The
	(numeric) lines terminate at the boundary of applicability $(T/e_p)
	\ln(t/t_0) \approx (\kappa-1)^2/8$ and $\sim 1$ for marginally and
	very strong pinning, respectively. At very short times $t \sim t_0$,
	our creep analysis breaks down as the barriers $U$ vanish.  The
	persistent current density  $j(t,T)$ (green) decreases monotonically
	with time for all three values of $\kappa$, approaching the TAFF
	region at $t \to t_\mathrm{eq}$ and vanishing in the long-time limit.
	On the contrary, the curvature $\alphaSP(t,T)$ approaches a finite
	value $\alpha_0$ for $t \rightarrow t_\mathrm{eq}$ due to the finite
	value of the force jump $\Delta f_\mathrm{pin}^\mathrm{fp}(x_0)$ at
	$x_0$ and then crosses over to the TAFF skin effect.}
     \label{fig:rescaled_curvatures}
\end{figure*}

A numerical evaluation of the Campbell curvature $\alphaSP(t,T)$, Eqs.\
\eqref{eq:alpha_finite_T} and \eqref{eq:alpha_finite_T_sk}, as a function of
the creep parameter $\mathcal{T} = (T/e_p)\ln(t/t_0)$ at different pinning
strengths $\kappa$ is shown in Fig.\ \ref{fig:rescaled_curvatures} (blue
lines). For comparison, we also show the decaying persistent current density
\eqref{eq:pers_curr} in the region $j < j_c$.  The plots show that
$\alphaSP(t,T)$ first increases with time and decreases at long times (but not
in the marginal case with $\kappa$ close to unity). From a phenomenological
perspective, this can be understood as a change in the relative occupation of
shallow and deep pinning wells in the course of relaxation.  Furthermore, we
find that, while the persistent current density $j(t,T)$ ultimately vanishes
on approaching the equilibrium state, the Campbell curvature $\alphaSP(t,T)$
remains finite and large. We will discuss these findings in detail later;
before, we analyze their physical origin with the help of analytic
considerations in the limits of moderately strong ($\kappa -1 \ll 1$) and very
strong ($\kappa \gg 1$) pinning.

\subsubsection{Short and intermediate times}\label{sec:st}

Following Eq.\ \eqref{eq:alpha_finite_T}, we have to determine the
thermally renormalized jumps $\Delta f^\mathrm{fp,jp}_\mathrm{pin}$ and
$\Delta f^\mathrm{pf,jp}_\mathrm{pin}$ as well as the corresponding trapping
parameters. Here, we first analyze the situation at short times $t \ll
t_\mathrm{eq} = t_0 \exp(U_0/T)$, where the jump positions $-x_-^\mathrm{jp}$
and $x_+^\mathrm{jp}$ remain close to the edges of the bistability intervals
$|x| \in [x_-,x_+]$,
\begin{equation}\label{eq:dx_pm}
   x^\mathrm{jp}_\pm = x_\pm \mp \delta x_\pm,
\end{equation}
with small asymptotic shifts $\delta x_\pm > 0$. Furthermore, we begin our
discussion with a study of the very strong pinning regime, where the bistable
interval $[x_-,x_+]$ is well separated from the defect potential, as $x_- \sim
\kappa^{1/(n+2)} \xi$ and $x_+ \sim \kappa \xi$ are both much larger than
$\xi$, see Eqs.\ \eqref{eq:x_+} -- \eqref{eq:x_-} for more precise
expressions for $x_\pm$.

In this limit, Eq.\ \eqref{eq:alpha_finite_T_lk} for $\alpha_\mathrm{sp}$ is
applicable, with the renormalized trapping length straightforwardly relating to
$\delta x_-$, $t_\perp^\mathrm{jp} = 2 x_-^\mathrm{jp} = t_\perp + 2 \delta
x_-$. Since $\delta x_- > 0$, we find that $t_\perp^\mathrm{jp} > t_\perp$,
with thermal fluctuations assisting vortex trapping. Hence, the task of
finding $t_\perp^\mathrm{jp}$ is reduced to the calculation of the shift
$\delta x_-$ in the jump position.

Next, we focus on the total force jump $\Delta f^\mathrm{jp}_\mathrm{pin} =
(\pi/4) \Delta f^\mathrm{pf,jp}_\mathrm{pin} + \Delta
f^\mathrm{fp,jp}_\mathrm{pin}$.  It is convenient to determine the difference
in force jumps between the shifted and original jump positions $\pm
x_\pm^\mathrm{jp}$ and $\pm x_\pm$ in the form
\begin{eqnarray}\label{eq:df_total}
   \Delta f^\mathrm{jp}_\mathrm{pin} - \Delta f_\mathrm{pin} &\approx&
   \rho[\delta f^\mathrm{p}_\mathrm{pin}(-x_-^\mathrm{jp})
   - \delta f^\mathrm{f}_\mathrm{pin}(-x_-^\mathrm{jp})] \\ \nonumber
   &&\quad + \delta f^\mathrm{f}_\mathrm{pin}(x_+^\mathrm{jp})
    - \delta f^\mathrm{p}_\mathrm{pin}(x_+^\mathrm{jp}),
\end{eqnarray}
with $\delta f^\mathrm{i}_\mathrm{pin}(x_+^\mathrm{jp}) =
f^\mathrm{i}_\mathrm{pin} (x_+^\mathrm{jp}) - f^\mathrm{i}_\mathrm{pin} (x_+)$
and $\mathrm{i} = \mathrm{p},\mathrm{f}$, as well as corresponding expressions
at $-x_-^\mathrm{jp}$. Above, we have introduced the ratio
\begin{equation}
   \rho={\pi}/{4}
\end{equation}
between the semi-circular and rectangular areas appearing in the trapping
geometry of Fig.\ \ref{fig:geometry}, see Eq. \eqref{eq:alpha_sp_lk}.  Taking
a closer look at Fig.\ \ref{fig:strong_pinning}, we see that the differences
in \eqref{eq:df_total} involve a large term $\delta
f^\mathrm{p}_\mathrm{pin} (x_+^\mathrm{jp})$ on the linear pinning branch near
$x_+$, a corresponding term $\delta
f^\mathrm{p}_\mathrm{pin}(-x_-^\mathrm{jp})$ near $-x_-$, as well as a term
$\delta f^\mathrm{f}_\mathrm{pin} (-x_-^\mathrm{jp})$ on the (curved, see
Fig.\ \ref{fig:strong_pinning}(d) inset) free branch near $-x_-$, the
remaining term $\delta f^\mathrm{f}_\mathrm{pin}(x_+^\mathrm{jp})$ being
obviously small at large $\kappa$.

The shifts $\delta f^\mathrm{p}_\mathrm{pin}$ at $-x_-^\mathrm{jp}$ and
$x_+^\mathrm{jp}$ are easily obtained from combining Eqs.\
\eqref{eq:force_balance} and \eqref{eq:fo_pin_tot},
\begin{equation}\label{eq:fbx}
   \Cbar(r-x) = f_p(r) = f_\mathrm{pin}(x).
\end{equation}
With $r_\mathrm{p}(x) < \xi$ and $x \in [x_-,x_+] \gg \xi$, we find that
$f^\mathrm{p}_\mathrm{pin}(x) \approx -\Cbar x$ and
\begin{equation}\label{eq:dfp+}
   \rho\delta f^\mathrm{p}_\mathrm{pin}(-x_-^\mathrm{jp}) 
   - \delta f^\mathrm{p}_\mathrm{pin}(x_+^\mathrm{jp}) 
   \approx  \Cbar \, \left(\rho\,\delta x_- - \delta x_+\right)
\end{equation}
follows from the shifts $\delta x_\pm$.  

For the calculation of the curvature term $\delta f^\mathrm{f}_\mathrm{pin}
(-x_-^\mathrm{jp})$, see Fig.\ \ref{fig:strong_pinning}(d), we have to include
both shifts in $r$ and $x$ in \eqref{eq:fbx} and find that
\begin{eqnarray}\label{eq:dfp-}
   \rho\, \delta f^\mathrm{f}_\mathrm{pin}(-x_-^\mathrm{jp}) &\approx&
   -\rho\, \Cbar \, (\delta r_\mathrm{f-} - \delta x_-),
\end{eqnarray}
with
\begin{eqnarray}\label{eq:rf_easy}
   r_\mathrm{f}(x_-^\mathrm{jp}) \equiv r_\mathrm{f}(x_-) + \delta r_\mathrm{f-}.
\end{eqnarray}
Note that $r_\mathrm{f}(x)$ increases with $x$, see Fig.\
\ref{fig:self-cons-sol}, hence we have $\delta r_\mathrm{f-} >0$ in the above
equation.  As a result, we obtain the thermal change in force jumps
\begin{eqnarray}\label{eq:df_total_lk}
   \Delta f^\mathrm{jp}_\mathrm{pin} - \Delta f_\mathrm{pin} &\approx&
   \Cbar(\rho\,\delta r_\mathrm{f-} - \delta x_+),
\end{eqnarray}
leaving us with the task to find the shifts $\delta x_+$ and $\delta
r_\mathrm{f-}$ in the asymptotic and free tip positions, the terms $\pm \rho\,\Cbar
\delta x_-$ from Eqs.\ \eqref{eq:dfp+} and \eqref{eq:dfp-} cancelling out.

Next, we determine the shift $\delta r_\mathrm{f-}$ in the tip position by
expanding the microscopic force balance Eq.\ \eqref{eq:force_balance} around
the branch endpoint $x_-$ (where $f_p'[r_\mathrm{f}(x_-)] = \Cbar$) and find
that $\delta x_- \approx (|f_p^{\prime\prime}[r_\mathrm{f}(x_-)]|/2\Cbar) \>
\delta r_{\mathrm{f}-}^{\> 2}$, hence the tip shift
\begin{eqnarray}\label{eq:dr_dx_easy}
   \delta r_\mathrm{f-} &\approx& \left(\frac{2\Cbar}{|f_p^{\prime\prime}[r_\mathrm{f}(x_-)]|}
   \>{\delta x_-}\right)^{1/2}
\end{eqnarray}
scales with the square root of the asymptotic shift $\delta x_-$. The
corresponding result for $\delta r_\mathrm{p+} = r_\mathrm{p}(x_+) -
r_\mathrm{p}(x_+^\mathrm{jp}) > 0$ involves $f_p^{\prime\prime}
[r_\mathrm{p}(x_+)]$.

At large $\kappa$, we can evaluate this expression within the tail of the
defect potential; assuming, as before, an algebraic decay $e_p(R) \approx 2 e_p\,
(\xi/R)^n$, we find that $|f_p^{\prime\prime}[r_\mathrm{f}(x_-)]| \approx
\Cbar (n+2)/r_\mathrm{f}(x_-)$ and relating $r_\mathrm{f}(x_-)$ to $x_-$ via
Eq.\ \eqref{eq:x_-}, we obtain the shift in the tip position
\begin{eqnarray}\label{eq:dr_f-}
   \delta r_{\mathrm{f}-} \approx \frac{\sqrt{2(n+1)}}{n+2} \sqrt{x_- \delta x_-}.
\end{eqnarray}
Approximating the force jump $\Delta f_\mathrm{pin} \approx \Cbar (x_+ +
\rho\, x_-) \approx \Cbar x_+$ in the absence of fluctuations by its leading
term, we arrive at a compact result for the scaled Campbell curvature at large
$\kappa$,
\begin{eqnarray}\label{eq:resc_curv_large_kappa_gen}
   &&\frac{\alpha_\mathrm{sp}(t,T)}{\alpha_\mathrm{sp}}
   \approx \biggl(1+ \frac{\delta x_-}{x_-} \biggr)
   \biggl(1 + \frac{\pi}{4}\frac{\delta r_\mathrm{f-}}{x_+}
                        - \frac{\delta x_+}{x_+} \biggr)
   \\ \nonumber
   &&\quad \approx \biggl(1+ \frac{\delta x_-}{x_-} \biggr)
   \biggl(1 
            + \frac{\pi}{4}\frac{\sqrt{2(n+1)}}{n+2}\frac{x_-}{x_+}
                    \sqrt{\frac{\delta x_-}{x_-}}
                        - \frac{\delta x_+}{x_+} \biggr).
\end{eqnarray}
Here, the linear terms $\propto \delta x_-/x_-$ and $\propto \delta x_+/x_+$
are universal, while the square-root- or curvature term $\propto \sqrt{\delta
x_-/x_-}$ depends on the shape of the defect potential, with the numerical
prefactor describing a potential with an algebraic tail decaying as $1/R^n$.
The result \eqref{eq:resc_curv_large_kappa_gen} involves several competing
elements: The first factor with a linear correction $\delta x_-/x_-$ is due to
the enhanced trapping distance and always leads to an increase in the Campbell
curvature $\alpha_\mathrm{sp}(t,T)$. On the other hand, the second factor
originating from the renormalized force jump contributes with competing terms:
While the positive term $\propto \sqrt{\delta x_-/x_-}$ arising from the
curvature in $f_\mathrm{pin}^\mathrm{f}$ at $-x_-$, see Fig.\
\ref{fig:strong_pinning}(d), is the dominating one at small shifts, i.e.,
short times, the negative contribution $- \delta x_+/x_+$ deriving from the
pinned branch $f_\mathrm{pin}^\mathrm{p}$ at $x_+$ becomes relevant at
intermediate and larger times.

Note that this competition between trapping area and force jumps appears as
well in the discussion of the pinning force density $F_\mathrm{pin}(t,T)$ in
Eq.\ \eqref{eq:F_pin_tT_num}, but with the force jump $\Delta f_\mathrm{pin}$
replaced by the jump $\Delta e_\mathrm{pin}$ in pinning energy.  This
competition has been encoded in the pre\-factor $g(\kappa)$ of
\eqref{eq:F_pin_tT} that involves the corresponding two factors related to
trapping and energy jumps. However, this time, the total energy jump misses
the positive square-root term present in \eqref{eq:resc_curv_large_kappa_gen}
and involves terms linear (negative) and quadratic (positive) in $\delta x_+$
(since $e_\mathrm{pin} \approx \Cbar\, (x_+-\delta x_+)^2$ for large
$\kappa$).  It turns out that the negative linear term in the total energy
jump dominates over the positive correction in the trapping area, up to very
large $\kappa$-values beyond $\kappa \sim 10^2$, such that
$F_\mathrm{pin}(t,T)$ decreases monotonously. Increasing $\kappa$ beyond this
very large value, the situation gets reversed and we find a regime where creep
{\it enhances} the pinning force density, with quite interesting new
observable effects that will be discussed in a separate paper
\cite{Gaggioli_2022}.

Going beyond small values of $\delta x_-$, the result
\eqref{eq:resc_curv_large_kappa_gen} has to be modified, since the square root
approximation for $\delta r_\mathrm{f-} \propto \sqrt{x_- \delta x_-}$ breaks
down.  This is easily seen when considering Fig.\ \ref{fig:strong_pinning} for
larger values of $\delta x_-$, where the curvature in the free branch
$f_\mathrm{pin}^\mathrm{f} (x)$ flattens out and $f_\mathrm{pin}^\mathrm{f}
(-x_-^\mathrm{jp}) \to 0$ vanishes. In this regime, we have $\delta
f_\mathrm{pin}^\mathrm{f}(-x_-^\mathrm{jp}) \approx -
f_\mathrm{pin}^\mathrm{f}(-x_-) = \Cbar[r_\mathrm{f}(x_-) - x_-] \approx
-\Cbar x_-/(n+2)$, where we have used the force balance equation
\eqref{eq:fbx} as well as the result \eqref{eq:x_-} for $r_\mathrm{f}(x_-)$
and $x_-$. The result \eqref{eq:resc_curv_large_kappa_gen} then is replaced by
\begin{eqnarray}\label{eq:resc_curv_large_kappa_gen_l}
   &&\frac{\alpha_\mathrm{sp}(t,T)}{\alpha_\mathrm{sp}}
   \approx \biggl(1+ \frac{\delta x_-}{x_-} \biggr)
   \\ \nonumber
   &&\qquad\qquad \times \biggl(1
            +\frac{\pi}{4} \frac{x_-}{x_+}\Bigl(\frac{1}{n+2} 
            + \frac{\delta x_-}{x_-}\Bigr)
            - \frac{\delta x_+}{x_+} \biggr)
\end{eqnarray}
at large $\delta x_- \gg x_-/2(n+1)$. Note that, while $\delta x_-$ is
parametrically small in $\kappa$ as compared to $\delta x_+$, this is not the
case for the ratios ${\delta x_-}/{x_-}$ and ${\delta x_+}/{x_+}$, since
pinning and depinning appear on very different scales $x_-$ and $x_+$,
respectively.

Next, we wish to evaluate the expressions \eqref{eq:resc_curv_large_kappa_gen}
and \eqref{eq:resc_curv_large_kappa_gen_l} in terms of experimental
parameters, i.e., as a function of temperature $T$ and waiting time $t$.  We
first find the shifts $\delta x_\pm^\mathrm{jp}$ in the jump positions for the
free and pinned branches near $-x_-$ and $x_+$, respectively. These are
determined by the condition \eqref{eq:barriers}, telling us that we have to
evaluate the depinning and pinning barriers \cite{Buchacek_2019} as given by
Eq.\ \eqref{eq:U_creep} close to $-x_-$ and $x_+$, respectively, see Fig.\
\ref{fig:strong_pinning}(b).  The expressions for $U_\mathrm{p}$ and
$U_\mathrm{dp}$ involve the free and pinned tip position
$r_\mathrm{f}(x_-^\mathrm{jp})$ and $r_\mathrm{p}(x_+^\mathrm{jp})$ discussed
above, cf.\ Eq.\ \eqref{eq:rf_easy}, as well as the unstable positions
$r_\mathrm{us}$ at $x_\pm^\mathrm{jp}$. The latter are arranged symmetrically
with respect to $r_\mathrm{f} (x_-)$ and $r_\mathrm{p}(x_+)$,
$r_\mathrm{us}(x_-^\mathrm{jp}) = r_\mathrm{f} (x_-) - \delta r_{\mathrm{f}-}$
and $r_\mathrm{us} (x_+^\mathrm{jp}) = r_\mathrm{p}(x_+) + \delta
r_{\mathrm{p}+}$.

While the shift $\delta r_\mathrm{f-}$ is given in Eq.\ \eqref{eq:dr_dx_easy},
the shift $\delta r_\mathrm{p+}$ involves the derivative $f_p^{\prime\prime}
[r_\mathrm{p}(x_+)]$ at short distances that depends on the details of the
pinning potential. Using the dimensional estimate $f_p^{\prime\prime}
[r_\mathrm{p}(x_+)] \sim e_p/\xi^3$, we find that
\begin{eqnarray}\label{eq:dr_p+}
   \delta r_{\mathrm{p}+} \sim [(\xi/\kappa) \> \delta x_+]^{1/2}.
\end{eqnarray}
For a Lorentzian potential, we have the more precise results
$f_p^{\prime\prime} [r_\mathrm{p}(x_+)] \approx (3/2)^{7/2} e_p/4\xi^3$ and
$\delta r_{\mathrm{p}+} \approx (\sqrt{2}/\kappa) (2/3)^{5/2} \sqrt{x_+ \delta
x_+}$. Expanding the pinning/depinning barriers $\Up$ and $\Udp$ away from
$-x_-$ and $x_+$, we find the barriers (in compact notation)
\begin{eqnarray} \label{eq:barrier_gen}
   U \approx \frac{4}{3} \Cbar \xi^2 \sqrt{2\Cbar/f_p^{\prime\prime} \xi}\>
   (\delta x/\xi)^{3/2}
\end{eqnarray}
where the second derivative $f_p^{\prime\prime}$ has to be evaluated at
$r_\mathrm{f}(-x_-)$ ($r_\mathrm{p}(x_+))$ for the pinning barrier $\Up$ (the
depinning barrier $\Udp$).  Focusing on a Lorentzian potential, we arrive at
\begin{eqnarray} \label{eq:barrier_large_v}
   \Up(-x_-^\mathrm{jp}) &\approx&
   \frac{e_p}{\sqrt{3}\,\kappa^{7/8}} \left(\frac{1}{3}\right)^{3/8} 
   (\delta x_-/\xi)^{3/2},
   \\ \label{eq:barrier_large_v+}
   \Udp(x_+^\mathrm{jp}) &\approx& \frac{e_p}{\sqrt{3}\, \kappa^{3/2}}
   \left(\frac{2}{3}\right)^{9/4} 
   (\delta x_+/\xi)^{3/2},
\end{eqnarray}
with contributions from $e^\prime_\mathrm{pin} \delta r \propto \bar C \,
\delta x\, \delta r$ and $e^{\prime\prime\prime}_\mathrm{pin} \propto \Cbar\,
\delta r^3$.  After a waiting time $t$, thermal fluctuations at temperature
$T$ can overcome barriers of size $T \ln (t/t_0)$, rendering smaller barriers
ineffective. Making use of Eq.\ \eqref{eq:barriers} as well as the creep
parameter $\mathcal{T}$ defined in \eqref{eq:def_U}, we obtain the final
results for the thermal shifts
\begin{eqnarray}\label{eq:delta_x_pm_fin}
   \delta x_+ &\approx& (3/2)^{3/2} \kappa \xi \> (\sqrt{3}\mathcal{T})^{2/3}, 
   \\ \nonumber
   \delta x_- &\approx& 3^{1/4} \kappa^{7/12} \xi \> (\sqrt{3}\mathcal{T})^{2/3}, 
   \\ \nonumber
   \delta r_{\mathrm{f}-} &\approx& 3^{1/4} \kappa^{5/12} \xi \> (\sqrt{3}\mathcal{T})^{1/3}.
\end{eqnarray}
Indeed, $\delta x_- \sim \kappa^{-5/12} \delta x_+ \ll \delta x_+ $ is small,
but the ratio $\delta x_-/x_- \approx (3/8) \kappa^{1/3}
(\sqrt{3}\mathcal{T})^{2/3}$ dominates $\delta x_+/x_+ \approx
(\sqrt{3}\mathcal{T})^{2/3}$ for $\kappa > 19$.

Returning back to the evaluation of the Campbell curvature
\eqref{eq:resc_curv_large_kappa_gen}, we find the renormalized trapping length
(with the numericals appertaining to a Lorentzian potential)
\begin{eqnarray}\label{eq:tp_lkappa}
   t_\perp^\mathrm{jp}(t,T) &\approx& 
   t_\perp + 2 \delta x_- \\ \nonumber
   &\approx& t_\perp \bigl[1+ (3/8) \kappa^{1/3}\, (\sqrt{3}\mathcal{T})^{2/3}\bigr],
\end{eqnarray}
where we have used that $t_\perp = 2 x_- \approx (16/3) (3\kappa)^{1/4} \xi$
in the last expression. Combining Eqs.\ \eqref{eq:df_total_lk} and
\eqref{eq:delta_x_pm_fin}, the final result for the renormalized total force
jump is
\begin{eqnarray}\label{eq:df_lkappa_dx}
   \Delta f^\mathrm{jp}_\mathrm{pin} -\Delta f_\mathrm{pin}
   &\approx& \kappa\xi\Cbar\>
   \Bigl[\rho\, 3^{1/4} \kappa^{-7/12} (\sqrt{3}\mathcal{T})^{1/3} 
   \\ \nonumber
   && \qquad -(3/2)^{3/2} (\sqrt{3}\mathcal{T})^{2/3}\Bigr].
\end{eqnarray}
These results then provide us with an expression for the (scaled) Campbell
curvature in the large-$\kappa$ -- small-time limit
\begin{eqnarray}\label{eq:resc_curv_large_kappa}
   &&\frac{\alpha_\mathrm{sp}(t,T)}{\alpha_\mathrm{sp}}
   \approx \biggl(1+ \frac{3}{8} \kappa^{1/3}
   (\sqrt{3}\mathcal{T})^{2/3}\, \biggr) 
   \\ \nonumber
   &&\qquad\qquad \times 
   \biggl(1 + \frac{\pi}{6}\frac{(4/3)^{1/4}}{\kappa^{7/12}}
                                (\sqrt{3}\mathcal{T})^{1/3}
                              - (\sqrt{3}\mathcal{T})^{2/3}\biggr),
\end{eqnarray}
where we have used that $\Delta f_\mathrm{pin} \approx
\Cbar(\rho\, x_- + x_+) \approx (3/2)^{3/2} \kappa \xi \Cbar$,
keeping only the leading term in $\kappa$. Equation
\eqref{eq:resc_curv_large_kappa} expresses the generic large-$\kappa$ result
\eqref{eq:resc_curv_large_kappa_gen} in terms of the creep parameter
$\mathcal{T}(t,T)$ with the numericals describing the situation for a
Lorentzian pinning potential.

Going beyond small values of $\mathcal{T}$, we have to use Eq.\
\eqref{eq:resc_curv_large_kappa_gen_l}; the condition $\delta x_- \gg x_-/6$
then translates to a creep parameter $\sqrt{3} \mathcal{T} > (2/3)^3
/\kappa^{1/2}$. Evaluating \eqref{eq:resc_curv_large_kappa_gen_l} using the
thermal shifts $\delta x_\pm$ in \eqref{eq:delta_x_pm_fin}, we find the result
for larger values of $\mathcal{T}$ to take the slightly modified form,
\begin{eqnarray}\label{eq:resc_curv_large_kappa_lT}
   &&\frac{\alpha_\mathrm{sp}(t,T)}{\alpha_\mathrm{sp}}
   \approx \biggl(1+ \frac{3}{8} \kappa^{1/3}
   (\sqrt{3}\mathcal{T})^{2/3}\, \biggr)
   \\ \nonumber
   &&\quad \times \biggl[1 + \frac{\pi}{4}\frac{(4/3)^{5/4}}{3\nu \kappa^{3/4}}
           - \biggl(1 - \frac{\pi}{6}\frac{(4/3)^{1/4}}{\kappa^{5/12}}
             \biggr) \frac{(\sqrt{3}\mathcal{T})^{2/3}}{\nu}\biggr].
\end{eqnarray}
In the above result, we have accounted for the improved normalization
$\Delta f_\mathrm{pin} \approx \Cbar x_+\,(1+ \rho\, x_- /x_+) = \Cbar x_+ \nu$
that includes the subdominant term $\rho\, x_-/x_+$ for better precision, $\nu =
1+ \rho\,(4/3\kappa^{1/3})^{9/4}$; it is this result that compares well with the
numerical result shown in Fig.\ \ref{fig:rescaled_curvatures}.  Finally, the
approach to equilibrium with $t \to t_\mathrm{eq} = t_0 \exp(U_0/T)$ and
beyond is discussed in Sec.\ \ref{sec:lt} below.

The above analysis applies to large $\kappa$, where the shift in the force
jump with its various contributions from free and pinned branches could be
physically well motivated. In the following, we focus on the opposite limit of
marginally strong pinning $\kappa \gtrsim 1$ with $\kappa - 1$ serving as the
small parameter (and setting $\kappa = 1$ otherwise). In this regime, the
Campbell curvature $\alpha_\mathrm{sp}$ is described through Eq.\
\eqref{eq:alpha_finite_T_sk}, which involves only the force jump $\Delta
f^\mathrm{fp,\, jp}_\mathrm{pin}$ at $x_+$.  Making use of the microscopic
force balance equation \eqref{eq:fbx}, we find the simple formula
\begin{align}\label{eq:df_gen_kappa_dr}
    \Delta f^\mathrm{fp,\,jp}_\mathrm{pin} -\Delta f^\mathrm{fp}_\mathrm{pin}
   &\approx
   \Cbar\bigl[
   (-\delta r_\mathrm{f+} + \delta x_+)
   -(-\delta r_\mathrm{p+} + \delta x_+)\bigr]\nonumber\\
   &\approx\Cbar\bigl[
   -\delta r_\mathrm{f+}
   +\delta r_\mathrm{p+}\bigr],
\end{align}
expressing the change in the total force jump by the shifts in tip positions
$\delta r$ alone.  The expressions \eqref{eq:dr_dx_easy} for the shift in the
vortex tip positions $\delta r_{\mathrm{p}+}$ and $\delta r_\mathrm{f-}$
involve the second derivative $f_p^{\prime\prime} \approx (e_p/\xi^3)
\sqrt{\gamma\,(\kappa -1)}$ at the edges $x_\pm$ (obtained with the help of 
the expansion Eq.\ \eqref{eq:f_p_exp_sk}) and we find  
\begin{eqnarray}\label{eq:dr_ppfm}
   \delta r_{\mathrm{p}+} = \delta r_{\mathrm{f}-}
   &\approx& \left[\frac{4\Cbar\xi^2}{e_p}\frac{\xi\> \delta x_\pm}
   {\sqrt{4\gamma(\kappa -1)}} \right]^{1/2},
\end{eqnarray}
symmetric at $r_\mathrm{p}(x_+)$ and $r_\mathrm{f}(x_-)$.

Besides the shift $\delta r_\mathrm{p+}$ associated with the edge of the
bistability region, we also need the free tip position near $x_+$, 
\begin{equation}\label{eq:r_hard}
   r_\mathrm{f}(x_+^\mathrm{jp}) \equiv r_\mathrm{f}(x_+) - \delta r_{\mathrm{f}+},
\end{equation}
that is not associated with a special point on the free branch. It is obtained
by evaluating the force-balance equation \eqref{eq:force_balance} on the free
branch close to $x_+$,
\begin{equation}\label{eq:dr_dx_hard}
   \Cbar[(r_\mathrm{f}(x_+) - \delta r_\mathrm{f+})-(x_+-\delta x_+)] 
   = f_p[r_\mathrm{f}(x_+) - \delta r_\mathrm{f+}],
\end{equation}
with the tip position $r_\mathrm{f}(x_+)$ given in \eqref{eq:rf+}. Solving
Eq.\ \eqref{eq:dr_dx_hard} with the help of the expansion
\eqref{eq:f_p_exp_sk}, we find the (symmetric) tip shifts expressed through
$\delta x_+$,
\begin{eqnarray}\label{eq:dr_pmfp}
   \delta r_{\mathrm{f}+}~[ = \delta r_{\mathrm{p}-}] &\approx&
   [3(\kappa-1)]^{-1} \> \delta x_+,
\end{eqnarray}
independent of $\gamma$.

Finally, we derive the asymptotic shifts $\delta x_\pm$ using the expansion
for the pinning/depinning barriers \eqref{eq:barrier_gen} and find $\delta
x_\pm / \xi \approx (e_p/4\Cbar \xi^2) [4 \gamma(\kappa-1)]^{1/6}
(3\mathcal{T})^{2/3}$.  This result simplifies considerably when focusing on
the Lorentzian potential, where $\delta x_\pm / \xi \approx
[3(\kappa-1)/2]^{1/6} \,(3\mathcal{T})^{2/3}$.

The renormalized total force jump is obtained by inserting the above tip
shifts into Eqs.\ \eqref{eq:df_gen_kappa_dr} and we obtain the shift in the
force jumps
\begin{eqnarray}\label{eq:df_s_kappa}
   \Delta f^\mathrm{fp,\,jp}_\mathrm{pin} \!-\! \Delta f_\mathrm{pin}^\mathrm{fp} &\approx& 
   \Cbar\xi\biggl[
   \frac{(2/3)^{1/6}(3\mathcal{T})^{1/3}}{(\kappa\!-\!1)^{1/6}}\\
   \nonumber
   && \qquad\qquad 
   -\frac{(3/2)^{1/6}(3\mathcal{T})^{2/3}}{3(\kappa\!-\!1)^{5/6}} \biggr],
\end{eqnarray}
where we have focused on the Lorentzian potential, the generalization to an
arbitrary potential being straightforward once the shape parameter $\gamma$ is
known.

To find the total force jump $\Delta f_\mathrm{pin}^\mathrm{fp} =
\Cbar[r_\mathrm{f}(x_+)-r_\mathrm{p}(x_+)]$ in the absence of fluctuations, we
need the free and pinned vortex tip positons at the edge $x_+$, see Eqs.\
\eqref{eq:delta_r_sk} and \eqref{eq:rf+}, that lead us to the result
\begin{eqnarray}\label{eq:df_tp_s_kappa}
   \Delta f_\mathrm{pin}^\mathrm{fp} &\approx& 
   3\Cbar \xi \left(\frac{4\Cbar\xi^2}{e_p}\right)^{1/2} 
   \left(\frac{\kappa-1}{4\gamma}\right)^{1/2}.
\end{eqnarray}
The trapping scale $x_0$ in Eq.\ \eqref{eq:alpha_finite_T_sk} depends on the
details of the potential, see Sec.\ \ref{sec:strong_pinning}.  For the
Lorentzian potential, we find the results $\Delta f_\mathrm{pin}^\mathrm{fp}
\approx 3 \Cbar \xi [2(\kappa - 1)/3]^{1/2}$ and $x_0 \approx \sqrt{2}\xi
[2+(\kappa -1)]$, see Eq.\ \eqref{eq:r_m-x_m}, that takes us to the final
result for the curvature at moderately strong pinning ${\kappa\gtrsim1}$, to
leading order in $\kappa - 1$,
\begin{eqnarray}\label{eq:resc_curv_small_kappa}
   \frac{\alpha_\mathrm{sp}(t,T)}{\alpha_\mathrm{sp}}
   &\approx& 1 + \frac{[(3/2)\mathcal{T}]^{1/3}}{[3(\kappa - 1)]^{2/3}}
   - \frac{[(3/2)\mathcal{T}]^{2/3}}{[3(\kappa - 1)]^{4/3}}.
\end{eqnarray}
Note that the last (negative) term is just the square of the second (positive)
contribution.

Let us discuss the results \eqref{eq:resc_curv_large_kappa},
\eqref{eq:resc_curv_large_kappa_lT}, and \eqref{eq:resc_curv_small_kappa} and
compare them with the numerical findings, with all of these shown in Fig.\
\ref{fig:rescaled_curvatures}.  First, we translate the time range $1 <t/t_0
\sim \exp(U_0/T)$ where our analysis is valid to the creep variable
$\mathcal{T} = (T/e_p) \ln(t/t_0)$, that results in the region $0 <
\mathcal{T} < U_0/e_p$. Note that going beyond $t > t_\mathrm{eq} = t_0
\exp(U_0/T)$, we have to include both terms in Eq.\ \eqref{eq:rate_equation};
instead, here, we simply terminate our approximate analysis with the same
result (up to irrelevant details in the form of the steps at $\pm x_0$) for
the equilibrium distribution $p_\mathrm{eq}$. 

The barriers $U_0$ at the branch crossing point $x_0$ have been derived both
for strong and marginal pinning in Sec.\ \ref{sec:strong_pinning}, see Eqs.\
\eqref{eq:U0_lk} and \eqref{eq:U0_sk}. We then find our analysis to be valid
for values of the creep variable $\mathcal{T}$ inside the ranges
\begin{eqnarray}\label{eq:T_range}
   & 0 < \mathcal{T} < 1 ~&\textrm{ for }~\kappa \gg 1,
   \\ \nonumber
   & 0 < \mathcal{T} < (\kappa - 1)^2/8~&\textrm{ for }~\kappa - 1 \ll 1,
\end{eqnarray}
resulting in very different relaxation ranges for the two situations.

Fig.\ \ref{fig:rescaled_curvatures} summarizes all results graphically:
the numerical evaluation of Eq.\ \eqref{eq:alpha_finite_T} in blue, the
asymptotic expressions \eqref{eq:resc_curv_small_kappa} and
\eqref{eq:resc_curv_large_kappa_lT} for moderate and large $\kappa$ at short
times in orange, and the long-time asymptotics discussed below, see Eq.\
\eqref{eq:resc_curv_large_kappa_slow}, in red. Starting with the simpler
result \eqref{eq:resc_curv_small_kappa} for moderately strong pinning $\kappa
-1 \ll 1$, we see that the increase in the curvature
term in the renormalized force jump produces an increase in the Campbell
curvature. The negative contribution in $\Delta
f_\mathrm{pin}^\mathrm{fp,\,jp}$ formally dominates the curvature term only at
values of the creep parameter $\mathcal{T}$ that reside beyond the criterion
\eqref{eq:T_range}, and hence the ratio \eqref{eq:resc_curv_small_kappa}
increases monotonously.

At very strong pinning $\kappa \gg 1$, again, the increase in the trapping
area and the curvature term in the force jump jointly produce an increase in
the Campbell curvature at small values of the creep parameter. In this rise,
the curvature term is the dominant one only at very small values $\mathcal{T}
< (\pi/4)^3[(4/3)^{27/4}/\sqrt{3}]\> \kappa^{-11/4}$. The curvature term goes
over into the linear correction $\propto \delta x_-/x_-$ at $\mathcal{T}
\approx [(2/3)^3/\sqrt{3}]\> \kappa^{-1/2}$, but this crossover is hidden by
the dominant increase in the trapping area. The increasing Campbell curvature
implies a decreasing Campbell length $\lambda_\mathrm{ {\scriptscriptstyle C}}
\propto 1/\sqrt{\alpha_\mathrm{sp}}$.  

At larger values of $\mathcal{T}$, the competition is among the two
$\propto \mathcal{T}^{2/3}$ terms in Eq.\ \eqref{eq:resc_curv_large_kappa_lT},
positive and $\propto \kappa^{1/3}$ in the trapping area and negative
($\propto 1$) in the force jump, that describes an inverted parabola in
$\mathcal{T}^{2/3}$. At large $\kappa$, the positive correction in the
trapping area dominates and we obtain a maximum in the Campbell curvature at
$\mathcal{T} \approx [(1/2)^{3/2}/\sqrt{3}] (1-8/3\kappa^{1/3})^{3/2}$, where
we have focused on the leading order in $\kappa$ only. This saturates at large
$\kappa$ at a value $\mathcal{T} \approx 0.2$, i.e., within the relevant time
range $0 < \mathcal{T} < 1$ found in \eqref{eq:T_range}.  With decreasing
$\kappa$ the correction in the trapping factor diminishes and at $\kappa \sim
10$ the negative term in the force correction drives the initial slope of the
inverted parabola negative, resulting in a monotonically decreasing Campbell
curvature. However, we should not trust the large $\kappa$ result Eq.\
\eqref{eq:resc_curv_large_kappa_lT} at these intermediate values of $\kappa$
any more. Indeed, as shown in Fig.\ \ref{fig:rescaled_curvatures}, at $\kappa
= 5$, we have already crossed over from the monotonously increasing behavior
predicted at marginal pinning to the non-monotonic result typical for large
values of $\kappa$.

The above results can be compared with different experimental findings: The
increasing curvature $\alpha_\mathrm{sp}$ that we find at small times produces
a decreasing-in-time Campbell length $\lambda_\mathrm{{\scriptscriptstyle C}}
\propto 1/\sqrt{\alpha_\mathrm{sp}}$, in agreement with experimental results
on Bi$_2$Sr$_2$CaCu$_2$O$_8$ (BiSCCO) by Prozorov et al., see Ref.\
\onlinecite{Prozorov_2003}.  On the other hand, measurements on
YBa$_2$Cu$_3$O$_7$ (YBCO) by Pasquini et al.\ \cite{Pasquini_2005} show a
Campbell length that increases with time under the effect of creep; this is
consistent with our long-time decrease in $\alpha_\mathrm{sp}$ that appears
for intermediate and large values of the strong-pinning parameter $\kappa$.

In Fig.\ \ref{fig:rescaled_curvatures}, we complete these results with the
curves $j(t,T)$, Eq.\ \eqref{eq:pers_curr}, for the persistent current
densities. While $j(t \to \infty,T)$ vanishes on approaching equilibrium, this
is not the case for the Campbell curvature $\alphaSP(t,T)$.  This is due to
the vanishing jumps $\Delta e_\mathrm{pin}$ at $x_0$ in the pinning energy,
see Fig.\ \ref{fig:strong_pinning}, while the force jumps $\Delta
f_\mathrm{pin}$ remain large at $x_0$ and hence $\alphaSP(t \to \infty,T) \to
\alpha_0$, with the latter defined in Eq.\ \eqref{eq:alpha_0}. The
observation of a finite Campbell penetration depth above the irreversibility
line \cite{Prozorov_2003} in a BiSCCO sample confirms this finding.

\subsubsection{Long time limit $t \to t_\mathrm{eq}$}\label{sec:lt}

On a timescale $t \to t_\mathrm{eq} = t_0\exp(U_0/T)$, the jumps
$x^\mathrm{jp}_\pm$ shift close to $x_0$ and the branch occupation
$p_\mathrm{th} (\mathrm{R})$ approaches the radially symmetric equilibrium
distribution $p_\mathrm{eq}(R)$ with a jump at $R \approx x_0$.  For
marginally strong pinning with $\kappa - 1 \ll 1$, the maximal barrier
$U_0$ is small and the relaxation to equilibrium happens rapidly. The bistable
region is narrow, with $x_\pm$ and $x_0$ given in Sec.\
\ref{sec:strong_pinning}, Eqs.\ \eqref{eq:x_pm_sk} and \eqref{eq:x_0_sk}, and
hence the above evaluation of the force jumps at $x_\pm^\mathrm{jp}$ remains
accurate when $x_\pm^\mathrm{jp} \to x_0$.  As shown in Fig.\
\ref{fig:rescaled_curvatures}, Eq.\ \eqref{eq:resc_curv_small_kappa} then
captures the corresponding long-time limit successfully.

In the very strong pinning limit $\kappa \gg 1$, equilibrium is only slowly
approached and the bistable region starts out broad and asymmetric, leading to
stark changes in the trapping geometry and in the total force jump as the
branch occupation relaxes.  Our analysis then has to be adapted to cope with
this different situation.  We start by evaluating the asymptotic equilibrium
value $\alpha_0$, and then study how this is approached as $x_\pm^\mathrm{jp}
\to x_0$.

We simplify our previous result for $\alpha_0$ at equilibrium, Eq.\
\eqref{eq:alpha_0}, by making use of the force balance Eq.\ \eqref{eq:fbx} in
order to express the jumps through the free and pinned tip positions at $\pm
x_0$,
\begin{equation}\label{eq:alpha_0_Lor}
   \alpha_0=n_p\frac{\pi x_0}{a_0}\frac{\Cbar[r_\mathrm{f}(x_0)-r_\mathrm{p}(x_0)]}{a_0}.
\end{equation}
The branch cutting point $x_0$ has been determined in Sec.\
\ref{sec:strong_pinning}, Eq.\ \eqref{eq:x_0}.  For the Lorentzian potential,
we find $x_0 \approx 2\sqrt{2\kappa}\,\xi$.  Combining Eq.\
\eqref{eq:alpha_0_Lor} with the results for $x_0$ and the associated tip
positions $r_\mathrm{f}(x_0) \approx x_0$ and $r_\mathrm{p}(x_0) \ll \xi$, see
Eq.\ \eqref{eq:r_f-p}, we arrive at the result
\begin{equation}\label{eq:alpha_0_Lor_gen}
   \alpha_0 \approx 2\pi\, n_p (e_p/a_0^2)
\end{equation}
that depends only on the defect density $n_p$ and depth $e_p$. Quite
remarkably, while the persistent current density vanishes upon approaching
equilibrium, the Campbell curvature and penetration depth remain finite. This
is in agreement with the experimental findings in Ref.\
\onlinecite{Prozorov_2003}, where a finite Campbell length was observed above
the irreversibility line in a BiSCCO sample.

On approaching equilibrium, the thermal jumps reside close to $x_0$ and we can
write $x_\pm^\mathrm{jp} = x_0 \pm \delta x_{0\pm}$ with small corrections
$\delta x_{0\pm} >0$. Using the general result Eq.\ \eqref{eq:alpha_finite_T}
for the Campbell curvature with $\theta_+^\mathrm{jp} \approx \pi$, see Eq.\
\eqref{eq:angle_pi/2} and making use of the smallness of $\delta x_{0\pm}$,
we have to evaluate
\begin{multline}\label{eq:alpha_lt}
   \alpha_\mathrm{sp}(t,T)
   = n_p \frac{\pi}{2}\bigg[\frac{(x_0+\delta x_{0+})}{a_0}
   \frac{\Delta f^\mathrm{fp,jp}_\mathrm{pin}}{a_0}\\
   +\frac{(x_0-\delta x_{0-})}{a_0}\frac{\Delta f^\mathrm{pf,jp}_\mathrm{pin}}{a_0}\bigg].
\end{multline}
With $f_\mathrm{pin}^\mathrm{f} (x) \approx 0$ and
$f_\mathrm{pin}^\mathrm{p} (x) \approx - \Cbar x$ in the vicinity of $x_0$, we
find the changes in the force jump away from $x_0$
\begin{eqnarray}\label{eq:long_times_force_jump1}
   \Delta f^\mathrm{fp,jp}_\mathrm{pin}- \Delta f _\mathrm{pin}^\mathrm{fp}(x_0) 
   & \approx & \Cbar\delta x_{0+}, \\
   \Delta f^\mathrm{pf,jp}_\mathrm{pin} - \Delta f _\mathrm{pin}^\mathrm{fp}(x_0) 
   & \approx & -\Cbar \delta x_{0-}\label{eq:long_times_force_jump2}.
\end{eqnarray}
For long times $t \to t_\mathrm{eq}$, the relevant creep barriers
\eqref{eq:U_creep} are to be evaluated close to the equilibrium value $U_0$,
justifying the expansions
\begin{align}\label{eq:long_times_barriers1}
   \Udp(\xpjp) &\approx U_0 + \Udp'(x_0)\>\delta x_{0+},\\
   \Up(-\xmjp) &\approx U_0 + \Up'(-x_0)\>\delta x_{0-}\label{eq:long_times_barriers2},
\end{align}
with the total derivatives assuming the simple form
\begin{align}\label{eq:long_times_derivatives1}
   \Up'(-x_0)&\approx\Cbar(r_\mathrm{us}-r_\mathrm{f})|_{x_0} < 0,\\
   \Udp'(x_0)&\approx\Cbar(r_\mathrm{p}-r_\mathrm{us})|_{x_0} < 0,
   \label{eq:long_times_derivatives2}
\end{align}
as all derivatives $\partial_x r_\mathrm{f}(x)$ and $\partial_x
r_\mathrm{us}(x)$ cancel due to Eq.\ \eqref{eq:fbx}; indeed, both barriers
decrease when going away from equilibrium.  Combining the above relations, we
can express the change in the force jumps $\Delta f_\mathrm{pin}^\mathrm{dp,
jp}$ and $\Delta f_\mathrm{pin}^\mathrm{pf, jp}$ in terms of the barrier difference
\begin{equation}\label{eq:T_teq}
   U_0-U = T\log(\teq/t) \equiv e_p \mathcal{T}_\mathrm{eq}
\end{equation}
to find
\begin{eqnarray}\label{eq:long_times_delta_f1}
   \Delta f_\mathrm{pin}^\mathrm{fp, jp}(t,T) - \Delta f _\mathrm{pin}^\mathrm{fp}(x_0)
   &\approx& \frac{e_p}{r_\mathrm{us}-r_\mathrm{p}}\mathcal{T}_\mathrm{eq},\\
   \Delta f_\mathrm{pin}^\mathrm{fp, jp}(t,T) - \Delta f _\mathrm{pin}^\mathrm{fp}(x_0)
   &\approx& -\frac{e_p}{r_\mathrm{f}-r_\mathrm{us}}\mathcal{T}_\mathrm{eq}
   \label{eq:long_times_delta_f2}.             
\end{eqnarray}
Combining Eq.\ \eqref{eq:long_times_force_jump1} and
\eqref{eq:long_times_delta_f1} also provides us with the expressions for
$\delta x_{0\pm}$ that we need in \eqref{eq:alpha_lt}.  As a result, we have
reduced the problem to the determination of the three vortex tip positions
$r_\mathrm{i}(x_0)$, $\mathrm{i} = \mathrm{p,us,f}$, at the asymptotic vortex
position $x_0$. These have been found in Sec.\ \ref{sec:strong_pinning}, Eqs.\
\eqref{eq:r_f-p} and \eqref{eq:r_us}.

Inserting the results for the force jumps and jump points into Eq.\
\eqref{eq:alpha_lt} and focusing on a Lorentzian potential, we find the scaled
Campbell curvature near equilibrium,
\begin{equation}\label{eq:resc_curv_large_kappa_slow}
   \frac{\alpha_\mathrm{sp}(t,T)}{\alpha_0} \approx 
   \frac{(1 - \mathcal{T}_\mathrm{eq}/2)^2 + 
   (1 + (\kappa/2)^{1/3}\> \mathcal{T}_\mathrm{eq})^2}{2},
\end{equation}
where we have used the form \eqref{eq:alpha_0} for $\alpha_0$ with $\Delta
f_\mathrm{pin}^\mathrm{fp}(x_0)\approx 2\Cbar(2\kappa)^{1/2}\xi$ and
$x_0 \approx 2 \sqrt{2\kappa}\, \xi$. Again, we find a competition
between the trapping length and the force jump that act the same way as
before, with the opposite signs compensated by evaluating
$\mathcal{T}_\mathrm{eq}$ away from the longest time $\teq = t_0 \exp(U_0/T)$.
The result is shown in the large $\kappa$ panel of Fig.\
\ref{fig:rescaled_curvatures} and agrees well with the full numerical result
close to equilibrium, where $\delta x_{0\pm}/x_0\ll 1$.

Going to very large times $t > t_\mathrm{eq}$, we enter the TAFF regime
\cite{Kes_1989} with a diffusive vortex motion characterized by the TAFF
resistivity, $\rho_{\rm\scriptscriptstyle TAFF} \propto \rho_\mathrm{flow}
\exp(-U_0/T)$ and $\rho_\mathrm{flow} = (B/H_{c2}) \, \rho_n$, cf.\ the
corresponding discussion of the asymptotic decay of the persistent current
density $j(t,T)$ in Sec.\ \ref{sec:creep_pc} above.  The Campbell penetration
depth $\lambda_{\rm\scriptscriptstyle C}$ then transforms into the dispersive
TAFF-skin depth $\lambda_{\rm\scriptscriptstyle TAFF}(\omega) \sim \sqrt{c^2
\rho_{\rm\scriptscriptstyle TAFF}/\omega}$: In the Campbell regime, vortices
displace by $U \sim jB/\alpha c$ at frequency $\omega$, resulting in a typical
velocity $v_{\rm\scriptscriptstyle C} \sim \omega U \sim (4\pi \omega
\lambda_{\rm\scriptscriptstyle C}^2/B c) \> j$. The typical velocity due to
the dissipative motion follows from Faraday's law, $E = vB/c$, combined with
Ohm's law $E = \rho_{\rm\scriptscriptstyle TAFF} j$, hence
$v_{\rm\scriptscriptstyle TAFF} \sim (c\rho_{\rm\scriptscriptstyle TAFF}/B) \>
j$. Equating the two, we find the crossover frequency
$\omega_{\rm\scriptscriptstyle TAFF} = c^2 \rho_{\rm\scriptscriptstyle TAFF}/
4\pi \lambda_{\rm\scriptscriptstyle C}^2$ where 
$\lambda_{\rm\scriptscriptstyle C} \sim \lambda_{\rm\scriptscriptstyle TAFF}$
and we enter the dispersive skin-effect regime at low frequencies $\omega <
\omega_{\rm\scriptscriptstyle TAFF}$.  Physically, as
$v_{\rm\scriptscriptstyle C}$ drops below $v_{\rm\scriptscriptstyle TAFF}$,
the ac oscillation of the vortex in the well is prematurely (i.e., before
completion of one cycle) terminated by an escape out of the well.

Note that the Campbell response requires sufficiently small frequencies
as well: Following the derivation of Eq.\ \eqref{eq:l_C_phen}, the Campbell
penetration physics requires frequencies $\omega < \alpha/\eta$ that
transforms to the condition $\omega < \omega_\mathrm{flow}$ with
$\omega_\mathrm{flow} =  c^2 \rho_\mathrm{flow}/ 4\pi
\lambda_{\rm\scriptscriptstyle C}^2$. As a result, we find the bounded regime
$\omega_{\rm\scriptscriptstyle TAFF} < \omega < \omega_\mathrm{flow}$ for the
application of the Campbell response at very long times $t > t_\mathrm{eq}$,
with a crossover to the usual skin-effect (with $\rho = \rho_{\rm\scriptscriptstyle
TAFF} \propto \rho_\mathrm{flow} \exp(-U_0/T)$) at very low and at very high
frequencies (with $\rho = \rho_\mathrm{flow}$).

\subsection{Campbell penetration depth $\lambda_{\rm\scriptscriptstyle
C}$ in \textrm{FC} state} \label{sec:FC}

The \textrm{FC} state is characterized by a homogeneous distribution of the
magnetic field inside the sample and hence is associated with a vanishing
current- and pinning-force density.  Correspondingly, at $\kappa > 1$, the
branch occupation $p_{\rm\scriptscriptstyle FC}(R)$ is rotationally symmetric
assuming a value $p_{\rm\scriptscriptstyle FC} \approx 1$ in a disk with
radius $x^\mathrm{jp} \in [x_-,x_+]$ centered around the defect (we note
that in Ref.\ \onlinecite{Willa_2016}, the trapping area for $\lambda_{\rm
\scriptscriptstyle C}$ was handled the same way as for $j_c$; it was described
by the transverse trapping length $t_\perp = 2x_- \sim 2\xi$ and its circular
geometry was ignored). The determination of the jump position $x^\mathrm{jp}$
as a function of the FC state preparation and its relaxation through creep is
the main objective of this section.

Ignoring the initialization process\cite{Willa_2015_PRB}, as we did in the
discussion of the \textrm{ZFC} case above, we determine the restoring force
$\delta F_\mathrm{pin}(U)$ using Eq.\ \eqref{eq:dFpin_ZFC} with $p_c(R)$
replaced by $p_{\rm\scriptscriptstyle FC}(R) = \Theta(x^\mathrm{jp}-R)$;
accounting for the radial symmetry of the problem, we find the restoring force
density
\begin{align}\label{eq:dFpin_FC}
   \delta F_\mathrm{pin}(U)
   &\approx -n_p \frac{\pi x^\mathrm{jp}}{a_0}
   \frac{\Delta f^\mathrm{fp}_\mathrm{pin}(x^\mathrm{jp})}
   {a_0}\,U,
\end{align}
directed along the displacement $\mathbf{U}$ parallel to the $x$-axis,
$\mathbf{U} = (U,0)$; the force jump $\Delta
f^\mathrm{fp}_\mathrm{pin}(x^\mathrm{jp}) =
f_\mathrm{pin}^\mathrm{f}(x^\mathrm{jp}) -
f_\mathrm{pin}^\mathrm{p}(x^\mathrm{jp})$ is to be evaluated at the radial
jump position $x^\mathrm{jp}$.  The Campbell curvature
then reads
\begin{equation}\label{eq:alpha_FC}
   \alpha^\mathrm{\scriptscriptstyle FC}_\mathrm{sp} = n_p\frac{\pi\,
   x^\mathrm{jp}}{a_0} \frac{\Delta f_\mathrm{pin}^\mathrm{fp}(x^\mathrm{jp})}{a_0}
\end{equation}
and depends on temperature through the position of the jump $x^\mathrm{jp}$ in
multiple ways: first, the pinning parameter $\kappa$ determining the shape of
the pinning force landscape $f_\mathrm{pin}(x)$ in Fig.\
\ref{fig:strong_pinning} depends on $T$ and $B$ through the parameters of the
mean-field Ginzburg-Landau theory which are functions of $(1-T/T_c)$ and
$(1-T/T_c - B/H_{c2(0)})$ with $H_{c2}$ the upper critical field
\cite{Willa_2016}---this dependence generates interesting hysteretic phenomena
in cyclic measurements of the Campbell penetration depth with varying
temperature $T$ \cite{Willa_2015_PRL, Willa_2016}
$\lambda_\mathrm{\scriptscriptstyle C}(T)$ and will be the subject of Sec.\
\ref{sec:FC_hysteresis}. Second, thermal fluctuations, i.e., creep, drive the
vortex state towards equilibrium, that corresponds to a relaxation of the
initial jump position $x^\mathrm{jp}$ after field cooling towards the
equilibrium position $x_0$ characterizing $p_\mathrm{eq}$---these creep
phenomena associated with the relaxing FC state will be discussed in Sec.\
\ref{sec:creep_FC}.

\begin{figure}
        \includegraphics[width = 1.\columnwidth]{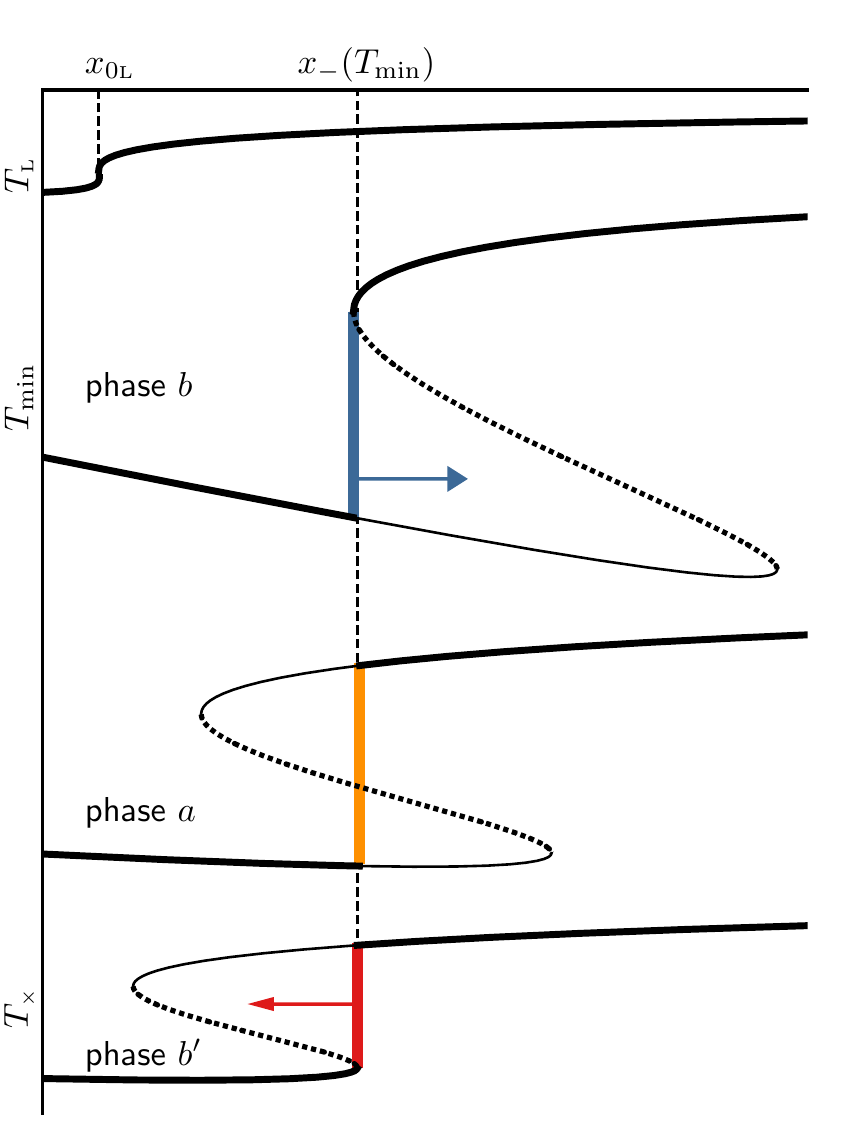}
	\caption{Illustration of the evolution of the force profile
	$f_\mathrm{pin}(x)$ and of the branch occupation along a temperature
	cycle $T_{\rm\scriptscriptstyle L} \to T_\mathrm{min} \to
	T_{\rm\scriptscriptstyle L}$. Top curve: Strong pinning turns on at
	the Labusch temperature $T_{\rm\scriptscriptstyle L}$ where
	$\kappa(T_{\rm\scriptscriptstyle L}) = 1$, with an infinite slope in
	$f_\mathrm{pin}(x)$ appearing at $x_\mathrm{0{\scriptscriptstyle L}}$.
	Lowering the temperature $T < T_{\rm\scriptscriptstyle L}$, a bistable
	region opens up, widens, and moves with respect to first instability
	point $x_\mathrm{0{\scriptscriptstyle L}}$. Shown here is the case for
	an insulating defect (or $\delta \ell$-pinning), where the bistable
	interval moves to the right/left upon cooling/heating. Bottom three
	curves: The three phases $b$ (cooling), $a$ (heating), and $b'$
	(heating) of the hysteresis loop are characterized by different branch
	occupations with jump positions within $f_\mathrm{pin}(x)$ marked in
	blue, orange, and red.  Upon cooling below $T_{\rm\scriptscriptstyle
	L}$, the branch occupation (thick solid line) changes as parts of the
	force profile become unstable. The jump $x^\mathrm{jp}$ in the branch
	occupation first follows the inner endpoint (phase $b$),
	$x_b^\mathrm{jp}(T) = x_-(T)$ (blue solid line, drawn at
	$T_\mathrm{min}$).  Reversing the change in temperature at
	$T_\mathrm{min}$, the bistable region shrinks while heating and moves
	leftwards, leaving the jump position $x^\mathrm{jp}(T)$ unchanged,
	$x_a^\mathrm{jp} = x_-(T_\mathrm{min})$ (orange solid line in phase
	$a$). When reaching the right end of the $S$-shaped force-profile at
	the crossover temperature $T_\times$, $x_+(T_\times) =
	x_a^\mathrm{jp}$, the jump point gets pinned to the boundary $x_+$ of
	the bistable region, $x_{b'}^\mathrm{jp}(T)=x_+(T)$ (red solid line in
	phase $b'$, drawn at $T_\times$). If at some temperature $T$ the cycle
	is interrupted, the vortex lattice relaxes towards equilibrium with
	$x_{b,b'}^\mathrm{jp}(T) \to x_0(T)$, as indicated by the horizontal
	arrows in phases $b$ and $b'$. In the intermediate phase $a$,
	relaxation is weak and  depends on the relative position between
	$x_a^\mathrm{jp}(T)$ and the edges $x_\pm(T)$.
}
     \label{fig:hysteresis_scheme}
\end{figure}

\subsubsection{Hysteresis of penetration depth $\lambda_{\rm\scriptscriptstyle
C}$ in \textrm{FC} state} \label{sec:FC_hysteresis}

We start with the discussion of the field-cooled state and the appearance of
hysteretic effects.  Here, we have in mind a setup where we fix the magnetic
field $H$ and vary the temperature $T$, typically in a (repeated)
cooling--warming cycle.  By changing the temperature $T$, the vortex lattice
elastic constant $\Cbar$ and the pinning energy $e_p$ are modified through
their dependence on the Ginzburg-Landau parameters $\lambda(T,B) \propto
(1-\tau-b_0)^{-1/2}$ and $\xi(T) \propto (1-\tau)^{-1/2}$ on the reduced
temperature $\tau = T/T_c$ and reduced field $b_0 = B/H_{c2}(0)$ (we ignore
rounding effects appearing at small temperatures).  These mean-field
dependencies on $T$ and $B$ entail a change in the pinning parameter $\kappa$
within the $B$--$T$ phase diagram that varies with the nature of the defect,
\begin{equation}\label{eq:kappa_evolution}
   \kappa(\tau,b_0) \sim \frac{k}{\sqrt{b_0}} \> 
   \left(1-\tau\right)^{\alpha} \left(1-\tau-b_0 \right)^{\beta},
\end{equation}
with the prefactor $k$ and exponents $\alpha$ and $\beta$ depending on the
type of defect/pinning (the dependence $\propto 1/\sqrt{b_0}$ derives from the
field dependence in the effective elasticity; at very small fields, we enter
the single vortex strong pinning regime where our analysis has to be adapted,
see Ref.\ \onlinecite{Koopmann_2004}).  The cases of small insulating ($k =
(\rho/\xi_0)^3$, $\alpha = 3/2$, $\beta = 1/2$) or metallic defects ($k = 1$,
$\alpha = 0$, $\beta = 1/2$), of $\delta T_c$-pinning (with local changes
$\delta T_c$ in the transition temperature $T_c$ and $k =
(\rho/\xi_0)^3(\delta T_c/T_c)$, $\alpha = 3/2$, $\beta = -1/2$) or $\delta
\ell$-pinning (with local changes $\delta \ell$ in the mean free path $\ell$
and $k = (\rho/\xi_0)^3(\delta \ell/\ell)$, $\alpha = 5/2$, $\beta = -1/2$)
have been discussed in Ref.\ \onlinecite{Willa_2016}, but other situations may
produce alternative dependencies (here, $\rho$ is the size of the defect and
$\xi_0$ the coherence length of the superconducting material at $T=0$).
Depending on the type of defect, strong pinning may turn on smoothly at the
phase boundary $H_{c2}(T)$ (this is the case for insulating and metallic
defects with $\beta > 0$) or collapse from infinity (this is the case for
$\delta T_c$- and $\delta \ell$-pinning with $\beta < 0$).  However, owed to
the factor $(1-\tau)^{\alpha}$ with a large exponent $\alpha = 3/2,~5/2$
characterizing $\delta T_c$- and $\delta \ell$-pinning, this divergence at
$H_{c2}(T)$ is strongly suppressed at the small fields where our analysis is
valid. As a result, at small fields, strong pinning always turns on smoothly
upon decreasing the temperature $T$ below the phase boundary $T_{c2}(H)$.
Also note that thermal fluctuations near the transition shift the onset of
pinning to below $T_{c2}(H)$.

In a first step, we establish the onset of strong pinning upon decreasing the
temperature $T$ below $T_{c2}(H)$ at a given field value $H$. Figure
\ref{fig:hysteresis_scheme} shows the pinning force profile
$f_\mathrm{pin}(x)$ at the onset of strong pinning where the function
$f_\mathrm{pin}(x)$ develops an infinite slope at the point
$x_\mathrm{0{\scriptscriptstyle L}}$. This happens at the Labusch temperature
$T_{\rm\scriptscriptstyle L}$ that is defined through the condition
$\kappa(T_{\rm \scriptscriptstyle L}) = 1$. The point
$x_\mathrm{0{\scriptscriptstyle L}}$ corresponds to the asymptotic position
$x_m$ associated with the inflection point $r_m$ of $f_p(r)$ at the Labusch
point $\kappa = 1$, hence $x_\mathrm{0 {\scriptscriptstyle L}} = x_m = r_m
-f_p(r_m)/\Cbar$ with $f''_p(r_m) = 0$ and $f_p'(r_m) = \Cbar$, as discussed
in Sec.\ \ref{sec:strong_pinning}, Eqs.\ \eqref{eq:f_p_exp_sk} and
\eqref{eq:r_m-x_m}. At $\kappa \to 1$, this point coincides with the branch
crossing point $x_0$, see Eq.\ \eqref{eq:x_0_sk}.

Decreasing the temperature $T$ below $T_{\rm\scriptscriptstyle L}$, the
singularity at $x_\mathrm{0 {\scriptscriptstyle L}}$ develops into the finite
bistable interval $[x_-,x_+]$ with a width $x_+ - x_- \sim (\kappa - 1)^{3/2}
\xi$ initially centered around $x_0 \approx (x_+ + x_-)/2 \sim
[\mathrm{const.} + (\kappa - 1)] \xi$. Depending on the relative increase in
$\kappa(\tau)$ and the decrease in $\xi(\tau) \approx \xi_0/\sqrt{1-\tau}$,
the bistable interval may grow to the right (or left) of $x_\mathrm{0
{\scriptscriptstyle L}}$ with $x_\mathrm{0 {\scriptscriptstyle L}} < x_-$
($x_\mathrm{0 {\scriptscriptstyle L}} > x_+$) or enclose the initial
instability at $x_\mathrm{0 {\scriptscriptstyle L}}$ with $x_- < x_\mathrm{0
{\scriptscriptstyle L}} < x_+$. Going to smaller temperatures, pinning becomes
stronger and the bistable interval $[x_- \sim \kappa^{1/4} \xi, x_+ \sim
\kappa\xi]$ increases asymmetrically around $x_0 \sim \sqrt{\kappa}\xi$;
again, the competition between the growing $\kappa(\tau)$ and the decreasing
$\xi(\tau)$ determines the evolution of the bistable interval with decreasing
temperature.

\begin{figure}[t]
    \subfloat{
      \includegraphics[width=.9\columnwidth]{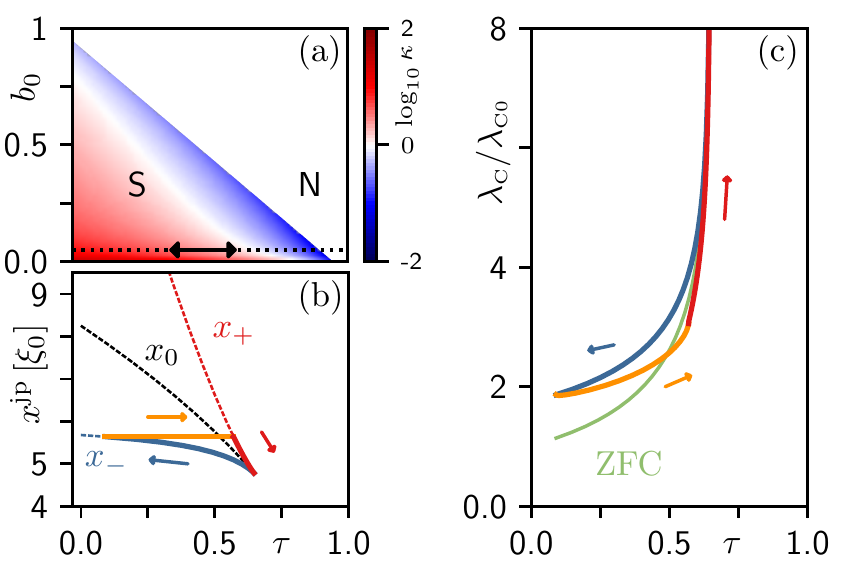}
   }

   \subfloat{
      \includegraphics[width=.9\columnwidth]{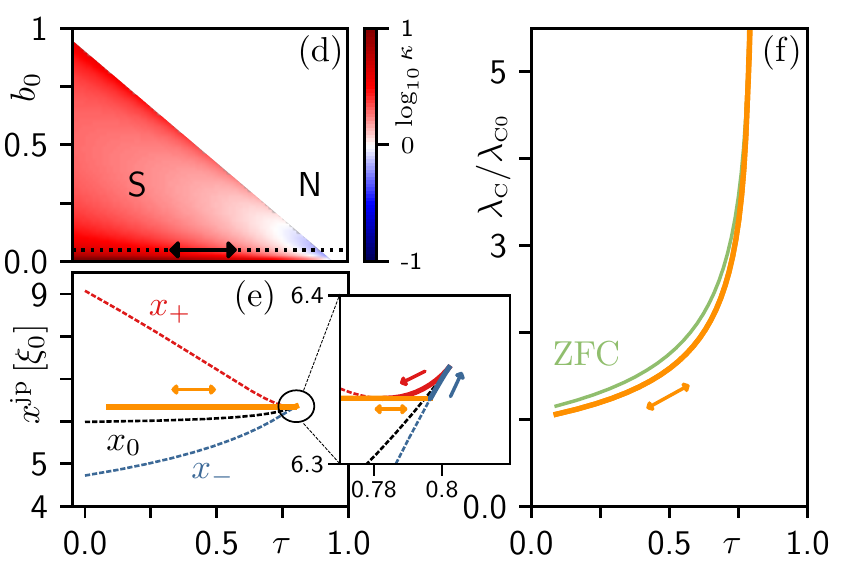}
      }
	\caption{Characteristics of strong pinning for insulating defects
	(top, relative defect size $\rho /\xi_0 = 1.25$) and for $\delta
	T_c$-pinning (bottom, same ratio $\rho /\xi_0$ and $\delta T_c/T_c =
	0.5$) assuming a Lorentzian potential. \textrm{(a,d)} show maps of the
	pinning parameter $\kappa$ over the $b_0$--$\tau$ phase diagram (N and
	S denote normal and superconducting phases); we focus on small fields
	$b_0 = 0.05$, with strong pinning (red) turning on at $\kappa = 1$
	(white color) when decreasing temperature.  \textrm{(b,e)} Evolution
	of the edges $x_+$ (red dots) and $x_-$ (blue dots) of the bistable
	pinning region, as well as branch crossing point $x_0$ (black dots),
	with decreasing temperature $\tau$.  During a cooling--warming cycle,
	the jump position $x^\mathrm{jp}$ follows the phases $b$ (blue line),
	$a$ (orange), $b'$ (red) in (b) but remains fixed in the $a$ phase
	(orange) near $x_\mathrm{0{\scriptscriptstyle L}}$ except for a small
	region near the strong-pinning onset, see (e) and expanding inset.
	\textrm{(c,f)} Hysteretic trace of the Campbell length $\lambda_{\rm
	\scriptscriptstyle C}(\tau)$ for a cooling--warming cycle, normalized
	with respect to $\lambda_\mathrm{{\scriptscriptstyle C}0} = (B_0^2/
	4\pi\alpha_0)^{1/2}$, $\alpha_0$ the equilibrium Campbell curvature at
	$T=0$. Colors indicate different phases $a$ (orange), $b$ (blue), $b'$
	(red) assumed along the cooling--warming cycle; the ZFC trace (green)
	is shown for comparison.  Insulating defects produce a hysteretic
	trace with $x^\mathrm{jp}$ following $x_-$ on cooling that is changing
	over via the $a$ phase to $x_+$ on heating, see (b).  For $\delta
	T_c$-pinning, no memory effect shows up away from the Labusch point,
	that is owed to the fixed jump position $x^\mathrm{jp} \approx
	x_\mathrm{0{\scriptscriptstyle L}}$ in (e).}
     \label{fig:ins-dTc-defects}
\end{figure}

Depending on which of the above scenaria is realized, the system will exhibit
quite a different behavior.  To fix ideas, let us start with the specifc case
of an insulating defect with $\alpha = 3/2$, $\beta = 1/2$ and a small field
$b_0$; the evolution of the Labusch parameter $\kappa(\tau, b_0)$ within the
$B$--$T$ phase diagram is shown in Fig.\ \ref{fig:ins-dTc-defects}(a).
Decreasing $T$ below $T_{\rm\scriptscriptstyle L}$, one finds that the slope
$\partial_\tau x_\pm|_{T_{\rm\scriptscriptstyle L}} < 0$ and hence the
bistable interval $[x_-,x_+]$ moves to larger values of $x$, $x_\mathrm{0
{\scriptscriptstyle L}} < x_- < x_+$. Going to smaller temperatures with
larger values of $\kappa$, one finds that $x_- \sim \mathrm{const.}$ and $x_+
\propto (1-\tau)^{3/2}$ increases with decreasing temperature, hence the
interval $[x_-,x_+]$ continues shifting to the right as $T$ goes down, see
Fig.\ \ref{fig:ins-dTc-defects}(b).

This is the situation illustrated in Fig.\ \ref{fig:hysteresis_scheme}: With
parts of the $S$-shaped force profile $f_\mathrm{pin}(x)$ turning unstable,
the jump position $x^\mathrm{jp}$, starting out at
$x_\mathrm{0{\scriptscriptstyle L}}$, gets pinned to the left (pinning) edge
of the $S$-profile, $x_b^\mathrm{jp}(T) = x_-(T)$---we name this the phase
$b$.  This behavior continues until the further decrease in temperature is
stopped at the minimal temperature $T_\mathrm{min}$ of the cycle where
$x^\mathrm{jp}$ reaches its maximal value $x_b^\mathrm{jp}(T_\mathrm{min})$.
Upon raising the temperature, we enter phase $a$ of the cycle where the
bistable region shrinks and moves leftward; the jump position then stays fixed
at its maximum, $x^\mathrm{jp}_a = x_-(T_\mathrm{min})$ until the right end
$x_+(T)$ of the bistable region coincides with the jump at $x^\mathrm{jp}_a$
at the crossover temperature $T_\times$, $x_-(T_\mathrm{min}) =
x_+(T_\times)$.  From $T_\times$ onwards, the jump position is pinned to the
right edge of the $S$-profile, with $x^\mathrm{jp}_{b'}(T)=x_+(T)$; this is
denoted as phase $b'$ in Fig.\ \ref{fig:hysteresis_scheme}. The complete
hysteretic loop traced out by $x^\mathrm{jp}(T)$ over the three phases $b$,
$a$, and $b'$ is shown in Fig.\ \ref{fig:ins-dTc-defects}(b). Finally, the
changeover between phases $b$, $a$, and $b'$ produces a pronounced hysteresis
in the Campbell penetration depth $\lambda_{\rm \scriptscriptstyle C}(\tau)$,
as illustrated in Fig.\ \ref{fig:ins-dTc-defects}(c).

Having established the cycle $T_{\rm\scriptscriptstyle L}\to T_\mathrm{min}
\to T_\times \to T_{\rm\scriptscriptstyle L}$ for the insulating defect, let
us briefly discuss other possibilities. Another typical situation is shown in
Figs.\ \ref{fig:ins-dTc-defects}(d--f), where we show the Labusch parameter
$\kappa(\tau, b)$ for the case of $\delta T_c$-pinning in (d), together with
the evolution of $x_-$, $x_+$, and $x_0$ as well as the resulting cycle in
(e), and the  Campbell penetration depth $\lambda_{\rm \scriptscriptstyle
C}(\tau)$ in (f). In this case, we find that $\partial_\tau
x_\pm|_{T_{\rm\scriptscriptstyle L}} >0$, hence the $S$ shaped instability in
$f_\mathrm{pin}(x)$ initially moves to the left, $[x_-,x_+] < x_\mathrm{0
{\scriptscriptstyle L}}$. However, as shown in Fig.\
\ref{fig:ins-dTc-defects}(e) and the expanded box, the upper edge $x_+$
quickly turns around with further decreasing $T$ and we change over from a
narrow $b'$ phase to an $a$ phase that completely dominates the cycle. With
the jump $x^\mathrm{jp}(\tau)$ remaining fixed close to $x_\mathrm{0
{\scriptscriptstyle L}}$ deep in the bistable regime over the entire cycle, we
find essentially no hysteresis for the case of $\delta T_c$-pinning, except
for a narrow region close to $T_{\rm\scriptscriptstyle L}$.

\begin{figure}[t]
    \subfloat{
      \includegraphics[width=.9\columnwidth]{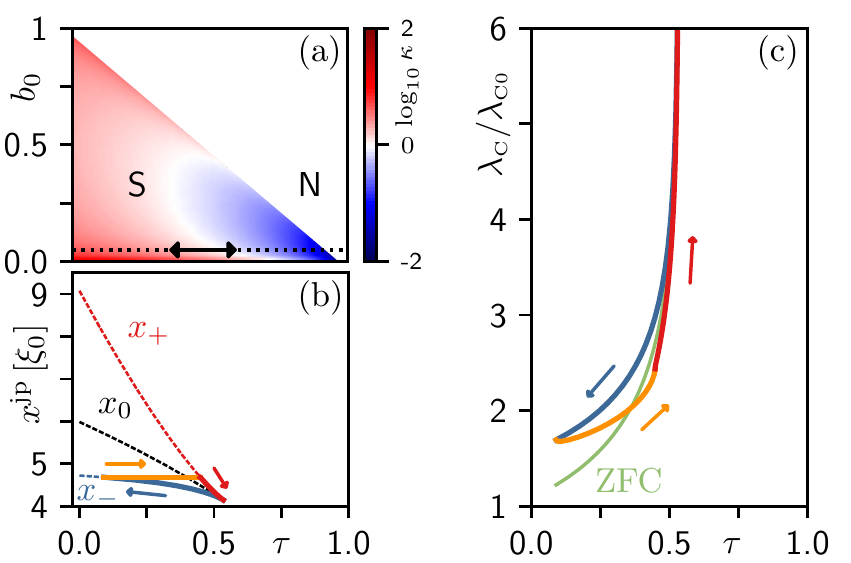}
   }

   \subfloat{
      \includegraphics[width=.9\columnwidth]{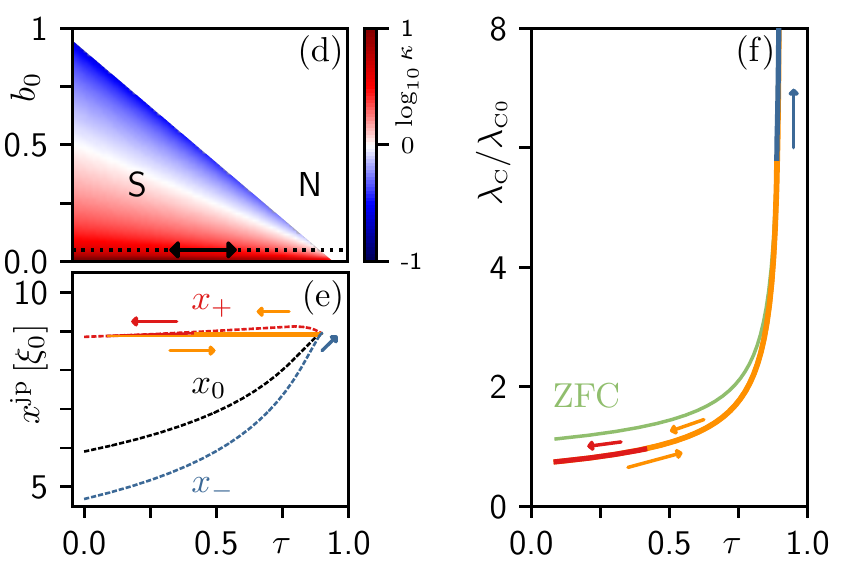}
      }
	\caption{Characteristics of strong pinning for $\delta \ell$-pinning
	(top, with $\delta \ell/\ell = 0.5$) and for metallic defects (bottom)
	assuming a Lorentzian potential.  \textrm{(a,d)} show maps of the
	pinning parameter $\kappa$ over the $b_0$--$\tau$ phase diagram (N and
	S denote normal and superconducting phases); we focus on small fields
	$b_0 = 0.05$, with strong pinning (red) turning on below $\kappa = 1$
	(white color) when decreasing temperature.  \textrm{(b,e)} Evolution
	of the edges $x_+$ (red dots) and $x_-$ (blue dots) of the bistable
	pinning region, as well as branch crossing point $x_0$, with
	decreasing temperature $\tau$.  During a cooling--warming cycle, the
	jump position $x^\mathrm{jp}$ follows the phases $b$ (blue line), $a$
	(orange), $b'$ (red) in (b) but remains close to
	$x_\mathrm{0{\scriptscriptstyle L}}$ and $x_+$ in (e), following a
	thin cycle $a$ (orange), $b'$ (red), $a$ (orange), $b$ (blue).
	\textrm{(c,f)} Hysteretic trace of the normalized Campbell length
	$\lambda_{\rm \scriptscriptstyle C}(\tau)$ for a cooling--warming
	cycle. Colors indicate different phases $a$ (orange), $b$ (blue), $b'$
	(red) assumed along the cooling--warming cycle; the ZFC trace (green)
	is shown for comparison. $\delta\ell$ pinning produces a hysteretic
	trace with $x^\mathrm{jp}$ following $x_-$ on cooling that is changing
	over via the $a$ phase to $x_+$ on heating, see (b). For metallic
	defects, memory effects are small, see (e) and (f), as the jump point
	always remains close to $x_\mathrm{0{\scriptscriptstyle L}}$ and
	$x_+$.}
     \label{fig:dl-met-defects}
\end{figure}

The two other cases, $\delta \ell$-pinning (with $\partial_\tau
x_\pm|_{T_{\rm\scriptscriptstyle L}} <0$) and the metallic defect (with
$\partial_\tau x_\pm|_{T_{\rm\scriptscriptstyle L}} = 0$) closely resemble the
behavior of the insulating defect and of $\delta T_c$-pinning, respectively,
see Fig.\ \ref{fig:dl-met-defects}, with an important difference remaining,
though. Indeed, focusing on the metallic defect and the $\delta T_c$-pinning,
with both not developing a hysteresis, we notice that for $\delta T_c$-pinning
the $a$ phase is realized deep in the bistable interval with
$x^\mathrm{jp}(\tau)$ far away from the edges $x_\pm(\tau)$, while for the
metallic defect, the $a$ phase resides close to the edge with $x^\mathrm{jp}
(\tau) \approx x_+(\tau)$, i.e., close to phase $b'$.  As we have already
learnt in Sec.\ \ref{sec:creep_l_C_th}, creep is strong when the barriers are
small, which is the case when the jumps at $x^\mathrm{jp}$ are close to the edges
$x_\pm$, see Fig.\ \ref{fig:strong_pinning}. On the other hand, creep is weak
when the jump $x^\mathrm{jp}$ resides deep within the bistable interval, e.g.,
away from the edges $x_\pm(T)$ where barriers become large. Hence, we
conclude that creep is strong in phases $b$ and $b'$ where the jump
$x^\mathrm{jp}$ is pinned to the edges, but is weak, deep in the phase $a$.
And hence, we expect that for $\delta T_c$-pinning creep will be small, while
the metallic defect will exhibit stronger creep.

We thus conclude, that `reading' a temperature cycle of the Campbell
penetration depth $\lambda_{\rm \scriptscriptstyle C}(\tau)$ allows us to gain
quite some insights into the pinning mechanism:

For an insulating defect and for $\delta \ell$ pinning, the cycle is
hysteretic with strong creep upon cooling in the $b$-phase, weak creep upon
heating in the $a$-phase, and again strong creep in the final heating phase
close to $T_{\rm\scriptscriptstyle L}$ where the $b'$-phase is realized.  For
a metallic defect, the cycle is non-hysteretic but creep is reasonably strong
since the system straddles the regime at the edge of the $a$-phase/$b'$-phase.
Finally, for $\delta T_c$-pinning, the cycle is again non-hysteretic but with
weak creep as the system resides deep in the $a$-phase. We note, that other
pinning types may occur exhibiting cycles that are yet different from those
analyzed here.

Below, we determine the Campbell curvatures $\alpha_\mathrm{sp}^\mathrm{
\scriptscriptstyle FC}$ for the different phases $a$, $b$, $b'$.  These
results then produce the Campbell penetration depth $\lambda_{\rm
\scriptscriptstyle C}(\tau)$ shown in figures \ref{fig:ins-dTc-defects}(c) and
(f) and \ref{fig:dl-met-defects}(c) and (f).  While for the $b$ and $b'$
phases the jump positions at the edges $x_\pm$ are well defined, for the $a$
phase the jump position depends on the way the phase is entered. E.g., for
the cooling-warming loop with underlying insulating defects or with $\delta
\ell$-pinning, we have $x_a^\mathrm{jp} = x_-(T_\mathrm{min})$ since we enter
the $a$ phase from the $b$ phase, while for metallic defects, we enter the $a$
phase upon the onset of strong pinning and hence $x_a^\mathrm{jp} =
x_\mathrm{0{\scriptscriptstyle L}}$. In a third case, realized for $\delta
\ell$-pinning, the $a$ phase is entered through the $b'$ phase with
$x_a^\mathrm{jp} = x_+(T_\mathrm{inf})$ and $T_\mathrm{inf}$ the temperature
where $\partial_\tau x_+(\tau)$ changes sign; this situation is realized for
$\delta T_c$-pinning close to $T_\mathrm{\scriptscriptstyle L}$, see Fig.\
\ref{fig:ins-dTc-defects}(e).  All these different cases produce different
values for $x_a^\mathrm{jp}$.

\subsubsection{Campbell curvatures for \textrm{FC} phases}
\label{sec:FC_phases}

We now determine the Campbell curvatures $\alpha_\mathrm{sp}^\mathrm{
\scriptscriptstyle FC}$ for the different phases $a$, $b$, $b'$ potentially
appearing in a hysteresis loop,  first in the marginal strong pinning
situation $\kappa - 1 \ll 1$ valid close to $T_{\rm\scriptscriptstyle L}$ and
thereafter at large $\kappa$ potentially realized at small temperatures. For
convenience, we scale the results for the curvatures
$\alpha_\mathrm{sp}^\mathrm{ \scriptscriptstyle FC}$ using the corrresponding
ZFC results $\alpha_\mathrm{sp}$, where we denote field-cooled results via an
upper index $^\mathrm{\scriptscriptstyle FC}$ or with specific phase indices
$^{a,b,b'}$, while the ZFC expressions remain without upper index.
Furthermore, while the results for the $b$ and $b'$ phases can be pushed to
closed expressions, this is not the case for the $a$ phase, as for the latter
the jump position $x_a^\mathrm{jp}$, while constant in temperature, resides
somewhere within the bistable interval $[x_-,x_+]$, as discussed above.

Let us start close to the Labusch point $\kappa - 1 \ll 1$ where the
\textrm{ZFC} result for the Campbell curvature follows from Eq.\
\eqref{eq:alpha_sp_sk} with the trapping scale $x_0$ and the force jump
\eqref{eq:df_tp_s_kappa}, providing the result
\begin{align}
\alpha_\mathrm{sp} \approx \frac{3 \pi x_0 \xi}{a_0^2}
   \left(\frac{4\Cbar\xi^2}{e_p}\right)^{1/2}
   \left(\frac{\kappa-1}{4\gamma}\right)^{1/2} n_p \Cbar.
\end{align}
For the Lorentzian-shaped potential, this reduces to the simple expression
$\alpha_\mathrm{sp} \approx \sqrt{3}\pi\sqrt{\kappa-1}\, n_p\,(e_p/a_0^2)$,
where we have used that $x_0\approx 2\sqrt{2}\xi$ to leading order in
$\kappa-1$.

For the \textrm{FC} state at marginal pinning $\kappa - 1 \ll 1$, we evaluate
the Campbell curvature \eqref{eq:alpha_FC} using the jump radius $x_0$ that is
same to leading order in $\kappa - 1$ for all three phases $a$, $b$, $b'$.  We
find the force jumps with the help of the force balance equation
\eqref{eq:fbx}, $\Delta f_\mathrm{pin}^\mathrm{fp}(x)= \Cbar
\left[r_\mathrm{f}(x)-r_\mathrm{p}(x)\right]$, and make use of Eqs.\
\eqref{eq:rp+f-} and \eqref{eq:rf+} for the tip positions at the
characteristic points $x_\pm$ relevant in phase $b$ and $b'$ of the cycle.
Furthermore, at small values $\kappa - 1 \ll 1$, the bistable interval is
narrow and symmetric around $x_0$; we then choose $x_a^\mathrm{jp} = x_0$ as a
representative point (with the largest force jump) and make use of
\begin{align}\label{eq:rf0}
   r_\mathrm{f}(x_0) \!-\! r_m & =  - r_\mathrm{p}(x_0) \!+\! r_m 
   \approx \sqrt{\frac{3\Cbar\xi^4}{\gamma e_p}}(\kappa-1)^{1/2}.
\end{align}
As a result, we find the force jumps,
\begin{align}\label{eq:dffppm}
   \Delta f_\mathrm{pin}^\mathrm{fp}(x_\pm)&\approx
   \frac{3}{2}\Cbar\xi\sqrt{\frac{4\Cbar\xi^2}{\gamma e_p}}(\kappa-1)^{1/2},
   \\ \label{eq:dffp0}
   \Delta f_\mathrm{pin}^\mathrm{fp}(x_0)&\approx \sqrt{3}\Cbar\xi\sqrt{\frac{4\Cbar\xi^2}
   {\gamma e_p}}(\kappa-1)^{1/2},
\end{align}
relevant, respectively, for the $b$, $b'$, and $a$ phases of the temperature
cycle. For a Lorentzian potential, $\gamma =3/8$ provides the force jumps
$\Delta f^\mathrm{fp}_\mathrm{pin}(x_\pm)\approx \bigl[\sqrt{3/8}\,
(\kappa-1)^{1/2}\bigr]e_p/\xi$ and $\Delta f^\mathrm{fp}_\mathrm{pin}
(x_0)\approx \left[(\kappa-1)^{1/2}/\sqrt{2}\right]e_p/\xi$. Inserting these
results in the expression \eqref{eq:alpha_FC} for the Campbell curvature, we
find, to leading order in $\kappa-1$,
\begin{align}\label{eq:alpha_FC_small_kappa_1}
   \alpha_\mathrm{sp}^{b,b'}&\approx \sqrt{3}\pi (\kappa-1)^{1/2}n_p(e_p/a_0^2),\\
   \alpha_\mathrm{sp}^{a}&\approx  2 \pi(\kappa-1)^{1/2}n_p(e_p/a_0^2)
   \label{eq:alpha_FC_small_kappa_2}.
\end{align}
Finally, comparing the \textrm{FC} and \textrm{ZFC} results, we obtain the ratios
\begin{equation}\label{eq:FC-ZFC}
   \frac{\alpha_\mathrm{sp}^{b,b'}}{\alphaSP} 
   \approx 1 ~\textrm{ and }~ \frac{\alpha_\mathrm{sp}^a}{\alphaSP}
   \approx \frac{2}{\sqrt{3}} \approx 1.15
\end{equation}
valid in the vicinity of the Labusch temperature $T \lesssim T_{\rm
\scriptscriptstyle L}$.  Given the symmetry between $x_-$ and $x_+$ of the
bistable region within the marginally strong pinning limit, the force jumps in
the $b$ and $b'$ phases are equal to the force jump \eqref{eq:df_tp_s_kappa}
in the \textrm{ZFC} state, and hence the FC Campbell curvature is identical to
the ZFC result in the limit $\kappa \to 1$. For the representative point
$x_a^\mathrm{jp} = x_0$ in the $a$ phase, the force jump is larger by a factor
$2/\sqrt{3} \approx 1.15$, resulting in a larger ratio for the Campbell
curvature.

Including the next (4th) order term $\tau (e_p/4\xi^5) \, \delta r^{\,4}$ in
the expansion \eqref{eq:f_p_exp_sk} for $f_p(r_m + \delta r)$, we find an
asymmetric correction to $\delta r_\mathrm{max}$ in Eq.\
\eqref{eq:delta_r_sk},
\begin{equation}\label{eq:delta_r_sk4}
   \delta r_\mathrm{max}(x_\pm) \approx \delta r_\mathrm{max} \mp
   \frac{\xi}{8} \frac{\tau}{\gamma^2}\frac{4\bar{C}\xi^2}{e_p} (\kappa -1),
\end{equation}
where the indices $\pm$ refer to pinned and free branches, respectively.  For
a Lorentzian potential, we have $\tau = 15/(24\sqrt{2})$. Accounting for these
4th order corrections in the evaluation of the force jumps \eqref{eq:dffppm},
the results $\alpha_\mathrm{sp}^{b,b'}$ for the $b$ and $b'$ phases separate,
in fact, symmetrically with respect to the ZFC result, $\alpha_\mathrm{sp} (t)
= [\alpha_\mathrm{sp}^b(t) + \alpha_\mathrm{sp}^{b'}(t)]/2$, as the latter
involves jumps both at $x_+$ and $x_-$. In Fig.\ \ref{fig:FC-rel-vs-t}, we
show our analytic results for $\alpha_\mathrm{sp}^{b,b'}$ at marginally strong
pinning and find that they compare well with the numerical results in the
limit $t\to t_0$ discussed here.

At smaller temperatures, the pinning parameter $\kappa$ grows larger and the
Campbell curvature has to be evaluated in the $\kappa \gg 1 $ limit. Using the
expressions \eqref{eq:x_+}--\eqref{eq:x_-} for the
endpoints $x_+$ and $x_-$, we find that the Campbell curvature in the
\textrm{ZFC} state scales as
\begin{equation}
   \alpha_\mathrm{sp}\approx n_p \frac{2x_-}{a_0}\frac{\Cbar x_+}{a_0}
   \sim n_p\, \kappa^{1/(n+2)} (e_p/a_0^2)
\end{equation}
for an algebraically decaying potential and $\alpha_\mathrm{sp}\approx
\sqrt{6}\,(3\kappa)^{1/4}\,n_p\,(e_p/a_0^2)$ for the Lorentzian. For the
\textrm{FC} state, we approximate the relevant force jumps in the phases
$a$, $b$, and $b'$ as (cf.\ Eq.\ \eqref{eq:x_-}) 
\begin{eqnarray}\label{eq:FC_fj_lk}
   \Delta f_\mathrm{pin}^\mathrm{fp}(x_a^\mathrm{jp}) &\approx& \Cbar r_\mathrm{f}(x_a^\mathrm{jp}) 
   \approx \Cbar x_a^\mathrm{jp}, \\ \nonumber
   \Delta f_\mathrm{pin}^\mathrm{fp}(x_-) &\approx& 
   \Cbar r_\mathrm{f}(x_-)\approx \frac{n+1}{n+2} \Cbar x_-, \\ \nonumber
   \Delta f_\mathrm{pin}^\mathrm{fp}(x_+) &\approx& \Cbar x_+,
 \end{eqnarray}
with $x_a^\mathrm{jp}$ depending on the specific situation, see the discussion
above.  We then arrive at the following results for the curvatures in phases
$a$, $b$, and $b'$ of the hysteretic temperature cycle,
\begin{eqnarray}\label{eq:alpha_FC_lk}
   \alpha_\mathrm{sp}^a &\approx& \pi \left(\frac{x_a^\mathrm{jp}}{a_0}\right)^2 n_p \Cbar,
   \\ \nonumber
   \alpha_\mathrm{sp}^b &\approx& \pi \frac{n+1}{n+2} \left(\frac{x_-}{a_0}\right)^2 n_p \Cbar,
   \\ \nonumber
   \alpha_\mathrm{sp}^{b'} &\approx& \pi \left(\frac{x_+}{a_0}\right)^2 n_p \Cbar.
\end{eqnarray}
Making use of the large-$\kappa$ expressions \eqref{eq:x_+L} and
\eqref{eq:x_-L} for $x_\pm$ and focusing on a Lorentzian potential, we find
\begin{align} \label{eq:alpha^b}
   \alpha_\mathrm{sp}^\mathrm{b}&\approx (16/3\kappa)^{1/2}\pi\,n_p (e_p/a_0^2),\\ 
   \label{eq:alpha^b'}
   \alpha_\mathrm{sp}^\mathrm{b'}&\approx (3/2)^3(\kappa/4)\,\pi\, n_p (e_p/a_0^2),
\end{align}
resulting in the following ratios valid at large $\kappa$ 
\begin{align}\label{eq:a_ratio_b}
   \frac{\alpha_\mathrm{sp}^b}{\alphaSP}&\approx
   2\sqrt{\frac{2}{3}}\frac{\pi}{(3\kappa)^{3/4}}, \\
   \label{eq:a_ratio_bp}
   \frac{\alpha_\mathrm{sp}^{b'}}{\alphaSP}&\approx
   \frac{3}{32}\sqrt{\frac{3}{2}}(3\kappa)^{3/4}\pi.
\end{align}
The expression in \eqref{eq:alpha_FC_lk} for the $a$ phase cannot be brought
to a simpler closed form as $x_a^\mathrm{jp}$ depends on the preparation. The
above results differ from those in Ref.\ \onlinecite{Willa_2016} due to the
more accurate handling of the (circular) trapping geometry in the FC
situation. At small temperatures and large $\kappa$, the equivalence between
the Campbell curvature in the phases $b$ and $b'$ is broken, and the
corresponding results are substantially different from the \textrm{ZFC} ones,
a consequence of the asymmetric nature of the bistable region at large
$\kappa$. Note the result \eqref{eq:a_ratio_bp} that turns out large; the
scaling $\propto \kappa^{3/4}$ follows from the different trapping areas,
$x_+^2 \propto \kappa^2$ for the FC case and $x_+ x_- \propto \kappa
\kappa^{1/4}$ for the ZFC state.

In Figs.\ \ref{fig:ins-dTc-defects}(c,f) and \ref{fig:dl-met-defects}(c,f), we
translate the curvatures $\alpha^{\mathrm{{\scriptscriptstyle FC}}}$ to the
Campbell penetration depth $\lambda_{\rm\scriptscriptstyle C} \propto
1/\sqrt{\alpha_\mathrm{sp}^{\mathrm{{\scriptscriptstyle FC}}}}$ and illustrate
typical hysteresis loops as expected in materials with different types of
pinning centers, where we normalize our results with
$\lambda_\mathrm{{\scriptscriptstyle C}0} = (B_0^2/ 4\pi\alpha_0)^{1/2}$,
$\alpha_0$ the equilibrium Campbell curvature at $T=0$.

\subsection{Creep effect on the hysteresis loop in \textrm{FC} state}\label{sec:creep_FC}

\begin{figure}[t]
        \includegraphics[width = 1.\columnwidth]{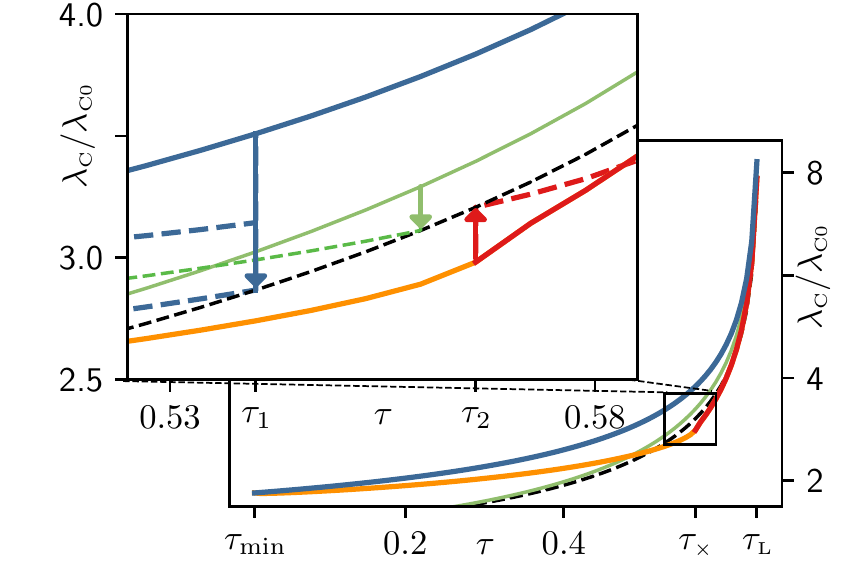}
	\caption{Hysteresis loop for the Campbell length $\lambda_{\rm
	\scriptscriptstyle C}$ measured along a temperature cycle
	$\tau_{\rm\scriptscriptstyle L} \rightarrow \tau_\mathrm{min}
	\rightarrow \tau_{\rm\scriptscriptstyle L}$. We assume insulating
	defects of Lorentzian form with a defect size $\rho / \xi_0 = 1.25$,
	that produces a maximum pinning parameter $\kappa(\tau_\mathrm{min},
	b_0) \approx 8.3$, see Eq.\ \eqref{eq:kappa_evolution}, and
	$\tau_{\rm\scriptscriptstyle L} \equiv T_{\rm\scriptscriptstyle L}/T_c
	\approx 0.65$. We choose $\tau_\mathrm{min} = 0.01$ and a reduced
	magnetic field $b_0 = 0.05$.  The Campbell length $\lambda_{\rm
	\scriptscriptstyle C}(\tau)$ is normalized with respect to
	$\lambda_\mathrm{{\scriptscriptstyle C}0} = (B_0^2/
	4\pi\alpha_0)^{1/2}$, $\alpha_0$ the equilibrium Campbell curvature at
	$T=0$. The cycle follows the phases $b$ (blue) on cooling, $a$
	(orange) and $b'$ (red) on heating, see also Fig.\
	\ref{fig:ins-dTc-defects}. For comparison, the Campbell length in the
	\textrm{ZFC} state (green solid line) is shown, as well as the
	equilibrium value (black dashed line).
        Inset: Effect of creep on the Campbell length in the \textrm{FC} and
        \textrm{ZFC} states. The temperature range corresponds to the small
        black box in the main figure. At a temperature $\tau_1$ during
        cooling, the \textrm{FC} state is relaxed (blue arrow) and approaches
        the equilibrium state, with the Campbell length gradually decreasing
        towards $\lambda_0 (\tau)$. Resuming the cooling process at a later
        time, the Campbell length follows different trajectories (thick dashed
        blue) for different waiting times.  Repeating this procedure at a
        temperature $\tau_2$ along the warming branch, analogous results are
        found, with the Campbell length now increasing under the effect of
        creep (shown in red).  Letting the system relax from the \textrm{ZFC}
        state (green arrow), the Campbell length approaches the equilibrium
        value $\lambda_0$ and behaves similar to the \textrm{FC} state upon
        further cooling (green dashed).}
     \label{fig:hysteresis_start_stop}
\end{figure}

We now proceed to include the effect of thermal fluctuations or creep on the
Campbell length $\lambda_{\rm\scriptscriptstyle C}$ in the FC case. Thermal
fluctuations drive the vortex state towards equilibrium, that corresponds to
shifting the original position of the force jump towards $x_0$, thereby
approaching the equilibrium distribution $p_\mathrm{eq}(R) = \Theta(x_0 - R)$.
This relaxation can be experimentally observed at any place along the
temperature cycle, with a specific example (assuming insulating defects) shown
in Fig.\ \ref{fig:hysteresis_start_stop}, by interrupting the temperature
sweep and letting the system relax. Alternatively, creep tends to close the
hysteresis loop when cycling the temperature at an ever slower rate.

The relaxation of the FC states is largely different in the three phases $a$,
$b$, and $b'$ of the cycle: when the force jump is pinned to the edges $x_-$
or $x_+$ in the phases $b$ and $b'$, the activation barriers are initially
small (as they vanish at $x_\pm$) and hence relaxation is large. On the other
hand, deep in the $a$ phase, the jump location $x_a^\mathrm{jp}$ resides away
from these edges and the initial barriers $U(x_a^\mathrm{jp})$ are already large
to start with, resulting in a slow relaxation. Nevertheless, there is an
interesting crossover to the fast creeping $b$ and $b'$ phases at the edges of
the $a$ phase; in the following, we first focus on relaxation of the phases
$b$ and $b'$ and discuss the slow relaxation of phase $a$ and its relation to
the $b$ and $b'$ phases at the end.

Before deriving the expressions for the time evolved curvatures
$\alpha^{\rm\scriptscriptstyle FC} (t,T)$, we summarize our results in Fig.\
\ref{fig:hysteresis_start_stop} on the example of a hysteretic cooling/warming
hysteretic cycle as it appears for the insulating defect or for $\delta
\ell$-pinning.  Depending on which part of the temperature cycle the
relaxation process takes place (by stopping the change in temperature), the
Campbell length $\lambda_{\rm \scriptscriptstyle C}(t,T)$ either grows or
decreases in approaching the equilibrium state.  The inset in Fig.\
\ref{fig:hysteresis_start_stop} shows three cases, relaxation during the $b$ phase
(on cooling, in blue, with decreasing $\lambda_\mathrm{\scriptscriptstyle C}^b$),
at the end of the $a$ phase/start of the $b'$ phase (upon heating, in red,
with increasing $\lambda_\mathrm{\scriptscriptstyle C}^{b'}$), and relaxation
of the ZFC state in green (decreasing $\lambda_\mathrm{\scriptscriptstyle
C}$). The various relaxation traces approach the equilibrium value
$\lambda_\mathrm{C}^0 \propto 1/\sqrt{\alpha_0}$ shown as a black dashed line.
Stopping the relaxation by further cooling/heating, the penetration depth
$\lambda_\mathrm{\scriptscriptstyle C}$ returns back to the FC lines (dashed
blue/red lines).  The relaxation in the $a$ phase is slow away from its edges
at $T_\mathrm{min}$ and $T_\times$ as the activation barriers become large.
Hence, we find that combining cooling, heating, and relaxation allows us to
install numerous different vortex states, permitting us to spectroscopize the
pinning force and thus probe the pinning potential of defects in a material.

\begin{figure*}
        \includegraphics[width = 1.\textwidth]{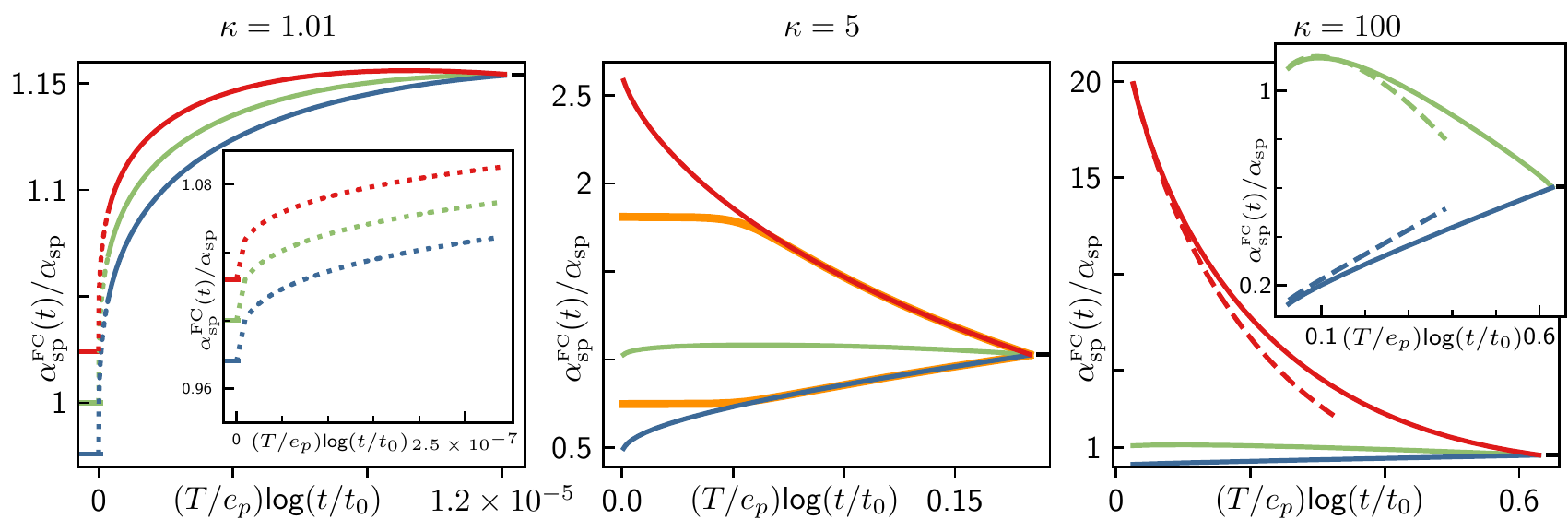}
\caption{Relaxation of the scaled Campbell curvature
	$\alpha_\mathrm{sp}^\mathrm{\scriptscriptstyle FC}
	(t)/\alpha_\mathrm{sp}$ versus creep parameter $\mathcal{T}  = (T/e_p)
	\ln(t/t_0)$. Shown are results for the field cooled (FC) phases $b$
	(in blue), $b'$ (in red), and $a$ (in orange, middle panel), as well
	as the zero-field cooled result (ZFC, green) for comparison. Assuming
	a Lorentzian pinning potential, we show results for marginally strong
	($\kappa = 1.01$), intermediate ($\kappa = 5$), and very strong
	($\kappa = 100$) pinning.  Numerical results are shown as continuous
	lines, dashed lines (at large $\kappa$) describe analytic results at
	small values of $\mathcal{T}$.  At marginally strong pinning ($\kappa
	= 1.01$, left panel), the relaxation curves for the phases $b$ and
	$b'$ separate symmetrically away from the ZFC result, a result
	obtained analytically only when going beyond leading order, see Eq.\
	\eqref{eq:delta_r_sk4}; the thick colored ticks marking the analytic
	results at $t = t_0$ agree well with the end points of the numerical
	curves (dotted lines show the extrapolation of the results to $t =
	0$). The middle panel ($\kappa = 5$) also shows results for the $a$
	phase with a flat initial regime (slow relaxation) joining the trace
	(at time $t_a$, see Eq.\ \eqref{eq:t_a}) for the $b'$ (red) or $b$
	(blue) phase (depending on the starting point $x_0< x_a^\mathrm{jp} <
	x_+$ or $x_- < x_a^\mathrm{jp} < x_0$, respectively) at large times.
	Analytic and numerical results agree well in the large $\kappa$ limit
	and at small $\mathcal{T}$, with the results for the $b$ ($b'$) phases
	increasing (decreasing) monotonously with time, while the ZFC result
	is non-monotonous. Note that the large values of
	$\alpha_\mathrm{sp}^{b'} /\alpha_\mathrm{sp}$ (red curve) require a
	large $\kappa$ that is realized at low temperatures, typically. Thick
	ticks at large times mark the asymptotic value $\alpha_0/\alphaSP$
	close to equilibrium $t \sim t_\mathrm{eq}$ before entering the TAFF
	region, see text. The (numerical) lines terminate at the boundary of
	applicability $(T/e_p) \ln(t/t_0) \approx (\kappa-1)^2/8$ and $\sim 1$
	for marginally strong and very strong pinning, respectively. At very
	short times $t \sim t_0$, our creep analysis breaks down as the
	barriers $U$ vanish.
}
     \label{fig:FC-rel-vs-t}
\end{figure*}
 
We now return back to the general discussion and derive the time evolution of
the Campbell curvatures $\alpha^{\rm \scriptscriptstyle FC}(t,T)$ for the
phases $b$ and $b'$.  In describing the effect of creep on the Campbell
curvature, we make heavy use of the results obtained in Sec.\
\ref{sec:creep_l_C_th}; specifically, we can make use of the shifts $\delta
x_\pm$ and $\delta r_\mathrm{p\pm}$, $\delta r_\mathrm{f\pm}$ of the vortex
asymptotic and tip position as determined through the dimensionless thermal
barrier $\mathcal{T}$ defined in Eq.\ \eqref{eq:def_U}.

Very close to the Labusch point $\kappa - 1 \ll 1$, the bistable region is
symmetric around $x_0$ and the force jumps at $x_-$ and $x_+$ are equal,
hence, relaxation of the Campbell curvature along phases $b$ and $b'$ of the
temperature cycle are the same to leading order and identical with the
result Eq.\ \eqref{eq:resc_curv_small_kappa} for the ZFC state with the
properly chosen Labusch parameter $\kappa(T)$. The beyond leading-order
correction \eqref{eq:delta_r_sk4} to the tip positions shifts the $b$ and $b'$
curves for the curvatures symmetrically down and up with respect to the ZFC
result, see Fig.\ \ref{fig:FC-rel-vs-t}.

Going to lower temperatures, the pinning parameter $\kappa$ grows larger and
the equivalence between the FC curvatures $\alpha_\mathrm{sp}^{b}$ and
$\alpha_\mathrm{sp}^{b'}$ and the ZFC result $\alpha_\mathrm{sp}$ is lifted.
Using the expression $\Delta f^\mathrm{fp}_\mathrm{pin}(x) =
\Cbar\left[r_\mathrm{f}(x)-r_\mathrm{p}(x)\right]$ to evaluate the force
jumps, and dropping the pinned against the free vortex tip position, i.e.,
$r_\mathrm{f}(x) \gg r_\mathrm{p}(x)$ for $\kappa\gg 1$, we find that 
\begin{eqnarray}\label{eq:FC_df_gen_kappa_dr1}
   \Delta f^\mathrm{jp,b}_\mathrm{pin} - \Delta f^\mathrm{fp}_\mathrm{pin}(x_-) 
   &\approx& \Cbar \, \delta r_\mathrm{f-},\\
   \Delta f^\mathrm{jp,b'}_\mathrm{pin} - \Delta f^\mathrm{fp}_\mathrm{pin}(x_+) 
   &\approx& 
   -\Cbar \delta r_\mathrm{f+} \label{eq:FC_df_gen_kappa_dr2},
\end{eqnarray}
where $\Delta f^\mathrm{jp,b}_\mathrm{pin} = \Delta
f^\mathrm{fp}_\mathrm{pin}(\xmjp) $ and $\Delta f^\mathrm{jp,b'}_\mathrm{pin}
= \Delta f^\mathrm{fp}_\mathrm{pin}(\xpjp)$.  Combining these results for the
force jumps with the renormalized trapping radii
\begin{align}
   x_b^\mathrm{jp}&=x_-(1\,+\,\delta x_-/x_-),\\
   x_{b'}^\mathrm{jp}&=x_+(1\,-\,\delta x_+/x_+),
\end{align}
the general expressions for the relaxation of the Campbell curvature
during the phases $b$ and $b'$ of the temperature cycle assume the form
(cf.\ Eq.\ \eqref{eq:resc_curv_large_kappa_gen} for the ZFC and note that
$\delta r_\mathrm{f+} \approx \delta x_+$; here, $\kappa = \kappa(T)$ is
always chosen at the appropriate temperature $T$)
\begin{align}\label{eq:FC_creep_alpha_b_gen}
   \frac{\alpha_\mathrm{sp}^{b}(t,T)}{\alpha_\mathrm{sp}^{b}}
   &\approx \biggl(1+\frac{\delta x_-}{x_-}\biggr)
   \biggl(1+\frac{\delta r_\mathrm{f-}}{r_\mathrm{f-}}\biggr),\\
   \frac{\alpha_\mathrm{sp}^{b'}(t,T)}{\alpha_\mathrm{sp}^{b'}}
   &\approx \biggl(1-\frac{\delta x_+}{x_+}\biggr)
   \biggl(1-\frac{\delta r_\mathrm{f+}}{x_+}\biggr).\label{eq:FC_creep_alpha_b'_gen}
   \end{align}
We observe that, contrary to the situation in the ZFC state where we found a
competition between an increasing trapping area and a decreasing force jump,
in the FC states, the changes in the trapping area and in the force jumps work
together. In particular, the trapping area {\it shrinks} with time in the $b'$
phase. Hence, relaxing the state during phase $b$ always leads to an
increasing curvature, larger then the ZFC result, i.e., a decreasing Campbell
penetration depth $\lambda_{\rm \scriptscriptstyle C}$, while the opposite
applies during phase $b'$, see Fig.\ \ref{fig:FC-rel-vs-t}.

Focusing on the phase $b$ for a Lorentzian potential, we can make use of the
results \eqref{eq:delta_x_pm_fin} for $\delta x_-$ and $\delta r_\mathrm{f-}$
and find that
\begin{multline}\label{eq:FC_creep_alpha_b}
   \frac{\alpha_\mathrm{sp}^{b}(t,T)}{\alpha_\mathrm{sp}^{b}}
   \approx \biggl(1+\frac{3}{8}\kappa^{1/3}(\sqrt{3}\mathcal{T})^{2/3}\biggr)\\
                \times\biggl(1+\frac{1}{2}\kappa^{1/6}(\sqrt{3}\mathcal{T})^{1/3}\biggr).
\end{multline}
This result is valid at short times, where $\delta r_\mathrm{f-}$ is dominated
by the square root behavior of the force profile close to $x_-$. Going beyond
short times, such that $\delta x_- \gg x_-/6$, the deviation $\delta
r_\mathrm{f-}\approx r_\mathrm{f-}/3 + \delta x_-$ is approximately linear, as
discussed in the derivation of Eq.\ \eqref{eq:resc_curv_large_kappa_gen_l},
and the Campbell curvature reads
\begin{multline}\label{eq:FC_creep_alpha_b_intermediate}
   \frac{\alpha_\mathrm{sp}^{b}(t,T)}{\alpha_\mathrm{sp}^{b}}
   \approx \biggl(1+\frac{3}{8}\kappa^{1/3}(\sqrt{3}\mathcal{T})^{2/3}\biggr)\\
   \times\biggl(1+\frac{1}{3}+\frac{1}{2}\kappa^{1/3}(\sqrt{3}\mathcal{T})^{2/3}\biggr).
\end{multline}

In the analysis of the relaxation during phase $b'$, we proceed the same way:
we make use of the result \eqref{eq:delta_x_pm_fin} for $\delta x_+$ and note
that $\delta r_\mathrm{f+} \approx \delta x_+$ to find the result
\begin{align}\label{eq:FC_creep_alpha_b'}
   \frac{\alpha_\mathrm{sp}^{b'}(t)}{\alpha_\mathrm{sp}^{b'}}
   &\approx \biggl(1-\frac{\delta x_+}{x_+}\biggr)
   \biggl(1-\frac{\delta x_+}{x_+}\biggr)\nonumber\\
   &\approx\bigl(1-2(\sqrt{3}\mathcal{T})^{2/3} + (\sqrt{3}\mathcal{T})^{4/3}\bigr)
\end{align}
for the Lorentzian potential. With the above results, we find that the
relaxation of the $b$ and $b'$ phases under FC conditions is always faster
than the relaxation for a ZFC experiment, see Eqs.\
\eqref{eq:resc_curv_large_kappa} and \eqref{eq:resc_curv_large_kappa_lT}.

The upward (downward) relaxation of the Campbell curvatures
$\alpha_\mathrm{sp}^{b(b')}(t,T)$ is illustrated in Fig.\
\ref{fig:FC-rel-vs-t}, middle and right panels, with the curves
$\alpha_\mathrm{sp}^{b,b'}(t,T)$ (blue and red) enclosing the ZFC result
$\alpha_\mathrm{sp}(t,T)$ (green). This implies a reverse behavior in the
Campbell penetration length $\lambda_{\rm \scriptscriptstyle C}$, which
decreases with time in the $b$ phase and increases in the $b'$ phase, see the
example of a cooling--warming cycle for the case of insulating defects in
Fig.\ \ref{fig:hysteresis_start_stop}.

Finally, we discuss the behavior of the $a$ phase upon relaxation. The $a$
phase can be entered either via the $b$ phase (see, e.g., Fig.\
\ref{fig:ins-dTc-defects}(b)), via the $b'$ phase (see, e.g., Fig.\
\ref{fig:ins-dTc-defects}(e) very close to $T_{\scriptscriptstyle L}$, or
directly from $x_\mathrm{0 {\scriptscriptstyle L}}$ at $T_{\scriptscriptstyle
L}$ (see Fig.\ \ref{fig:dl-met-defects}(e)).  To fix ideas, here we focus on
the situation where the $a$ phase is entered from the $b'$ phase; the jump
location $x^\mathrm{jp}_a$ then resides in the interval $[x_0,x_+]$ and the
(depinning) barrier $U_\mathrm{dp} (x^\mathrm{jp}_a)$ is finite right from the
start at $t_0$.  Rather then reanalyzing this new situation `microscopically',
let us consider a substitute process where we start in the $b'$ phase and let
it decay in time (at the same point $(\tau,b_0)$ in phase space). Then, after
a time $t_a$, the jump location $x_{b'}^\mathrm{jp}(t_a)$ will reach
$x^\mathrm{jp}_a$ and from there on, the further decay of the $b'$ phase
traces the decay of the $a$ phase.  As a result, we find that
\begin{equation}\label{eq:alpha_a_b'} 
   \alpha_\mathrm{sp}^a (t) = \alpha_\mathrm{sp}^{b'} (t+t_a)
\end{equation}
with the waiting time $t_a$ given by the usual estimate $T \ln (t_a/t_0)
\approx U_\mathrm{dp}(x_a^\mathrm{jp})$ or
\begin{equation}\label{eq:t_a}
   t_a \approx t_0 \exp[U_\mathrm{dp}(x_a^\mathrm{jp})/T].
\end{equation}
For the case where we enter the $a$ phase from the $b$ phase, we have to shift
$\alpha_\mathrm{sp}^{b}$ instead, $\alpha_\mathrm{sp}^a (t) =
\alpha_\mathrm{sp}^{b} (t+t_a)$, and substitute $U_\mathrm{dp} (x^\mathrm{jp}_a)$ by
$U_\mathrm{p}(x^\mathrm{jp}_a)$. In general, when $x_0 < x_a^\mathrm{jp} < x_+$,
we shift $\alpha_\mathrm{sp}^{b'}$, while $\alpha_\mathrm{sp}^{b}$ is to be shifted
when $x_- < x_a^\mathrm{jp} < x_0$.

Translating the seemingly trivial linear-in-time shift $t \to t + t_a$ to the
$\log(t/t_0)$ plot of Fig.\ \ref{fig:FC-rel-vs-t} produces an interesting
outcome, see the orange line in the middle panel. With $x^\mathrm{jp}_a <
x_+$, we have a smaller force jump $\Delta f_\mathrm{pin}$ and hence
$\alpha^a_\mathrm{sp}$ starts out at a lower value then the
$\alpha^{b'}_\mathrm{sp}(t)$ curve, $\alpha^a_\mathrm{sp}(t_0) <
\alpha^{b'}_\mathrm{sp}(t_0)$. Next, the slope $\partial_{\log(t/t_0)}
\alpha^a_\mathrm{sp}\big|_{t}$ at small times $t \ll t_a$ relates to the slope of
$\alpha^{b'}_\mathrm{sp}$ at $t_a$, $\partial_{\log(t/t_0)}
\alpha^{b'}_\mathrm{sp}\big|_{t_a} \equiv - \alpha'$, via
\begin{eqnarray}\label{eq:slope_a^a}
   \partial_{\log(t/t_0)} \alpha^a_\mathrm{sp}\big|_{t} &=& 
   t\, \partial_{t} \alpha^a_\mathrm{sp}\big|_{t}  \\ \nonumber
   &=& \frac{t}{t_a}\, \big[t\, \partial_{t} \alpha^{b'}_\mathrm{sp}\big]_{t_a} 
   = - \frac{t}{t_a} \alpha'
\end{eqnarray}
and hence is small by the factor $t/t_a \ll 1$. As a result, we find that
$\alpha^a_\mathrm{sp}(t)$ evolves flat in $\log(t/t_0)$ and then bends over to
$\alpha^{b'}_\mathrm{sp}(t)$ at $t \sim t_a$, see Fig.\ \ref{fig:FC-rel-vs-t},
within a $\log$-time interval of unit size or a $\mathcal{T}$-interval of
order $T/e_p$, which is small on the extension $\mathcal{T} \approx 1$ of the
creep parameter.

In an experiment, where the relaxation of the Campbell curvature (or length)
is plotted versus $\log$-time, the $a$ phase will start out with a seemingly
slow decay (a flat curve) as compared to the decay of the $b'$ phase, see
Fig.\ \ref{fig:FC-rel-vs-t}. This is owed to the vanishing of the barrier for
the $b'$ phase at small times, hence the $b'$ phase decays much faster than
the $a$ phase. Once the waiting time $t_a$ is reached, the decay of the $b'$
phase has slowed down such as to catch up with the decay of the $a$ phase.
The detection of the $a$ phase in an experiment then depends on its time
resolution: this will be successful if $t_a$ resides within the observable
time window of the relaxation experiment. If $t_a$ is too large (note that
$t_a \propto \exp[U(x_a^\mathrm{jp})/T]$ exponentially depends on the barrier
and the temperature $T$) as compared to the time window of the measurement,
only the flat part of the curve will be observed, with apparently no
relaxation of the Campbell length. On the other hand, if $t_a$ is too short to
be caught by the experiment then one will resolve a phase $b'$ type relaxation
and the feature appertaining to the $a$ phase (flat part) is lost. The
absence of relaxation in the Campbell length observed \cite{Prozorov_2003} in
a BiSCCO sample finds a simple explanation in terms of large barriers that are
present in the $a$ phase of the hysteresis loop.

\section{Summary and outlook}
\label{sec:sum}

Strong pinning theory delivers a quantitative description of vortex pinning in
the dilute defect limit. This fact is particularly prominent in the context of
the Campbell ac response: not only can we describe a multitude of different
vortex states, the zero-field cooled state and various types of field cooled
states, we also can accurately trace the time evolution of these states and
their signatures in Campbell penetration depth measurements. The strong
pinning theory thus provides access to hysteretic and relaxation effects in
the ac response that are otherwise, e.g., via weak collective pinning theory,
at least so far, not available.

In this work, we have studied the effects of thermal fluctuations at finite
temperatures $T$, or creep, on the Campbell penetration depth
$\lambda_\mathrm{\scriptscriptstyle C} \propto 1/\sqrt{\Delta f_\mathrm{pin}}$
that tracks the force jumps $\Delta f_\mathrm{pin}$ in the strong pinning
landscape of Fig.\ \ref{fig:strong_pinning}. The proportionality
$\alpha_\mathrm{sp} \propto \Delta f_\mathrm{pin}$, first found in Ref.\
\onlinecite{Willa_2015_PRL}, provides a satisfying connection to the curvature
$\alpha$ appearing in Campbell's original \cite{Campbell_1969}
phenomenological description: the jump $\Delta f_\mathrm{pin}$ effectively
averages the curvatures in the pinning landscape.  Remarkably, ac penetration
experiments provide new information on the pinning landscape, different from
standard critical current density $j_c \propto \Delta e_\mathrm{pin}$
measurements that tell about the jumps $\Delta e_\mathrm{pin}$ in energy.

In our analysis of the zero-field cooled (ZFC) state, we found an interesting
relaxation behaviour of the Campbell curvature $\alpha_\mathrm{sp}(t,T)$ (or
penetration depth $\lambda_\mathrm{\scriptscriptstyle C} \propto
1/\sqrt{\alpha_\mathrm{sp}}$) with a non-monotonous time-evolution at medium
to large values of the pinning parameter $\kappa$, increasing first at small
waiting times $t$ and then decreasing towards a finite equilibrium value
$\alpha_0 > 0$.  At small values of $\kappa -1 \ ll 1$, the marginal strong
pinning situation, we found the curvature $\alpha_\mathrm{sp}(t,T)$ rising
monotonously; numerical analysis shows that non-monotonicity appears at still
rather small values of $\kappa \approx 2$.  The decay to a finite value
$\alpha_0$ in the Campbell curvature is very different from the decay to zero
of the persistent current density $j(t,T)$, a fact owed to the different
limits of $\Delta f_\mathrm{pin} > 0$ and $\Delta e_\mathrm{pin} = 0$ at the
branch crossing point $x_0$, see Fig.\ \ref{fig:strong_pinning}.

The relaxation of the field cooled (FC) states provides a rich variety of
results as well: first of all, we find numerous types of hysteresis loops,
depending on the characteristics of the defects, with insulating and metallic
point-like defects, $\delta T_c$- and $\delta \ell$-pinning studied in more
detail here, see also Ref.\ \onlinecite{Willa_2016}. This different behavior
is owed to the competition between an increasing $\kappa(T)$ and a decreasing
$\xi(T)$ as the temperature $T$ is decreased, with the scaling of $\kappa(T)$
determined by the type of pinning. Depending on the relative motion between
the bistable interval $[x_-,x_+]$ (with $x_- \sim \kappa^{1/4} \xi$ and $x_+
\sim \kappa \xi$ at large $\kappa$) and the initial instability point
$x_\mathrm{0{\scriptscriptstyle L}} = x_0(T_{\scriptscriptstyle L})$ upon
decreasing $T$, the jump location $x^\mathrm{jp}$ gets pinned at the edges
$x_\pm(T)$ or stays put somewhere in between---these three cases define the
phases $b$ and $b'$, as well as the $a$ phase, that appear in the hysteresis
loop when cycling the temperature down and up. The appearance of these phases
within a loop again depends on the defect type, with insulating defects and
$\delta\ell$-pinning exhibiting all phases in the sequence $b$ (cooling) to
$a$ (heating) to $b'$ (heating), while the loops for $\delta T_c$-pinning and
metallic defects are dominated by the $a$ phase.  The three phases behave
quite differently, with the $b'$ phase providing a smaller penetration depth
$\lambda_\mathrm{\scriptscriptstyle C}$ at large $\kappa$ (where
$\alpha_\mathrm{sp}^{b} /\alpha_\mathrm{sp}^{b'} \propto 1/\kappa^{3/2}$),
while in terms of creep, the $a$ phase sticks out by its slower decay.
Further experimental signatures for these phases are the decay (increase) in
magnitude of $\lambda_\mathrm{{\scriptscriptstyle C}}$ under creep for the $b$
($b'$) phases and a plateau, i.e., an initially much slower relaxation (both
up or down is possible) in the $a$ phase due to the presence of large thermal
barriers.  Such characteristic differences then allow to make conjectures
about the underlying pinning landscape.

In the present work, we have made an additional step towards better precision
in our analytic results.  Using the Lorentzian-shaped potential describing a
point-like defect as an example, we have provided analytic results including
numerical factors; this further illustrates the value of the strong pinning
concept as a quantitative theory. Furthermore, care has been taken to properly
treat the trapping geometry of strong pinning, see Fig.\ \ref{fig:geometry}.
It turns out, that this geometry affects transport and ac response in
different ways: while in transport only the transverse trapping length $t_\perp
= 2 x_-$ shows up, when dealing with the (ZFC) ac response, pinning (involving
a semi-circle of radius $x_-$) and depinning (at a circular segment of radius
$x_+$) are weighted with separate factors.  Furthermore, in the FC situation,
trapping always appears on a circle with a radius $R \in [x_-, x_+]$ spanning
the entire bistable region, depending on the induced vortex state; this
feature has been missed in our previous analysis \cite{Willa_2016}.

Several of the predictions made in the present work have been observed in
experiments measuring the Campbell penetration length. Examples are the
decreasing $\lambda_\mathrm{{\scriptscriptstyle C}}(t)$ in a BiSCCO sample
\cite{Prozorov_2003} that is consistent with an increasing Campbell curvature
at short times or marginally strong pinning, see Fig.\
\ref{fig:rescaled_curvatures}, the increasing
$\lambda_\mathrm{{\scriptscriptstyle C}}(t)$ in an YBCO superconductor
\cite{Pasquini_2005} that is consistent with the long time behavior of the
Campbell curvature at intermediate and very strong pinning, see again Fig.\
\ref{fig:rescaled_curvatures}, the finite equilibrium value of the Campbell
length $\lambda_\mathrm{{\scriptscriptstyle C}0}$, see Eq.\
\eqref{eq:alpha_0_Lor_gen}, that has been observed in a BiSCCO sample above
the irreversibility line, and the absence of creep in the field cooled
(FC) state of a BiSCCO single crystal \cite{Prozorov_2003}, here explained in
terms of an $a$ phase that is characterized by the presence of large barriers,
see also Fig.\ \ref{fig:FC-rel-vs-t}.

In the present study, we have focused on the low-field regime where the
trapping length $x_+$ stays below the vortex lattice constant, $x_+ < a_0/2$.
At larger field values, multiple vortices start competing for the same defect
and our single-pin--single-vortex description has to be extended to include
several vortices, see also Refs.\ \onlinecite{Willa_2018a, Willa_2018b}. This
becomes particularly relevant in very anisotropic and layered material with
$\varepsilon \ll 1$, where $\kappa \propto 1/\varepsilon$ can become large and
$x_+ \sim \kappa \xi$ easily goes beyond $a_0/2$ even at moderate field values
already. Related to the possibility of such very strong pinning is the
prediction \cite{Gaggioli_2022} of a creep-{\it enhanced} critical current due
to a dominant increase in the trapping area at very large $\kappa$.
Furthermore, since pinning remains active also beyond \cite{Thomann_2017}
$j_c$ it would be interesting to measure and analyze the Campbell penetration
physics in the dynamical vortex state.

\section*{Acknowledgments} 

We thank Martin Buchacek and Roland Willa for discussions and acknowledge
financial support of the Swiss National Science Foundation, Division II.

\bibliographystyle{apsrev4-1}
\bibliography{refs_vortices}

\end{document}